\documentclass[aps,prapplied,reprint,superscriptaddress]{revtex4-2}

\usepackage{amsmath}
\usepackage{bm}
\usepackage{amsfonts}
\usepackage{newtxtext}
\usepackage[varvw]{newtxmath}
\usepackage{graphicx}

\usepackage{comment}
\usepackage{xcolor}
\usepackage{siunitx} 
\DeclareSIUnit[quantity-product = ]\percent{\char`\%}
\usepackage{booktabs,dcolumn}
\usepackage[colorlinks=true,allcolors=blue,bookmarks=true,bookmarksnumbered=true,breaklinks=true]{hyperref}

\newcommand{\kB}{k_\mathrm{B}}
\newcommand{\Sr}{${}^{88}\mathrm{Sr}^+$}
\newcommand{\Dfh}{$D_{5/2}$} 
\newcommand{\Dth}{$D_{3/2}$} 
\newcommand{\Soh}{$S_{1/2}$} 
\newcommand{\Srtrans}{$5s\,{}^2\!S_{1/2} \rightarrow 4d\,{}^2\!D_{5/2}$}

\begin{document}

\title{\texorpdfstring{$^{\mathbf{88}}${Sr}$^{\mathbf{+}}$}{88Sr+} optical clock with \texorpdfstring{$\mathbf{7.9\times 10^{-19}}$}{7.9e-19} systematic uncertainty and measurement of its absolute frequency with \texorpdfstring{$\mathbf{9.8\times 10^{-17}}$}{9.8e-17} uncertainty}

\author{T.~Lindvall}
\email[]{thomas.lindvall@vtt.fi}

\author{T.~Fordell}

\author{K.~J.~Hanhijärvi}

\affiliation{VTT Technical Research Centre of Finland Ltd, National Metrology Institute VTT MIKES, P.O.\ Box 1000, FI-02044 VTT, Finland}

\author{M.~Dole\v{z}al}
\affiliation{Czech Metrology Institute, V Botanice 4, 150\,72 Prague, Czech Republic}

\author{J.~Rahm}

\author{S.~Weyers}
\affiliation{Physikalisch-Technische Bundesanstalt, Bundesallee 100, 38116 Braunschweig, Germany}

\author{A.~E.\ Wallin}
\affiliation{VTT Technical Research Centre of Finland Ltd, National Metrology Institute VTT MIKES, P.O.\ Box 1000, FI-02044 VTT, Finland}

\date{\today}

\begin{abstract}
We report on a \Sr\ single-ion optical clock with an estimated fractional systematic uncertainty of \num{7.9e-19}. The low uncertainty is enabled by small rf losses, a thorough evaluation of the blackbody-radiation temperature, and our recent measurement of the differential polarizability.  
A detailed uncertainty evaluation is presented. We also report on two absolute frequency measurements: one against a remote cesium fountain clock and one against International Atomic Time (TAI). 
The former lasted 12~days and resulted in a frequency value of \num{444779044095485.49(15)}\;Hz. 
The latter spanned ten months with monthly optical-clock uptimes between 68\% and 99\% and yielded a frequency value of \num{444779044095485.373(44)}\;Hz. With a fractional uncertainty of \num{9.8e-17}, it is, to our knowledge, the most accurate optical frequency measurement reported to date. Both frequency values are in agreement with other recent measurements, providing further evidence that the 2021 CIPM recommended frequency value is too high by 1.6 times its uncertainty.

Accepted to Phys.~Rev.~Applied, DOI: \href{https://doi.org/10.1103/czlf-bfvp}{10.1103/czlf-bfvp}.
\end{abstract}

\maketitle

\section{Introduction \label{sec:intro}}

Optical atomic clocks \cite{Ludlow2015b}, with record estimated systematic uncertainties below $10^{-18}$ \cite{Brewer2019a,Aeppli2024a,Marshall2025a,Zhang2025b}, have applications in fundamental physics such as tests of relativity \cite{Delva2017b,Sanner2019a,Lange2021a}, searches for variations of fundamental constants \cite{Lange2021a,Sherrill2023a}, and searches for dark matter \cite{Roberts2020a,Kennedy2020a,Beloy2021a,Kobayashi2022a,Filzinger2023a}. Resolving the gravitational red shift \cite{Chou2010a} has a field application in relativistic (chronometric) geodesy~\cite{Grotti2018a,Takamoto2020a,GRotti2024a}.

As optical clocks have systematic uncertainties two orders of magnitude lower than those of the best cesium-fountain microwave clocks, a roadmap towards a redefinition of the SI (Syst\`{e}me International) second using optical clocks has been prepared \cite{Dimarcq2024a}. It lists criteria that must be fulfilled before a redefinition can take place. The criterion on uncertainty budgets (criterion I.1) requires multiple optical clocks with evaluated uncertainties ${\lesssim}\num{2e-18}$. Among ion clocks, this uncertainty target has so far been met only by $^{27}$Al$^+$ \cite{Brewer2019a,Marshall2025a}, while $^{171}$Yb$^+$(E3) \cite{Sanner2019a,Tofful2024a} and $^{115}$In$^+$ \cite{Hausser2025a} clocks have reported uncertainties slightly above it. Among optical lattice clocks, there are Yb \cite{McGrew2018a} and Sr \cite{Bothwell2019a,Aeppli2024a,Lu2025a} clocks meeting the target. However, the criterion to have three clocks based on the same transition, in different institutes,  with uncertainties ${\lesssim}\num{2e-18}$ (criterion I.1.a) is yet to be fulfilled. Other mandatory criteria require absolute frequency measurements with uncertainties ${<}\num{3e-16}$ (criterion I.3) and regular TAI calibrations by optical clocks with uncertainties ${\lesssim}\num{2e-16}$, excluding the uncertainty of the recommended frequency value, (criterion I.4).
 
The \Sr\ ion clock transition, one of the secondary representations of the second (SRS), offers several advantages: the required laser system is simple, efficient schemes for canceling the electric quadrupole shift \cite{Dube2005a,Lindvall2022a} and the micromotion shifts \cite{Dube2014b} exist, the quadratic Zeeman shift is extremely small \cite{Gan2018a}, and we have recently measured the differential static scalar polarizability with a fractional uncertainty of \num{4.1e-4} \cite{Lindvall2025b}. This makes it an attractive candidate for laboratory, transportable \cite{Dube2021a}, space-deployable  \cite{Spampinato2024a}, and even commercial optical clocks \cite{Fairbank2025a}. Recently, also multi-ion operation without \cite{Steinel2023a} and with dynamic decoupling \cite{Akerman2025a} has been demonstrated, as has interrogation using an integrated waveguide cavity \cite{Loh2025a}.

In this work, we show that with a well-characterized  blackbody-radiation (BBR) temperature, a reduced differential polarizability uncertainty \cite{Lindvall2025b}, and a low vacuum pressure, the \Sr\ clock can reach an uncertainty of \num{7.9e-19}.
This is the third lowest uncertainty reported to date after \cite{Marshall2025a,Zhang2025b}, more than a factor of ten lower than those of previous \Sr\ clocks \cite{Steinel2023a,Jian2023a},  and a factor of 2.5 lower than the redefinition target (criterion I.1) \cite{Dimarcq2024a}. With technical improvements, the uncertainty could be further reduced to about \num{6e-19}. We also demonstrate self-comparison instabilities down to $\num{2.0e-15} \tau^{-1/2}$, limited by the performance of the clock laser and with no additional instability due to background magnetic-field noise, which has frequently been an issue in clocks based on the alkaline-earth-metal ions \cite{Barwood2015a,Steinel2023a,Zeng2023c}. 

We also report on absolute frequency measurements of the \Srtrans\ clock transition against the cesium fountain clock CSF2 at Physikalisch-Technische Bundesanstalt (PTB, Germany) \cite{Weyers2018a} using an Integer Precise Point
Positioning (IPPP) link \cite{Petit2015b} and against International Atomic Time (TAI). Spanning ten months with monthly optical-clock uptimes between 68\% and 99\%, the latter yields an uncertainty of \num{9.8e-17}, which, to our knowledge, is the lowest reported uncertainty to date and a factor of three lower than the roadmap target (criterion I.3).
In fact, only a handful of optical frequency measurements with uncertainties ${\lesssim}\num{2e-16}$ have been published  \cite{McGrew2019a,Schwarz2020a,Nemitz2021a,Lange2021a,Kobayashi2025a,Marceau2025a}. 
Our measurement includes seven months with TAI calibration uncertainties ${\lesssim}\num{2e-16}$, the roadmap target  (criterion I.4). 
Our absolute frequency values are in good agreement with other recent high-accuracy measurements \cite{Steinel2023a,Marceau2025a}, indicating that the 2021 CIPM (Comit\'e International des Poids et Mesures) recommended frequency value is too high by 1.6 times its uncertainty.

The paper is organized as follows: Section~\ref{sec:clock} describes the experimental setup and the working principle of the optical clock, Sec.~\ref{sec:syst} details the evaluation of the systematic shifts and uncertainty budget, and Sec.~\ref{sec:absfreq} describes the absolute frequency measurements. Section~\ref{sec:conclusions} summarizes the results.

\section{The \texorpdfstring{$\mathbf{^{88} Sr^+}$}{88Sr+} optical clock} \label{sec:clock}

\subsection{Experimental setup}

The optical clock is based on a single \Sr\ ion trapped in an ion trap of the endcap type, described in detail in \cite{Lindvall2022a}. It is designed to minimize rf heating  and thus the uncertainty of the BBR shift, which dominates the uncertainty of many optical clocks. A tunable helical resonator \cite{Lindvall2022a} is used to provide rf to the trap, which is typically operated at an rf voltage amplitude of 350--390\;V and a frequency of 14.4\;MHz. This results in secular frequencies of $\omega_x/2\pi \approx \omega_y/2\pi \approx \qty{1}{MHz}$ and $\omega_z/2\pi \approx \qty{2}{MHz}$ in the radial  and axial directions, with a radial splitting of $(\omega_y-\omega_x)/2\pi \approx \qty{50}{kHz}$. The trap is mounted in a compact aluminum vacuum chamber pumped by a combination pump containing a small sputter ion pump and a large non-evaporable getter (NEG) element  (SAES Getters NEXTorr D-200), resulting in pressure in the low $10^{-11}$\;mbar range. 

A three-layer magnetic shield encloses the trap. The attenuations of the middle and outer shields along the $(X, Y, Z)$ chamber axes, see Fig.~\ref{fig:BBR}(a) for coordinate system, were determined to be $(1330, 730, 380)$ by correlating the field magnitude measured by the ion with the external field measured by a three-axis magnetometer during an anomaly in the Earth's magnetic field on MJD 60372, in which all components of the field varied by hundreds of nanotesla over six hours. The combined attenuations of all three shields were measured by applying a modulated external field along each axis using a pair of large square-shaped coils. This yielded total attenuations on the order of $(80\,000, 16\,000, 12\,000)$.
The inside of the innermost shield is painted with high-emissivity paint to absorb scattered light and provide a well-defined blackbody radiation temperature.

The trap is loaded from a beam of neutral Sr atoms from a commercially available oven (Alvatec Alvasource AS-2-Sr-45-F).
Inside the trapping volume, the atoms are photoionized in a resonant two-step process \cite{Brownnutt2007a}. Here, the neutral atoms are first excited on the $5s^2 \,{}^1\!S_0 \rightarrow 5s5p\,{}^1\!P_1$ transition at 461\;nm using a frequency-doubled single-mode distributed Bragg reflector (DBR) laser (the reference laser in \cite{Fordell2019a}). This laser can be frequency referenced to an identical Sr oven in a small auxiliary vacuum cell, but no active frequency stabilization is used. From the $5s5p\, {}^1\!P_1$ state, the atoms are excited to the broad, autoionizing $(4d^2 + 5p^2) \,{}^1\!D_2$ state \cite{Haq2006a} using a free-running multimode diode laser at 405\;nm. With an oven current of 1.75\;A, it typically takes a few minutes to load a single ion. 

\begin{figure}[tb]
\includegraphics[width=0.85\columnwidth]{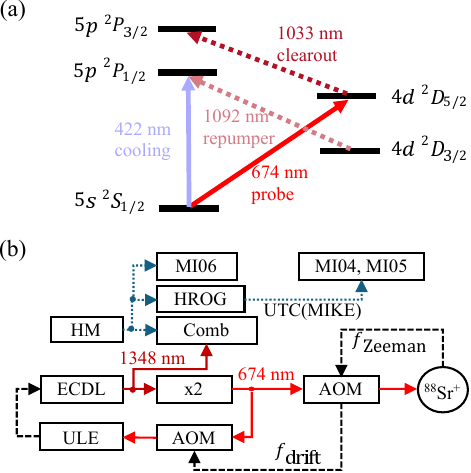}
\caption{(a) Partial energy level scheme of \Sr. Solid arrows indicate lasers, dashed arrows ASE sources.
(b) Clock-laser setup and frequency chain to the \Sr\ ion, frequency comb, hydrogen maser (HM), and geodetic GNSS receivers MI04--MI06 for time transfer. HROG---high resolution offset generator. Solid (dotted) lines indicate optical (rf) signals, while dashed lines indicate feedback loops for PDH locking (left), drift compensation of the cavity ($f_\text{drift}$, middle), and tracking the Zeeman components of the clock transition ($f_\text{Zeeman}$, right). AOMs for fiber noise cancellation are not shown.
\label{fig:energy_chain}}
\end{figure}

The ion is Doppler cooled on the $5s \,{}^2\!S_{1/2} \rightarrow 5p\,{}^2\!P_{1/2}$ transition, see Fig.~\ref{fig:energy_chain}(a), using a \qty{422}{nm} external-cavity diode laser (ECDL, Toptica DL pro). It is frequency stabilized to a rubidium transition, detuned from the cooling transition by 436\;MHz \cite{Shiner2007a}, using modulation transfer spectroscopy in a vapor cell. 
The cooling and state preparation (spin polarization) light is provided by two separate laser beams controlled by independent double-pass acousto-optic modulators (AOMs), while a common mechanical shutter (Stanford Research Systems SR475) ensures that the \qty{422}{nm} light is fully extinguished during clock interrogation. 
For repumping on the \qty{1092}{nm} $4d\,{}^2\!D_{3/2} \rightarrow 5p\,{}^2\!P_{1/2}$ transition and clock-state clearout on the \qty{1033}{nm} $4d\,{}^2\!D_{5/2} \rightarrow 5p\,{}^2\!P_{3/2}$ transition,  unpolarized, broadband amplified-spontaneous-emission (ASE) light sources \cite{Lindvall2013a,Fordell2015a} 
are used. They require no frequency stabilization and destabilize dark states \cite{Lindvall2012a} without external polarization modulation. In addition to the electronic switching demonstrated in \cite{Fordell2015a}, a fiber-coupled MEMS switch (Thorlabs OSW12-980-SM) has been added to reduce the fall time of the repumper, making its light shift negligible while reducing the dead time of the clock.

Ion fluorescence is detected using an objective (numerical aperture $\text{NA} = 0.26$) close to the antireflection-coated fused silica window of the vacuum system. A 70/30 beam splitter divides the collected photons between a photomultiplier tube (PMT, Hamamatsu H10682-210, total photon collection efficiency ${\approx}1/550$) and an sCMOS camera (ZWO ASI290MM). A \qty{400}{\micro\meter} pinhole in front of the PMT improves the signal-to-noise ratio (SNR) for ion state detection. A \qty{420}{nm} bandpass filter (Semrock Brightline 420BW10) in the detection objective enables fluorescence detection with good SNR during loading despite the intense photoionization light at \qty{405}{\nano\meter}.

The clock (probe) laser is a \qty{1348}{nm} ECDL (Toptica DL pro), which is frequency-doubled to 674\;nm and stabilized to a \qty{30}{cm} ultra-low-expansion (ULE) glass cavity of a design similar to that in \cite{Hafner2015a} using Pound-Drever-Hall (PDH) locking, see Fig.~\ref{fig:energy_chain}(b). The flicker floor of the clock laser is \num{\approx 4e-16} for integration times of 1--300\;s based on a comparison with a vertical \qty{18}{cm} ULE cavity. 
The \qty{30}{cm} cavity has a drift of around $+20\;\mathrm{mHz/s}$. The light to the ion and frequency comb is de-drifted  by a DDS-controlled (Direct
Digital Synthesizer) AOM based on feedback from the ion during clock operation. When feedback from the ion is not available, for example during ion temperature measurements and other diagnostics, the de-drifting can be done based on a moving fit of the laser frequency against a hydrogen maser (HM) or with a constant drift rate.

The laser frequency depends on the intracavity power, which is stabilized using the drive power of the drift-compensation AOM with feedback from a photodetector measuring the cavity transmission, where the sensitivity is \qty{27}{\Hz/\micro\W}.  
Fiber noise cancellation is used for the polarization-maintaining fibers from the cavity to the ion (\qty{674}{\nano\meter}) and from the cavity to the frequency comb (\qty{1348}{\nano\meter}). The cavity is operated at the zero-crossing temperature for thermal expansion, \qty{+32.0(2)}{\celsius}.
The frequency chain from the clock laser to the ion, frequency comb, hydrogen maser, and GNSS (global navigation satellite system) receivers is schematically shown in Fig.~\ref{fig:energy_chain}(b).

\subsection{Interrogation sequence} \label{sec:interr}

Figure~\ref{fig:timing}(a) shows the interrogation  pulse sequence as measured using a photodetector. After the probe pulse $n$ of duration $\tau_\mathrm{p}$, the cooling laser and repumper are switched on. Following a \qty{\approx3}{\milli\second} delay determined by the rise times of the cooling-laser mechanical shutter and the repumper,  fluorescence is counted for \qty{5}{\milli\second} to determine whether the clock excitation was successful. {Typical bright (dark) counts are 40 (0.1) photons/5\;ms, and the detection threshold is set close to the geometric mean of these. Then, a \qty{5}{\milli\second} clearout pulse is applied to return the ion to the Doppler-cooling cycle, partly overlapping with a second \qty{5}{\milli\second} state detection, which verifies whether the fluorescence has returned. If the ion is still dark, the servo cycle is invalidated. 

\begin{figure}[tb]
\includegraphics[width=1\columnwidth]{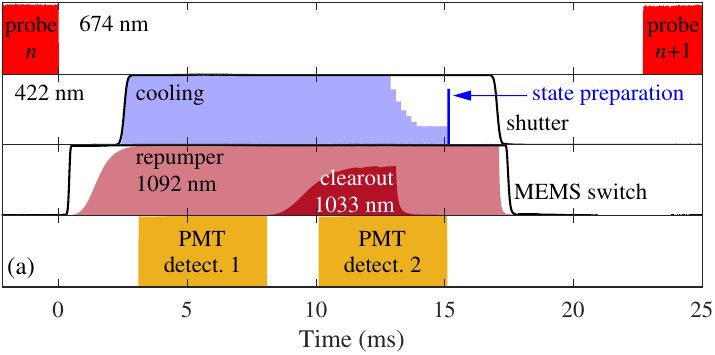}\\
\vspace{3pt}
\includegraphics[width=1\columnwidth]{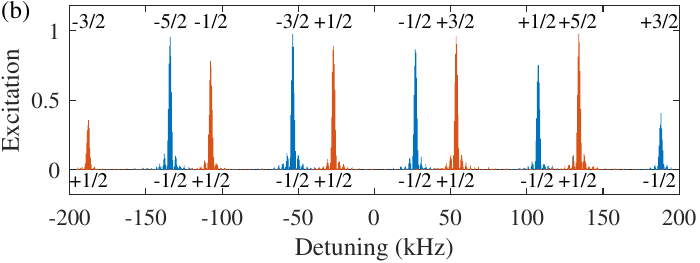}\\
\vspace{2pt}
\includegraphics[width=1\columnwidth]
{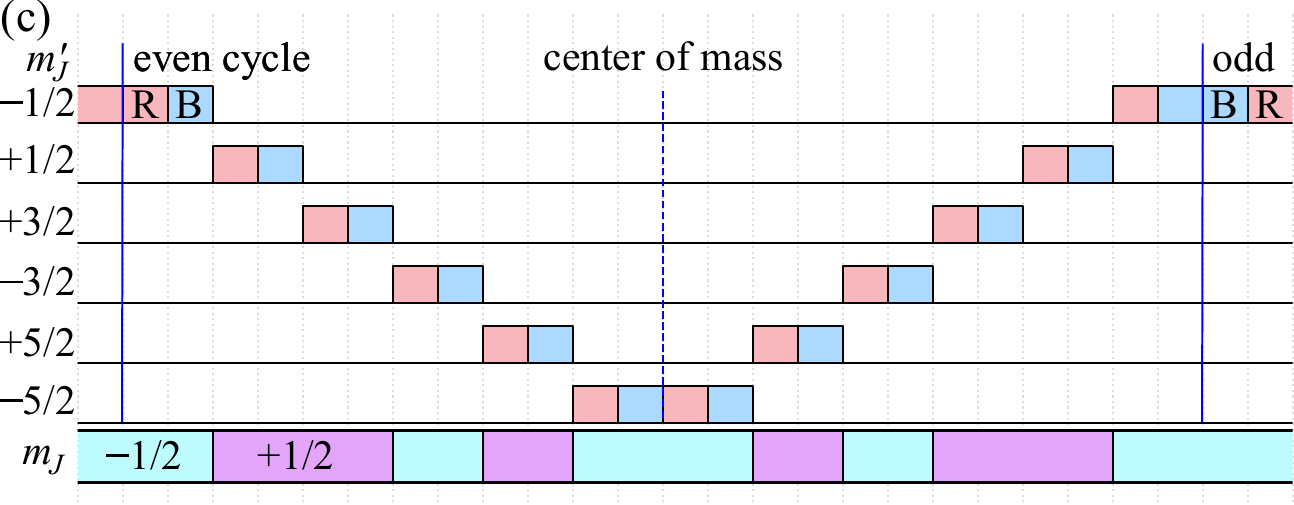}
\caption{(a) Interrogation pulse sequence measured using a fast photodetector. Light pulses and the PMT detection windows are plotted as colored areas (with arbitrary amplitude), while the solid lines indicate the transmission of the mechanical shutter and MEMS switch.
(b) Measured Zeeman spectra at a magnetic field of \qty{4.8}{\micro\tesla} with state preparation into either \Soh\ sublevel $m_J = \pm 1/2$, as labeled below the spectra. The corresponding \Dfh\ sublevels $m_J'$ are labeled above.
(c) Servo-cycle sequence for probing the red (R) and blue (B) sides of the Zeeman transitions $|S_{1/2}, m_J\rangle\rightarrow|D_{5/2}, m_{J}'\rangle$. Each of the 24 R or B interrogations consists of the pulse sequence in~(a).
\label{fig:timing}}
\end{figure}

At the end of the cooling period, the cooling laser power is stepped down using the rf power of the AOM to give a lower ion temperature. It is then switched off and the state-preparation beam, whose polarization can be controlled using a liquid-crystal phase retarder (Thorlabs LCC1413-A), is switched on for \qty{0.1}{\milli\second} with $\sigma^+$ or $\sigma^-$ polarization to transfer the ion into the desired ground-state sublevel with an efficiency of ${\approx}99\%$.
The \qty{422}{nm} mechanical shutter is then closed followed by switching off the repumper current and MEMS switch. A \qty{5}{\milli\second} delay before the next probe pulse $n+1$ ensures that the cooling and repumper light is fully extinguished and also that the small vibrations and magnetic field transient caused by the mechanical shutter have decayed. This sequence results in a dead time $\tau_\mathrm{d}=22.7\;\mathrm{ms}$ between probe pulses.

Figure~\ref{fig:timing}(b) shows clock-transition Zeeman spectra with state preparation into either ground-state sublevel, measured using a constant probe laser power and a probe time of \qty{0.37}{ms}.  
To cancel the linear Zeeman shift as well as the electric quadrupole shift and other tensor shifts \cite{Dube2005a}, we lock independent servos to six $|S_{1/2}, m_J\rangle\rightarrow|D_{5/2}, m_{J}'\rangle$ Zeeman components with $m_J \rightarrow m_J'$ equal to $\pm1/2 \rightarrow \pm1/2$, $\pm1/2 \rightarrow \pm3/2$, and $\pm1/2 \rightarrow \pm5/2$ (the first, second, and fourth pairs from the line center). Their frequencies are averaged at the end of each cycle for real-time processing, but are also stored individually for postprocessing.

To minimize the sensitivity to external magnetic-field variations, which are dominantly vertical, a nearly horizontal state preparation beam and applied magnetic field direction are used. For clock operation, a \qty{4.8}{\micro\tesla} magnetic field is applied with three orthogonal coil pairs and a precision constant-current driver (DM Technologies Multichannel Current Source). The field magnitude is chosen to keep all second-order (at $\pm (\omega_y - \omega_x)/2\pi \approx \pm 50$\;kHz) and third-order secular sidebands (at $\pm (\omega_z - \omega_x - \omega_y)/2\pi \approx \pm 100$\;kHz) more than \qty{1}{kHz} away from the interrogated transitions to ensure negligible coupling to the sidebands. 
The polarization direction of the clock-laser beam is chosen such that all six interrogated Zeeman components have similar strengths, see Fig.~\ref{fig:timing}(b), requiring ${<} 1$\;dB difference in the rf power to the AOM. 

During one servo cycle each Zeeman component is interrogated twice on the red and twice on the blue side of the transition. The 24 probe pulses are ordered in a forward-backward scheme \footnote{Martin Steinel (private communication).} where the red (blue) side is probed first for even (odd) servo cycles, see Fig.~\ref{fig:timing}(c). This has the benefit that the centers of mass of each Zeeman component coincide. It also provides rejection of servo errors,  while the R/B switching makes noise-induced servo errors negligible \cite{Lindvall2023a}.
The center frequency $f_0$ of each Zeeman component is tracked by probing the transition on the red (blue) side at $f_0-\Delta f/2$  ($f_0+\Delta f/2$) and determining the excitation probability  $p_\mathrm{R}$ ($p_\mathrm{B}$). Here $\Delta f \approx 0.8/\tau_\mathrm{p}$  is the FWHM for a Rabi line shape. After each servo cycle the center frequency is updated by
\begin{equation}\label{eq:servo}
f_0\rightarrow{f_0+g\frac{\Delta f}{3}\frac{p_\mathrm{B}-p_\mathrm{R}}{2p_0}},
\end{equation}
where $g$ is a nondimensional gain parameter, $\Delta f/3$ is the inverse of the line-shape slope at FWHM/2,  and $p_0$ is the average excitation probability determined from a moving average of servo cycles during the past \qty{900}{\second} \cite{Lindvall2023a}. A gain of $g=1$ would correct the frequency offset in a single step. Gain values $g=0.1\ldots1$ have been found to work well with a probe time of \qty{140}{ms}. For the current work, $g=0.3$ was used.

The clock is controlled using \textsc{artiq} \cite{Bourdeauducq2016a} software and Sinara hardware that ensures precise hardware timing. All data from a servo cycle are written to an InfluxDB database in real time, can be visualized with Grafana, and are archived as daily HDF5 files. 

The duration of unattended clock operation has been significantly increased by implementing an automatic recovery process if the ion fluorescence disappears. 
The cause of these dark-ion events is likely collisions with background molecules that leave the ion in a high motional  state, where the normal cooling pulse is insufficient to  recool the ion. When a dark ion is detected, the interrogation sequence is stopped and all cooling beams are turned on with a \qty{-76}{\mega\hertz} ($-3.5$ natural linewidths) red detuning and high intensity (\num{\approx 30} times the normal cooling and detection intensity of the main cooling beam). Ion fluorescence is usually recovered within \qty{10}{\second} and clock operation is resumed. These dark-ion events occur on average about four times a day,  but in some cases the fluorescence would likely return also without the recovery process. Since implementing the automatic recovery, the ion lifetime distribution resembles a log-normal distribution with values between 1\;d and 45\;d and a mean of 15\,d. The loss mechanism is believed to be formation of a molecular ion, most likely SrH$^+$, that is not dissociated by the laser wavelengths in use \cite{Wu2021b}.
Unattended clock operation is typically limited only by ion loss, which means that we achieve 99\% uptimes for periods of days to weeks. The 1\% downtime consists of micromotion minimization (see Sec.~\ref{sec:EMM}), the ion recovery events, and cycles invalidated by collisions.
By implementing automated ion reloading \cite{Tofful2024a}, the long-term uptime could be improved further.

\subsection{Instability}

The clock has been operated with probe times up to \qty{320}{ms}. Figure~\ref{fig:self}(a) shows self-comparison Allan deviations (ADEVs) (see Appendix~\ref{sec:self} for details) for probe times of \qty{137}{ms} and \qty{202}{ms}. Probe times around the former are typically used in order to be reasonably close to the quantum projection noise (QPN) limit. For longer probe times, the laser limits the instability, which does not decrease much below $\num{2.0e-15} \tau^{-1/2}$. 
During long measurements, fluctuations and drift in the electric quadrupole shift (EQS) at the mHz level stop the self-comparison from averaging down, but an EQS-free self-comparison solves this issue (Appendix~\ref{sec:self}).

\begin{figure}[tb]
\includegraphics[width=1\columnwidth]{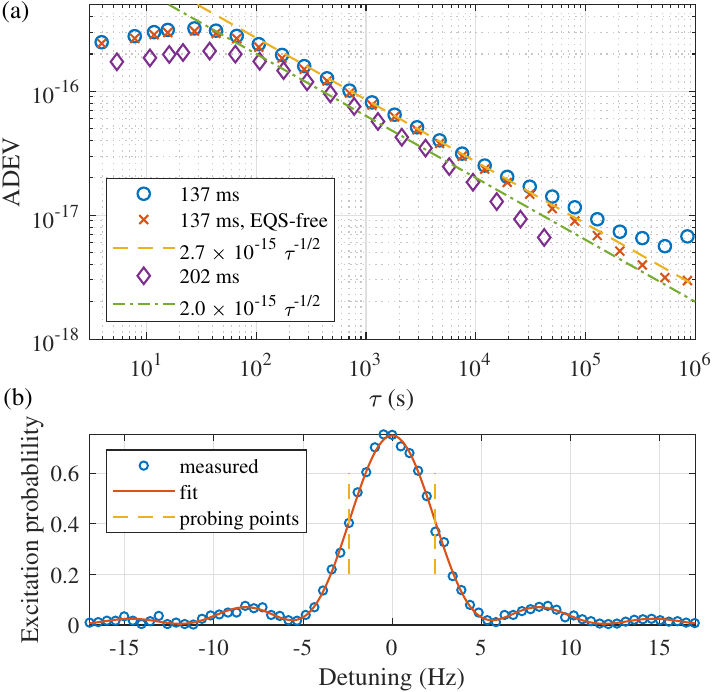}
\caption{(a) Measured clock self-comparison ADEV  (symbols) and $\tau^{-1/2}$ slopes (lines). (b) Measured (520 probe pulses per point) and fitted line shape with 165-ms probe time. The $\pm\Delta f/2$ probing points (B/R) are also shown.
\label{fig:self}}
\end{figure}

The instability of the individual Zeeman-pair line centers can be obtained from the line-center differences using a three-cornered hat calculation. Even for the longest probe times, there is no significant instability difference between the $\pm1/2 \rightarrow \pm1/2$ and $\pm1/2 \rightarrow \pm5/2$ pairs with magnetic-field sensitivities of \qty{5.6}{\hertz/\nano\tesla} and \qty{28}{\hertz/\nano\tesla}, respectively, indicating that magnetic field noise is not limiting the instability. This required a careful mitigation of ground loops. With a better clock laser, it should therefore be possible to reduce the instability to $\num{\approx 1.5e-15} \tau^{-1/2}$, close to the lifetime-limited instability of $\num{1.3e-15} \tau^{-1/2}$ \cite{Dube2015a}.

Figure~\ref{fig:self}(b) shows a clock-transition spectrum measured with a \qty{165}{ms} probe time. Here, servos tracked the $\pm1/2 \rightarrow \pm1/2$ Zeeman pair in order to provide drift compensation, while the spectrum of the $-1/2 \rightarrow -1/2$ transition was measured in an interleaved fashion. Also shown are the $\pm\Delta f/2$ probing points and a fit taking into account the ion temperature.

\section{Systematic frequency shifts} \label{sec:syst}

The evaluation of all known frequency shifts is presented in the following sections. Several of the shifts and uncertainties vary with the operational parameters, but typical values are summarized in an uncertainty budget in Table~\ref{tab:u-bud}. For an earlier, detailed \Sr\ systematics evaluation, see \cite{Dube2013a}.

\subsection{Electric quadrupole shift}

The electric quadrupole shift (EQS) is a tensor shift caused by the interaction between the electric quadrupole moment of the \Dfh\ clock state and the static electric field gradient at the position of the ion.
Our treatment of the EQS has been described in \cite{Lindvall2022a}. As the magnetic field direction is fixed by the state-preparation beam direction,
the trap dc voltage is used to minimize the shift so that $|\Delta\nu_{\mathrm{EQS}}(|m_J'|=5/2)| < 50$\;mHz. In \cite{Lindvall2022a}, a lower limit of 1070 was determined for the cancellation factor of the Zeeman-averaging EQS cancellation scheme \cite{Dube2005a} in an interleaved high-/low-EQS measurement. Since we have changed our interrogation sequence significantly since then, this measurement was repeated. With 45\;h of data, the frequencies of the two `clocks' with $\Delta\nu_{\mathrm{EQS}}(|m_J'|=5/2) = 10.6$\;Hz and $|\Delta\nu_{\mathrm{EQS}}(|m_J'|=5/2)| \leq 50$\;mHz showed a frequency difference of $-1.0(57)$\;mHz. This gives a lower limit for the EQS suppression factor of $10.6\;\mathrm{Hz}/5.7\;\mathrm{mHz} \approx  1900$, limited only by the statistical uncertainty.
In principle, the cancellation should be limited only by fluctuations in the electric field gradient and magnetic field direction on a time scale shorter than the servo cycle time (around 4\;s), but as these are difficult to assess, the experimental lower limit is used as a conservative estimate. The actual mean EQS is obtained from the clock data and can be used to evaluate the uncertainty, which is typically about $3\times 10^{-20}$ in fractional units.

\subsection{Excess micromotion} \label{sec:EMM}

Excess micromotion (EMM) can be caused by static stray electric fields that displace the ion from the rf null or by a phase difference between the endcap rf electrodes \cite{Berkeland1998a}. No signs of the latter have been observed in our trap due to its symmetric design.
The EMM causes a positive scalar quadratic Stark shift, $\Delta\nu_\mathrm{QS} = -\Delta\alpha_0/(2 h) \langle E^2 \rangle$, and a negative second-order Doppler shift,  $\Delta\nu_{\mathrm{D}2} = -(\nu_0/2) \langle v^2\rangle/c^2$,
which cancel at a `magic' rf frequency. Here, $h$ is the Planck constant, $E$ is the rf electric field, $\nu_0$ is the transition frequency, $c$ is the speed of light, and $v$ is the ion's velocity. We have recently measured the differential static scalar polarizability (DSSP) with low uncertainty, $\Delta\alpha_0 = -4.8314(20)\times 10^{-40}\;\mathrm{J\, m^2/V^2}$ \cite{Lindvall2025b}.

The EMM is measured and minimized using the rf photon correlation technique \cite{Keller2015a} with three noncoplanar cooling laser beams. The relation between the modulation index and the photon-correlation contrast is obtained by solving the time-dependent eight-level $S_{1/2}$--$P_{1/2}$--$D_{3/2}$ optical Bloch equations \cite{Lindvall2012a,Lindvall2013a} for an ion position sinusoidally modulated at the rf frequency. The three laser beams are not mutually orthogonal, but the rf electric field experienced by the ion can be obtained using the laser beam projections and the relative phases of the photon correlation signals. 

The EMM is minimized using Brent's root-finding algorithm for each beam individually, using voltage combinations that change the photon-correlation contrast for one beam only \cite{Lindvall2025b}. 
Typically, the rms rf electric field is $E_\text{rms} \lesssim 5\;\mathrm{V/m}$ after minimization for all three beams and increases by ${\lesssim}10\;\mathrm{V/m}$ per day during clock operation. With minimization every few days, a typical time-averaged rms field is \qty{\approx 20}{V/m}, corresponding to EMM shifts of $\pm\num{3e-19}$.

In a spherical Paul trap, the cancellation of EMM with an unknown direction is limited by the different magic frequencies for axial and radial motion. As in \cite{Dube2014b}, we operate at the mean of these, which to second order in the Mathieu $q_z$ parameter and for negligible Mathieu $a_i$ parameters is $\Omega_\mathrm{EMM} = \Omega_0^0 ( 1-15 q_z^2/2^{10})$, where $\Omega_0^0 = e/(m c)\sqrt{-h\nu_0/\Delta\alpha_0}$.
The minimum cancellation factor for EMM, defined as the ratio of the scalar Stark shift and the maximum residual shift, can then be estimated as $S = \left[ (9 q_z^2/2^9)^2 + (u_{\Delta\alpha_0}/\Delta\alpha_0)^2 \right]^{-1/2}$, which for $q_z=0.4$ is about 350. Even though the majority of the measurements presented here used a magic frequency based on the DSSP value from \cite{Dube2014a}, this still gives a suppression factor of $90$, which is sufficient to keep the EMM uncertainty at a negligible level even for less frequent minimization.

EMM also causes a tensor Stark shift, which is cancelled by the EQS cancellation scheme. The maximum shift is obtained with the rf field along the quantization axis, which for $m_J' = 5/2$ gives the ratio between the tensor and scalar shifts as $\Delta\nu_\mathrm{QS,t}/\Delta\nu_\mathrm{QS} = \alpha_2/\Delta\alpha_0 \approx -1.64$, where $\alpha_2 = 48.1(10)$\;a.u.~\cite{UDportal} is the tensor polarizability of the $D_{5/2}$ state. This corresponds to a negligible maximum tensor Stark shift uncertainty of $\num{\approx3e-21}$ for $E_\text{rms} = 20\;\mathrm{V/m}$.

\subsection{Thermal-motion shifts} \label{sec:thermal}

The thermal-motion shifts consist of the second-order Doppler shift due to secular motion and intrinsic micromotion and the quadratic Stark shift due to intrinsic micromotion~\cite{Dube2013a} 
\begin{subequations} \label{eq:thermal}
\begin{eqnarray} 
    \frac{\Delta\nu_{\mathrm{D}2,T}}{\nu_0} &=& -\frac{3 \kB T_\mathrm{ion}}{m c^2}, \\
   \frac{\Delta\nu_{\mathrm{QS},T}}{\nu_0} &=& -\frac{\Delta\alpha_0}{2 h\nu_0}  \frac{m\Omega^2}{e^2} 3 \kB T_\mathrm{ion}.
\end{eqnarray}
\end{subequations}
At the magic frequency, the micromotion shifts cancel (i.e., $\Delta\nu_{\mathrm{QS},T}$ cancels half of $\Delta\nu_{\mathrm{D}2,T}$) and the total thermal-motion shift is $\Delta\nu_T/\nu_0 \approx -3 \kB T_\mathrm{ion}/2 m c^2$, which for \Sr\ is $-\num{1.6e-18}\;T_\mathrm{ion}/\mathrm{mK}$. 
The ion temperature is the mean over the three vibrational modes, $T_\mathrm{ion} = (T_x+T_y+T_z)/3$. 

We measure the ion temperature using either carrier Rabi flopping, which only gives a mean temperature that is strongly weighted by the less strongly confined radial modes, or sideband-to-carrier ratios, which give the temperature separately for each mode. A typical value is \qty{0.7(2)}{mK}, where the uncertainty is dominated by the repeatability of the sideband-to-carrier temperatures, close to the Doppler limit of \qty{0.52}{mK}.

The effective ion temperature during interrogation depends on the initial ion temperature $T_\mathrm{ion,0}$ and the heating rate $dT_\mathrm{ion}/dt$ of the ion: $T_\text{ion} = T_\mathrm{ion,0} + (dT_\mathrm{ion}/dt) \tau_\mathrm{p}/2$. The delay between cooling and probing is the same during temperature measurements and clock operation and does not need to be included separately.

The heating rate of the ion is determined by measuring its temperature after a variable delay time. Between November 2020 and April 2024, the trap was typically operated at \qty{14.4}{MHz} and $q_z \approx 0.375$ and heating rates consistent with \qty{1.8(2)}{mK/s} were measured using carrier Rabi flopping. In November 2024, significantly higher heating rates around \qty{10}{mK/s} were observed for the same parameters. The heating rate was found to vary with the secular frequencies \cite{Lindvall2025b} and was likely caused by resonant heating when one of the radial secular frequencies coincided with an rf noise peak present in the dc bias voltages.
Because of this, a large heating rate of \qty{10(8)}{mK/s} is used to evaluate the thermal-motion shift for the 2024--2025 absolute frequency campaign, see Sec.~\ref{sec:TAI}.

Since then, the heating rate has been reduced to \qty{1.5(5)}{mK/s} by increasing $q_z$ to 0.422. Sideband-to-carrier ratios reveal that it is likely still dominated by technical noise heating the radial modes and not anomalous heating \cite{Brownnutt2015a}. This will be the topic of further studies.

\subsection{Blackbody radiation shift}

The electric field of the blackbody radiation (BBR) causes a Stark shift of the clock transition given by
\begin{equation} \label{eq:BBR}
\Delta \nu_\text{BBR} = -\frac{1}{2h} \langle E^2(T) \rangle \Delta\alpha_0 [1+\eta(T)],
\end{equation}
where the mean-square electric field is 
\begin{eqnarray}
\langle E^2(T) \rangle &=& \pi^2(\kB T)^4 / (15\varepsilon_0 \hbar^3 c^3) \nonumber \\
&\approx& (831.943\;\mathrm{V/m})^2 (T/300\;\mathrm{K})^4.
\end{eqnarray}
Here, $\kB$ is the Boltzmann constant and $\varepsilon_0$ the vacuum permittivity. 
The small dynamic correction can be evaluated as $\eta(T) = -0.008\,95(17)(T/295\;\mathrm{K})^2$ around our typical BBR temperature of 295\;K, see Appendix~\ref{sec:polarizability}.

For ion clocks, there are two main challenges in evaluating the BBR field seen by the ion: the ion trap, which spans a large solid angle around the ion, heats up due to Ohmic losses in conductors and dielectric losses in insulators, and the temperature of rf-carrying conductors cannot be measured using electrical sensors due to self-heating and interference. While the temperature of the vacuum chamber and other grounded parts is monitored continuously using calibrated Pt100 sensors, the effective BBR temperature seen by the ion was determined using a combination of thermal imaging and finite-element-method (FEM) simulations as described in \cite{Dolezal2015a,OC18-Guidelines}. 
In this method, (i) the temperature rise of the trap and chamber parts caused by the rf voltage is measured using a combination of sensors and thermal imaging, (ii) the model parameters in the FEM simulation are adjusted around typical values until the temperature distribution agrees with the measured one, and (iii) the temperature rise seen by the ion is determined from the temperature of a small ideal blackbody sphere at the position of the ion.
As detailed below, several measures were taken to allow reaching a lower temperature uncertainty than in previous work using this technique \cite{Dolezal2015a,Nisbet-Jones2016a,Nordmann2020a}. All the measurements were performed on the actual operational trap, not a dummy trap.

Inspired by a design from the National Physical Laboratory (NPL, UK) \cite{Nisbet-Jones2016a}, the trap was designed to have low intrinsic Ohmic and dielectric losses. Preliminary measurements verified that the heating of the trap was dominated by heat from the secondary coil of the helical resonator. To minimize the heat flow from the resonator to the rf vacuum feedthrough, these are connected by \qty{100}{\micro\meter} copper sheet, which provides good rf electric conductance but has significantly lower thermal conductance than a bulk copper connector. To carry away heat, the grounded shield and input endcap of the helical resonator are heat sinked to the optical breadboard using massive aluminum mounts, and a ceramic Shapal heat sink is installed between the alumina insulator of the rf feedthrough and the bottom of the middle magnetic shield, similarly to \cite{Nisbet-Jones2016a}, see Fig.~\ref{fig:BBR}(a). The heat sinks reduced the temperature rise of the trap by approximately a factor of two.

\begin{figure}[tb]
\includegraphics[width=1\columnwidth]{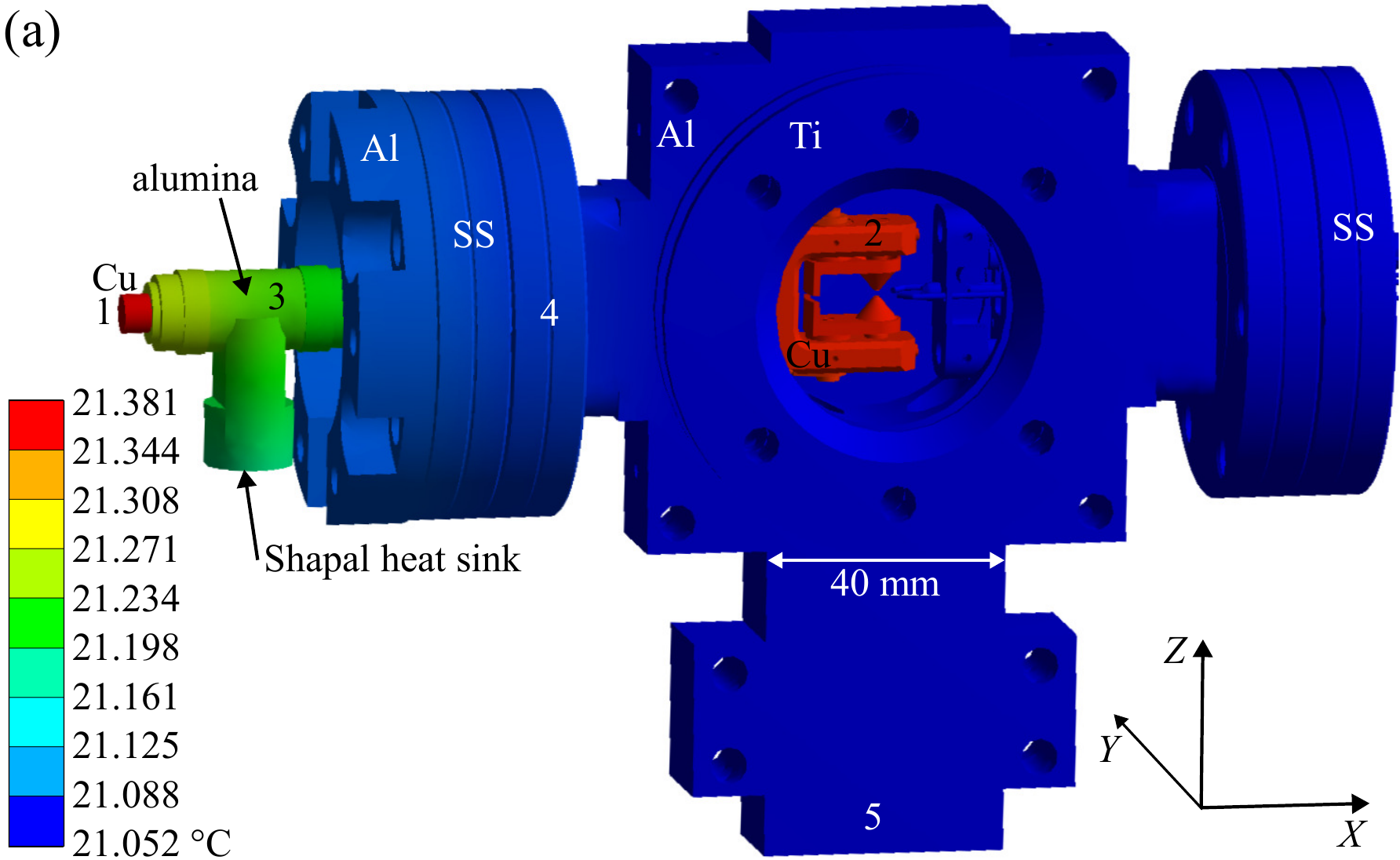}\\
\vspace{3pt}
\includegraphics[width=1\columnwidth]{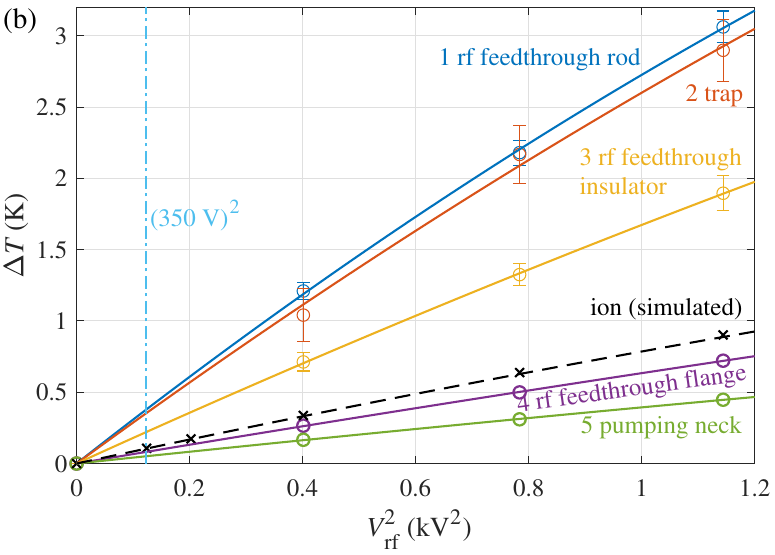}
\caption{(a) FEM model of the trap and vacuum chamber showing the simulated temperature distribution for an ambient temperature of $21\;^\circ$C and an rf voltage of 350\;V. Materials are labeled in the figure (SS---stainless steel). The Cu rf feedthrough rod (1) and the Shapal heat sink are longer in reality, but are in the FEM model truncated at the point where their temperature was measured to provide well-defined boundary conditions.
(b) Measured temperature rise (circles) of selected parts indicated by 1--5 in (a) as a function of the squared rf voltage. Error bars are shown for thermal camera data; for sensor data the uncertainties are smaller than the symbols. The simulated BBR temperature seen by the ion is plotted as crosses. The fit curves are of the form $\Delta T_i(V_\mathrm{rf}) = k_{2,i} V_\mathrm{rf}^2 + k_{4,i} V_\mathrm{rf}^4$. The vertical dash-dotted line indicates the nominal rf voltage.
\label{fig:BBR}}
\end{figure}

Using a thermal camera (FLIR A615 LW) and Pt100 sensors, the temperature rise above ambient of multiple points on the rf vacuum feedthrough, the trap, and the chamber was measured at rf voltages of 630\;V, 890\;V, and 1070\;V, significantly higher than the nominal trapping voltage of 350\;V, in order to reduce the relative uncertainty of the camera measurements. Details of the measurement and analysis techniques are given in \cite{OC18-Guidelines}. A Pt100 sensor covered by black tape of known, high emissivity was used to calibrate the camera readings, thereby reducing its absolute uncertainty. The temperature of the IR window and the camera sensor were continuously measured \cite{OC18-Guidelines} and changes in the reflected temperature were estimated.
The transmission of the CaF$_2$ window and the emissivities of materials used for the trap had beforehand been  measured using the same thermal camera. The emissivities are thus weighted by the window transmission and camera spectral response and are not total emissivities. As the same emissivities are used in the FEM simulation, which requires total emissivities, their uncertainties were conservatively chosen to cover literature values for total emissivities, e.g., \cite{Ablewski2020a}, see Table~\ref{tab:BBR}.

\begin{table}[tb]
\centering
  \caption{
    Uncertainty budget for the BBR temperature seen by the ion at the nominal rf voltage $V_0$ (corresponding to $q_z=0.376$ at \qty{14.4}{MHz}). The last column gives the rf voltage scaling of the different contributions.
   } \label{tab:BBR}
  \begin{ruledtabular}
  \begin{tabular}{l S S c} 
Contribution & {Value} & {$u_{T,\mathrm{BBR}}$ (mK)} & Scaling\\
\midrule
\emph{Thermal evaluation:} & & & \\
$\Delta T_\mathrm{trap}$ (mK)     & 356(61) & 9  & \\
$\Delta T_\mathrm{chamber}$ (mK)  & 70(10)  & 9  & \\
Emissivity, machined Mo  & 0.13(7) & 12  & \\
Emissivity, polished Mo  & 0.05(5) & 3  & \\
Emissivity, Cu           & 0.07(6) & 4  & \\
Emissivity, fused silica & 0.75(20) & 4 & \\
Emissivity, CaF$_2$\footnote{Effective emissivity, considering the high-emissivity magnetic shield outside the window. Actual values at 294\;K are emissivity 0.68, transmission 0.30, and reflectivity 0.02.}      & 0.80(20) & 4  & \\
Emissivities, other       &  & 1  & \\
rf voltage, absolute (V) & 350(20) & 1  & \\
rf voltage, relative (\%) & 2.2 & 5  & \\
Computational uncertainty & & 3  & \\
\multicolumn{2}{l}{Diffusive reflection approximation}  & 20  & \\
FEM model errors & & 20  & \\
\emph{Subtotal:}  & & 34  & $V_\mathrm{rf}^2$ \\
\midrule
\emph{Clock operation:} & &  & \\
Sensor calibration & & 8  & const.\\
Ambient $T$ gradient & & 10 & const.\\
$\Delta T$ repeatability (\%) & 20 & 21 & $V_\mathrm{rf}^2$ \\
\emph{Subtotal:}  & & 25 & \\
\midrule
Total: & & 42 \\
  \end{tabular}
     \end{ruledtabular}
\end{table}

To minimize heat conduction through the Pt100 leads, the last 5\;cm of these are taped to the chamber with Kapton tape and then covered with thin foam and aluminum tape to ensure thermal equilibrium. Furthermore, to minimize the effect of ambient temperature fluctuations, the measurements were carried out with remote operation during weekends after the laboratory temperature had stabilized for more than a day.

It was found that the threefold magnetic shields around the vacuum chamber significantly change the thermal properties of the system: the air temperature inside the innermost shield is very close to the chamber temperature. This is important as the windows, with much lower thermal conductivity than the aluminum chamber, span a large fraction of the solid angle seen by the ion. It also ensures that the thermal conductivity of the sensor leads causes a negligible error. 

For the thermal camera measurements, dummy magnetic shields with small imaging holes were used. To validate this approach, two reference measurements with the actual shields were carried out. No significant differences in the temperature rise recorded by the Pt100 sensors were observed between the reference and thermal imaging measurements. The temperature rise of all measured points $i$ (by both the Pt100 sensors and the camera) can be well fitted by the relation $\Delta T_i(V_\mathrm{rf}) = k_{2,i} V_\mathrm{rf}^2 + k_{4,i} V_\mathrm{rf}^4$, see Fig.~\ref{fig:BBR}(b). The second, small  term is believed to be due to radiative heat loss, as indicated by negative $k_{4,i}$ coefficients that increase in magnitude for increasing $k_{2,i}$, i.e., increasing temperature rise. This equation can be used to interpolate the measured $\Delta T_i$ values to the nominal  rf voltage (as $\Delta T_i(0) = 0$ by definition, this is interpolation rather than extrapolation). 

A temperature difference between the optical table and the air causes a gradient over the vacuum system, as the bottom of the chamber sits on the table and the helical resonator is exposed to the air (outside the magnetic shields). To show that the temperature rise of the individual parts is independent of the exact ambient temperature and the gradient, different arrangements of baffles were used to vary the gradient over the system (with the trap rf off) between 10\;mK and 140\;mK, which yielded temperature rise values consistent within the measurement uncertainty. The small temperature gradients are due to the compact aluminum vacuum system, see Fig.~\ref{fig:BBR}(a).

A FEM model of the trap and vacuum system was created in the \textsc{ansys} software, see Fig.~\ref{fig:BBR}(a). 
Model parameters, mainly the thermal contact conductivities between the different metal and insulator parts of the rf feedthrough and between the feedthrough and the Shapal heat sink,
were varied around typical values to reach agreement between the simulated temperature distribution and the temperature rise values measured at 1070\;V. Due to the magnetic shields, convection of the inner parts of the chamber had to be suppressed. Simulations at 630\;V and 890\;V then showed good agreement with the corresponding measurements with no free parameters. Finally, the nominal voltage of 350\;V was simulated using interpolated boundary temperatures as described above. The results show that the whole ion trap structure heats up uniformly by 360\;mK and the chamber by approximately 70\;mK, while the temperature rise seen by the ion is 105\;mK.

The simulation also estimates the power dissipated in the simulated parts: trap  5.0\;mW, rf feedthrough 12.4\;mW, vacuum chamber 1.4\;mW, and other simulated parts 0.2\;mW. This confirms the observation that most of the 250\;mW of rf power supplied to the helical resonator is dissipated in the resonator itself.

To make the uncertainty analysis more transparent, it is divided into two parts, see Table~\ref{tab:BBR}. The \emph{thermal evaluation} refers to the effective BBR temperature seen by the ion relative to the ambient temperature under the conditions during the thermal evaluation, whereas the contribution from \emph{clock operation} includes the uncertainty of the absolute ambient temperature (including gradients) and the uncertainty related to how well we can establish that the trap temperature rise, which is not measured continuously, corresponds to the value during the thermal evaluation.

The simulations show that the temperature rise seen by the ion can be written as $\Delta T_\mathrm{ion} = x \Delta T_\mathrm{trap} + (1-x) \Delta T_\mathrm{chamber}$, where $x \approx 0.14$ in a wide range around the nominal voltage. This allows us to write the uncertainty as a function of the uncertainties of the measured temperatures rather than those of the boundary conditions of the FEM simulation. The uncertainty of the trap temperature rise is dominated by the thermal camera uncertainty, in particular since the effective transmission of the CaF$_2$ window is only 30\%, but is reduced when interpolating to the nominal voltage. The uncertainty of the temperature rise of the chamber is dominated by the temperature gradient of ${\approx} 5$\;mK, which is multiplied by a factor of two as all parts of the chamber visible to the ion were not measured.

The uncertainty contribution from material emissitivities is dominated by the values for machined (shield electrodes and their mounting structures) and polished (endcap-electrode faces) molybdenum due to their large solid angles and the elevated trap temperature. The uncertainty of the absolute rf voltage has a negligible effect due to the small fraction of the power that is dissipated in the trap, but the relative voltage uncertainty affects the interpolation to the nominal rf voltage. 
In the simulations, the maximum observed computational uncertainty (deviation from equilibrium with all boundary conditions at the same temperature and no applied rf voltage) was 3\;mK for typical run times.

\textsc{Ansys} treats reflections as perfectly diffusive, while in reality they are partly specular, which affects the simulated BBR temperature seen by the ion. To estimate the uncertainty caused by this approximation, a simple analytical model was used. The fractional solid angle of the endcap faces, the gap between the endcap and outer electrodes (which was designed to be zero, but is nonzero in the final assembled trap due to tolerances), the outer electrodes, the Sr beam collimator, and `the rest' (mainly the windows) was calculated. Using the emissivity values in Table~\ref{tab:BBR} and the simulated temperature rise values, the temperature seen by the ion was calculated in two ways: (i) assuming that all reflected radiation is at the temperature of the windows and (ii) assuming one reflection of the outer electrodes from each surface. This gave temperature rise values of 89\;mK and 109\;mK, respectively. We take the difference between these, 20\;mK, as the uncertainty due to the assumption of diffusive reflections. Note that the latter value is close to the simulated temperature rise of 105\;mK.

The same model was used to estimate the influence of differences between the true geometry and the FEM model (measured dimensions were used for the model rather than design dimensions, but fabrication errors and tolerances were not studied in detail). Here, the most critical point is the gap between the electrodes, which can be considered to have an emissivity of unity due to multiple reflections. When its fractional solid angle, estimated to be 0.029, is changed to zero or twice this value, the temperature rise seen by the ion changes by $\mp 7$\;mK. Since this is expected to be the most significant source of error, we estimate an upper limit of 20\;mK for the total uncertainty due to model errors.
This gives a total uncertainty of 34\;mK for the thermal evaluation.

Among the uncertainty contributions related to clock operation, the sensor calibration contains significant contributions from the water-bath temperature homogeneity, the resistance measurement, sensor self-heating, and the calibration of the two reference standard platinum resistance thermometers (SPRTs).

During clock operation, three Pt100 sensors on the vacuum chamber around the position of the ion are used to evaluate the BBR shift dynamically.
The interpolated temperature rise of each part, see Fig.~\ref{fig:BBR}(b), is subtracted from the sensor readings to obtain corresponding ambient temperatures. The BBR temperature is then obtained by adding the interpolated BBR temperature rise to the mean ambient temperature.
The temperature is corrected for the difference between thermodynamic temperature and the International Temperature Scale ITS-90 \cite{Gaiser2022a}, although this correction is only ${\approx}3$\;mK.
During the thermal camera measurements, it was found that the temperature of the rf feedthrough rod followed that of the feedthrough flange quite well even when the laboratory temperature was intentionally made to oscillate. As an estimate of the temperature inhomogeneity, we take the maximum deviation among the three derived ambient temperatures from their mean, which typically has a time average of \qty{<10}{mK} (including the larger inhomogeneity after ion loading).

To validate that the thermal behavior has not changed since the evaluation, e.g., due to changes in the helical resonator $Q$ value or some contact resistance, the temperature rise measured by the Pt100 sensors at the nominal voltage is measured regularly (approximately once a year), again with remote control for  stable ambient temperature. The temperature rise of the chamber parts typically agrees with the thermal evaluation within 10\% if the ambient temperature is stable enough, but we use 20\% as a conservative upper limit since the trap temperature is not measured directly. 

An external hot spot, e.g., a person, visible through all three beam openings on the CaF$_2$ side of the magnetic shields could in principle cause an increase in the BBR temperature of order 10\;mK. This is prevented by black walls around the setup, which also reduce temperature variations due to convection.

The final temperature rise seen by the ion is thus 105(42)\;mK at the nominal rf voltage $V_0$. At a typical BBR temperature of 295\;K, the  BBR shift and uncertainty can be evaluated using Eq.~\eqref{eq:BBR} to be $525.69(37) \times 10^{-18}$.  For operation at other rf voltages, the BBR temperature relative to the mean of the Pt100 sensors can be evaluated as $T_\text{BBR} = \langle T_\text{Pt100}\rangle + \qty{322}{\milli\kelvin} (V_\mathrm{rf}/\mathrm{kV})^2 - \qty{43}{\milli\kelvin} (V_\mathrm{rf}/\mathrm{kV})^4$ and the uncertainty as $u_{T,\text{BBR}} = \{12.8^2 + [(V_\text{rf}/V_0)^2 40.3]^2\}^{1/2}$\;mK.

The BBR temperature uncertainty of 42--51\;mK for typical rf voltages is the lowest reported for an ion clock to date. 
As the heating is dominated by the helical resonator, the magnetic shields suppress the convection, and the heat conduction from the chamber to the breadboard is significant, we stress the importance of carrying out the measurements on the final system as it is used for clock operation, including the magnetic shields, rather than on a dummy setup, if this level of uncertainty is to be achieved.

Also the magnetic field of the BBR causes a small frequency shift through magnetic dipole (M1) coupling. This shift is strongly dominated by the transition between the \Dth\  and \Dfh\ states with frequency $\nu_{DD} = 8.4044(4)$\;THz \cite{Sansonetti2012a}. As this transition frequency is close to the peak frequency of the BBR spectrum, 17\;THz at 295\;K, the frequency shift must be numerically integrated over the BBR spectrum. Following \cite{Farley1981a,Porsev2006a,Gan2018a}, the shift can be written as
\begin{equation}
\Delta\nu_\mathrm{M1} = \frac{1}{36h \pi^2 c^5 \epsilon_0} \left(\frac{\kB T}{\hbar}\right)^3
|\langle D_{5/2}|| \mathbf{M} || D_{3/2}\rangle |^2 F\left(\frac{h\nu_{DD}}{\kB T}\right),
\end{equation}
where the function
\begin{equation}
F(y) = \int_0^\infty \left(\frac{1}{y+x} + \frac{1}{y-x}\right) \frac{x^3 dx}{e^x-1},
\end{equation}
introduced in \cite{Farley1981a}, should be evaluated as a Cauchy principal value.

For the reduced matrix element of the magnetic moment, we use the relativistic theory value $\langle D_{5/2}|| \mathbf{M} || D_{3/2}\rangle = 1.545(6) \mu_\mathrm{B}$ \cite{Arora2012a}, which agrees well with other relativistic calculations \cite{Safronova2011c}. Conservatively increasing the uncertainty to 1\%, $1.545(15) \mu_\mathrm{B}$, the shift can be evaluated and approximated around 295\;K as $\Delta\nu_\mathrm{M1} = -4.54(9)\;\mu\mathrm{Hz}\,(T/295\;\mathrm{K})^{3.467}$.
Accounting for the differences in the used transition frequency and matrix element, this agrees with the recent value of $-4.9\;\mu\mathrm{Hz}$ at 300\;K given in \cite{Tang2024a}.

At $-1\times 10^{-20}$ in fractional frequency, this shift is negligible at the current total uncertainty, but we note that it is significantly smaller in magnitude and of opposite sign to previous evaluations, where the BBR was incorrectly assumed to be static relative to the 8.4-THz transition frequency \cite{Dube2013a,Arora2012a}.

\subsection{Quadratic Zeeman shift}

Like the BBR M1 shift, the quadratic Zeeman shift is strongly dominated by the \Dth--\Dfh\ transition. It can be written as
\begin{equation} \label{eq:QZS}
\Delta\nu_\mathrm{QZ}(m_J') = |C_{m_J'}|^2 \frac{|\langle D_{5/2}|| \mathbf{M} || D_{3/2}\rangle |^2}{h^2 \nu_{DD}} B^2,
\end{equation}
where the transition strengths $|C_{m_J'}|^2$ are $1/10$, $1/15$, and $0$ for the \Dfh\ sublevels $|m_J'| = 1/2$, $3/2$, and $5/2$. After averaging over the Zeeman components, the mean value is $\langle |C_{m_J'}|^2\rangle = 1/18$. Using the same matrix-element value as above, the averaged quadratic Zeeman shift becomes $3.09(4)\;\mu\mathrm{Hz}/\mu\mathrm{T}^2$. This is the smallest quadratic Zeeman shift among the commonly used ions---only $^{115}$In$^+$ has a shift of similar magnitude  \cite{Hausser2025a}.

At a typical static magnetic field of $B_0 = 4.8\;\mu$T, the fractional shift is thus $1.6\times 10^{-19}$. As the magnetic field is measured continuously during clock operation based on the linear Zeeman shift, the relative uncertainty of this shift is $1\times 10^{-4}$, limited by the knowledge of the ground-state $g$ factor, for which we assume $g_S = 2.0023(2)$, and thus negligible.

Radio-frequency currents in the trap structure and vacuum chamber can create an rf magnetic field at the position of the ion~\cite{Gan2018a}.
In Eq.~\eqref{eq:QZS}, the total squared magnetic field is then $B^2 = B_0^2 + \langle B_\mathrm{rf}^2\rangle$, where $\langle B_\mathrm{rf}^2\rangle$ is the mean-square rf field.
The rf magnetic field was determined by measuring the Autler-Townes (AT) splitting of the clock transition when the applied field makes the ground-state Zeeman splitting resonant with the rf drive frequency (at \qty{515}{\micro\tesla}). The splitting is $\Omega_\mathrm{rf}' = (\delta_\mathrm{rf}^2 + \Omega_\mathrm{rf}^2)^{1/2}$, where $\delta_\mathrm{rf}$ is the rf detuning and the rf Rabi frequency is $\Omega_\mathrm{rf} = g_S\mu_\mathrm{B} B_{\perp}/2 \hbar$ \cite{Arnold2020a}, where $B_{\perp}$ is the rf field perpendicular to $B_0$.

\begin{figure}[tb]
\includegraphics[width=1\columnwidth]{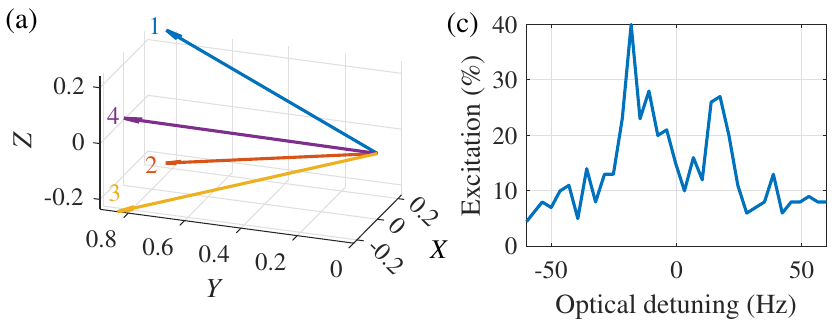}
\includegraphics[width=1\columnwidth]{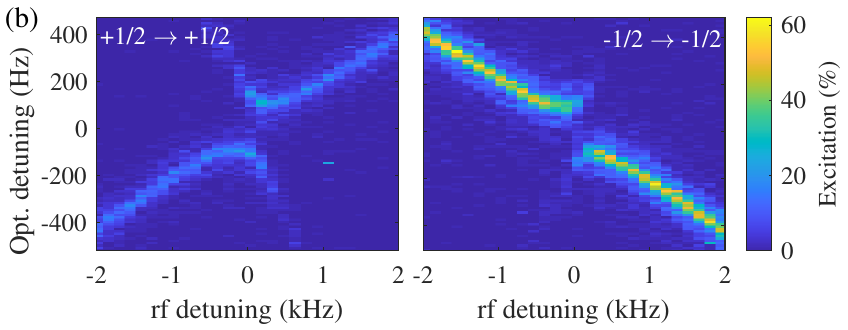}
\caption{(a) Static magnetic field directions used for the Autler-Townes (AT) measurements. Direction 4 is along $Y$, the others displaced by $22^\circ$ in different directions. See Fig.~\ref{fig:BBR}(a) for coordinate system. (b) Measured AT splittings of the least magnetically sensitive Zeeman pair for direction 1. By measuring a pair, there are four branches to fit, but only three free parameters: line center, Zeeman splitting, and rf Rabi frequency $\Omega_\mathrm{rf}$. The large Zeeman shifts cause optical pumping into $|{}^2\!S_{1/2}, m_J=-1/2\rangle$ by the cooling laser, which explains the difference in excitation probability (state preparation did not work with these field directions). (c) With direction 4, a splitting could barely be resolved at rf resonance with a probe time of 45\;ms. 
\label{fig:A-T}}
\end{figure}

A large magnetic field can be applied only along the $Y$ axis of the chamber, see Fig.~\ref{fig:BBR}(a), which has larger coils with heavier-gauge wire. We therefore measured with the static field tilted away from the $Y$ axis by $22^\circ$ towards the $(X+Z)$, $(X-Z)$ and $(-X-Z)$ directions, see Fig.~\ref{fig:A-T}(a). This resulted in on-resonance AT splittings ($\Omega_\mathrm{rf}/2\pi$) of \num{239.5(27)}, \num{232.1(25)} and \qty{198.9(45)}{Hz} with the static field along directions 1, 2 and 3, respectively, see Fig.~\ref{fig:A-T}(b) for an example. 
Numerically solving the rf magnetic field vector $\mathbf{B}_\mathrm{rf}$ from the measured splittings gave two possible solutions. Noting that these would correspond to vastly different splittings of \qty{236}{Hz} and \qty{26}{Hz} with the static field along $Y$, additional measurements were carried out with this field direction. No \qty{236}{Hz} splitting was observed, but increasing the probe time to 45\;ms, a double peak with a splitting of about 30\;Hz could just barely be resolved, see Fig.~\ref{fig:A-T}(c). This confirmed the rf field amplitude vector to be $\mathbf{B}_\mathrm{rf} = \pm [ -1.82(30), 42.00(80), -0.23(24) ]$\;nT, where the uncertainties were evaluated using a Monte Carlo simulation including the uncertainties of the measured splittings, the static field directions, the $g_S$ factor, and the relative rf voltage during the measurements. 
The rms field of \qty{29.73(56)}{nT} is, to our knowledge, among the lowest ever measured and corresponds to a completely negligible shift of $6\times 10^{-24}$. 

The FEM used for the BBR evaluation was also used to estimate the ac magnetic field, yielding $\mathbf{B}_\mathrm{rf} \approx \pm [ 0, 20, 0 ]$\;nT. The direction of the field agrees well with the measured, while the magnitude is a factor of two lower. This could be due to fabrication tolerances, rf-carrying parts not included in the simulation (e.g., the helical resonator), or some limitation of the magnetic-field simulation (the low $\mathbf{B}_\mathrm{rf}$ field makes this result much more prone to numerical errors than the BBR simulation). In the simulation, the field is within numerical precision caused completely by the titanium plate nuts and screws used to attach the trap body to the rf feedthrough rod. These are the closest parts to the ion breaking the mirror symmetry of the trap, which shows that by paying attention to symmetry in the trap design and fabrication, the ac magnetic field can be made negligible. 
An rf magnetic field of the same order as ours has recently been measured in a trap of similar design \cite{Curtis2024a}, which shows that this is a deterministic feature of this design. The importance of symmetry has also recently been observed in the context of microfabricated surface traps~\cite{Ivory2024a}.

\subsection{ac Stark shifts}

\subsubsection{\qty{674}{nm} E1 ac Stark shift \label{sec:674E1}}

The clock laser causes a nonresonant ac Stark shift during the Rabi interrogation by coupling to all electric-dipole (E1) allowed transitions sharing a level with the clock transition. Since the Zeeman components are probed with different laser intensities, the tensor shift is not fully cancelled.
For the angle $38^\circ$ between the polarization vector and the magnetic field direction, the differential polarizability calculation in Appendix~\ref{sec:polarizability} gives intensity shift coefficients of $(0.77, 0.74, 0.67)\;\mathrm{mHz/(W/m^2)}$ for the Zeeman pairs $(\pm1/2 \rightarrow \pm1/2,\, \pm1/2 \rightarrow \pm3/2,\, \pm1/2 \rightarrow \pm5/2)$.

The very low laser intensity required for long clock pulses is not straightforward to measure, but one can use the relation between intensity and Rabi frequency \cite{James1998a} to estimate the shift. For our angles between the laser beam direction, polarization plane and magnetic field, the intensity as a function of Rabi frequency can be expressed as $I = (2.6, 1.7, 1.9)\;\mathrm{(mW/m^2)} \times [\Omega_0/(2\pi \times 5\;\mathrm{Hz})]^2$ for the three Zeeman pairs, where we used the reduced electric quadrupole (E2) matrix element $13.747(51) e a_0^2$ \cite{Safronova2017a}.
Taking into account that a part of the laser intensity is in resolved sidebands that contribute to the Stark shift but not to driving the transition and that thermal dephasing decreases the effective Rabi frequency, and assuming a pulse area of $1.1\pi$, we estimate the mean fractional shift as a function of probe time as $6.6\times 10^{-21} (100\;\mathrm{ms}/\tau_\mathrm{p})^2$.

In an alternative approach, the maximum clock laser power after the chamber was measured and the intensity at the ion was calculated using the window transmission and the estimated beam waist. Using the measured AOM attenuation, intensities a factor 2 higher than those from the Rabi frequency approach were obtained, which is reasonable considering the large uncertainty of the beam waist. We take the full shift estimated from the Rabi frequency as the uncertainty.

\subsubsection{\qty{674}{nm} E2 ac Stark shift}

When a particular Zeeman transition is probed, electric quadrupole coupling to all other E2-allowed Zeeman transitions sharing a level with the probed transition will cause an E2 ac Stark shift in addition to the E1 shift discussed in Sec.~\ref{sec:674E1}. While the individual Zeeman components can experience shifts of ${\approx}500\;\mu$Hz, the shifts are symmetric for a perfectly linearly polarized laser beam and vanish when a Zeeman pair is averaged. However, if the laser beam has an elliptical polarization, the symmetry is broken and a net shift remains after Zeeman averaging. This has recently been studied for optical clocks based on a ${}^1\!S_0 \rightarrow {}^3\!P_0$ transition \cite{Yudin2023a}.

The shift can be estimated by calculating the Rabi frequencies for an elliptically polarized laser beam using the actual magnetic-field, laser-beam and polarization directions. Accounting for sidebands, ion temperature effects and pulse area as above gives $-3.4\times 10^{-21} (\eta_\mathrm{p}/1^\circ)(100\;\mathrm{ms}/\tau_\mathrm{p})^2$, where $\eta_\mathrm{p}$ is the polarization ellipticity. 

An elliptical polarization causes the Rabi frequencies for transitions with $\Delta m_J = \pm 1, \pm 2$ to differ, which should also result in an asymmetric excitation probability for these Zeeman pairs. Experimentally, the asymmetry is consistent with QPN for all pairs, showing no sign of elliptical polarization.
However, an ellipticity of $-2.1^\circ$ has been measured at the output of the fiber delivering light to the trap, so we take $5^\circ$ as a conservative uncertainty estimate, which gives a frequency uncertainty of $1.7\times 10^{-20} (100\;\mathrm{ms}/\tau_\mathrm{p})^2$.

\subsubsection{\qty{1092}{nm} ac Stark shift}

Following Appendix~\ref{sec:polarizability}, the differential scalar polarizability at 1092\;nm is $\Delta\alpha(274.59\;\mathrm{THz}) = 329(5)$\;a.u., which gives the intensity shift coefficient $-1.54(3)\;\mathrm{mHz/(W/m^2)}$. However, the ASE repumper has a broad pedestal, so its spectrum must be accounted for.
As the shift is several kHz with the repumper on, it can easily be measured \cite{Fordell2015a}.
Combining electronic switching of the repumper with a MEMS switch with a maximum switching time of 1\;ms and measured ${>}71$\;dB extinction allowed shortening the interrogation-cycle dead time while ensuring a negligible shift of magnitude ${<}10^{-23}$ at the beginning of the probe pulse.

\subsubsection{\qty{422}{nm} ac Stark shift}

As the \qty{422}{nm} cooling transition shares the ground state with the clock transition and is detuned by approximately half a linewidth, it causes a large ac Stark shift of the clock transition, which in the low-saturation limit is given by
\begin{equation}
\Delta\omega_{422} = - \frac{\delta (C \Omega_\mathrm{c}/2)^2}{\delta^2 +(\Gamma/2)^2}.
\end{equation}
Here $\delta$ is the laser detuning, $\Omega_\mathrm{c}$ is the cooling-laser two-level Rabi frequency \cite{Lindvall2012a}, $C^2=1/6$ is the mean transition strength for linearly polarized light, and $\Gamma$ is the natural linewidth. Using the relation between Rabi frequency and intensity $I_\mathrm{c}$, $\Omega_\mathrm{c} = \mu \sqrt{2I_\mathrm{c}/\epsilon_0 c}/\hbar$, where the reduced dipole moment of the $S_{1/2} \rightarrow P_{1/2}$ transition is $\mu = 3.0710(29) e a_0$ \cite{Roberts2023a}, intensity shift coefficients $\kappa = \Delta\omega_{422}/(2\pi \,I)$ of $2.2\;\mathrm{kHz/(W/m^2)}$ and $110\;\mathrm{Hz/(W/m^2)}$ are obtained for the AOM-shifted cooling light ($-14$\;MHz detuning) and the Rb-stabilized laser ($-436$\;MHz detuning), respectively.

With the AOMs off, all \qty{422}{nm} light is detuned by $-436$\;MHz. If the mechanical shutter is left open, the light leaking through the cooling AOM causes an ac Stark shift of several hertz. No collimated beam is expected to propagate through the closed shutter. A lower limit of 63\;dB was measured for its extinction, limited by the power meter, corresponding to a light shift ${<}10^{-20}$.
Concerning scattered light, if 1\% of the 15\;mW total laser power (\qty{-436}{MHz} detuned) were scattered into a full solid angle and had direct optical access to the ion a distance of 1\;m away, it would cause an ac Stark shift of \num{3e-18}. However, the laser, mechanical shutter, and AOMs are separated from the ion trap setup by multiple black low-reflectance walls to block scattered light (and BBR). In combination with the small holes in the threefold magnetic shielding around the trap, which are largely blocked by the beam objectives, any scattered light would have to undergo multiple diffusive reflections to reach the ion, and the \qty{422}{nm} ac Stark shift is estimated to be negligible.

\subsection{Collisional shift}

After activating the NEG element of the pump, the pressure indicated by the ion pump current dropped below the lowest nonzero reading, $2\times 10^{-11}$\;mbar, in about 100~minutes as the NEG element cooled down. Extrapolating the pressure drop indicates a final pressure at the low $10^{-12}$\;mbar level at the position of the ion pump.

To estimate the pressure at the position of the trapped ion, the vacuum system including trap structure was simulated in the \textsc{Molflow+} molecular-flow Monte Carlo simulation software \cite{Kersevan2019a}. The (linear) relation between the outgassing rate of the different parts of the system and the resulting partial pressure was obtained for H$_2$ and CO, the dominating background gases in UHV systems, including aluminum ones. The pumping speed at the position of the ion was found to be limited by the conductance of the pumping neck between the trap and pump parts of the chamber, so the exact pumping speed of the pump (nominally $200\;\mathrm{l/s}$ and $140\;\mathrm{l/s}$ for H$_2$ and CO, respectively) does not significantly affect the results. The geometry was slightly simplified in that the flanges for the electrical vacuum feedthroughs and the closed all-metal valve were replaced by blank surfaces. In this case the outgassing rates were increased to compensate for the reduced surface area.

The obtained outgassing rate versus pressure relations were then used as input for another Monte Carlo simulation, where the outgassing rates for different materials were drawn from log-normal distributions covering available literature values. For example, the outgassing rates $q_i$ for  stainless steel and aluminum were taken as $\log{[q_\mathrm{SS}/(\mathrm{mbar\, l/s\,cm^2})]} = \mathcal{N}(-11.5,0.5^2)$ and $\log{[q_\mathrm{Al}/(\mathrm{mbar\, l/s\,cm^2})]} = \mathcal{N}(-13.1,0.3^2)$, where $\mathcal{N}(\mu,\sigma^2)$ is a normal distribution with mean $\mu$ and standard deviation $\sigma$. The percentage of H$_2$ was varied between 70\% and 100\% with the rest being CO. No significant effect on the final total pressure was found, so 100\% H$_2$ is assumed in the following. The resulting pressure distribution is asymmetric with median $1.1\times 10^{-11}$\;mbar, mean $1.8\times 10^{-11}$\;mbar, and standard deviation $2.4\times 10^{-11}$\;mbar. The pressure is dominated by outgassing from the stainless steel parts: feedthrough flanges, all-metal valve, and one blank CF16 flange.
Helium diffusion through the fused silica window was considered, but was found to be negligible.

Background-collision rates have been measured during continuous laser cooling by detecting dark periods in the ion fluorescence.
Comparing the measured collision rates to those measured for \Sr\ \cite{Likforman2016a}, $^{40}$Ca$^+$ \cite{Barton2000a}, and $^{171}$Yb$^+$ \cite{Pagano2019a} indicates that the pressure indeed is at the low $10^{-11}$\;mbar level.

\citet{Hankin2019a} have theoretically studied the H$_2$ collisional shift for the $^{27}$Al$^+$ clock at the National Institute of Standards and Technology (USA). Estimates are given also for other clocks, including the $^{88}$Sr$^+$ clock at the National Research Council Canada (NRC)  \cite{Dube2013a,Jian2023a}. At a pressure of \qty{1.6e-10}{mbar} and with a 100\;ms probe time, the fractional shift for $^{88}$Sr$^+$ is $(-0.2 \pm 11.9) \times 10^{-19}$, where the two terms come from time dilation due to collisional heating and worst-case $\pm\pi/2$ phase shifts to the atomic superposition state, respectively. As our operational parameters are similar to those of NRC except for the background pressure and probe time $\tau_\mathrm{p}$, we introduce the shift as $\mathcal{N}(-0.2(\tau_\mathrm{p}/100\;\mathrm{ms}),11.9^2) \times 10^{-19}/\qty{1.6e-10}{mbar}$ in the Monte Carlo simulation used to estimate the pressure (the time dilation is proportional to the probe time as long as it is much shorter than the mean time between collisions \cite{Hankin2019a}). Despite the skewed pressure distribution, the shift distribution is symmetric and yields a collisional shift $0.00(22) \times 10^{-18}$ for typical probe times of 50--200\;ms.

If the ion is dark during the second fluorescence detection window after the clearout pulse, this is attributed to a collision and the servo cycle is flagged as invalid. This is expected to further suppress the collisional shift and the above value is considered a sufficient upper limit.

\subsection{Other shifts}

\subsubsection{Servo errors}

Noise-induced servo errors \cite{Lindvall2023a} are made negligible by normalizing the servo discriminator by a moving mean and by alternating the order of probing on the red and blue sides of the transitions as mentioned in Sec.~\ref{sec:interr}. The magnetic field noise level has also been reduced since the work in \cite{Lindvall2023a} by replacing the current source for the magnetic field coils with one with lower noise and by removing ground loops. The servo error caused by magnetic-field drift is typically small and is made negligible by the forward-backward probing, see Sec.~\ref{sec:interr}.

The servo error caused by linear residual laser drift can be canceled by using a unit gain that corrects the servo offset in a single step and adding the correction to the probing frequency, see Appendix A of \cite{Lindvall2023a}. With this gain, the right-hand side of Eq.~\eqref{eq:servo} predicts the 
line center exactly (apart from QPN) if the probing frequencies are on the linear slopes of the line shape.
As long as the servo tracks the transition well enough that one stays in the linear regime, the unbiased line center can be estimated also in the postprocessing by finding the gain $g$ that causes the self-comparison to follow a $\tau^{-1/2}$ slope from the first point. For short cycle times around \qty{4}{s}, this is valid even for residual drifts of 
order 100 mHz/s. The difference between the normal  line center and the postprocessed estimate can then be used as an estimate of the drift-induced servo error. An example of this is shown in Fig.~\ref{fig:servo_error}, where the drift was large and changing rapidly after work on the clock laser, and only a constant drift was removed. The servo error estimated using the clock data is shown in Fig.~\ref{fig:servo_error}(c), averaged over 100 cycles to reduce the QPN, together with the servo error calculated from the fitted drift $d$, Fig.~\ref{fig:servo_error}(b), using \cite{Lindvall2023a}
\begin{equation} \label{eq:servo_error}
\Delta\nu_\text{drift} = d T_\mathrm{s} \frac{1 - g}{g},
\end{equation}
where $T_\mathrm{s}$ is the servo (cycle) time and $g$ is the normalized gain.
This error can also be estimated from numerical servo simulations, but these require a realistic line-shape model or at least knowledge of the exact line-shape slope at the probing points as well as experimental input on the laser behavior. Contrary to this, the unit-gain method estimates the servo error due to laser drift directly from the clock data without other input than the unit gain that needs to be adjusted to give $\tau^{-1/2}$ behavior from the first ADEV point, see Fig.~\ref{fig:servo_error}(d). Thus the method requires some form of clock self-comparison (Appendix~\ref{sec:self}).

With drift compensation from the ion, significant servo errors occur only during the initial settling of the servo if the clock is started with an inaccurate line center or drift estimate. These events are easily detected and invalidated, and the remaining estimated servo error is then typically a factor of 20--100 lower than the clock instability at the total measurement time. As an upper limit of the uncertainty of the total servo error, we take $1\times 10^{-19}$ based on detailed servo simulations that included noise, drifts, and oscillations in the magnetic field and the laser frequency, intensity, and polarization.

\begin{figure}[tb]
\includegraphics[width=1\columnwidth]{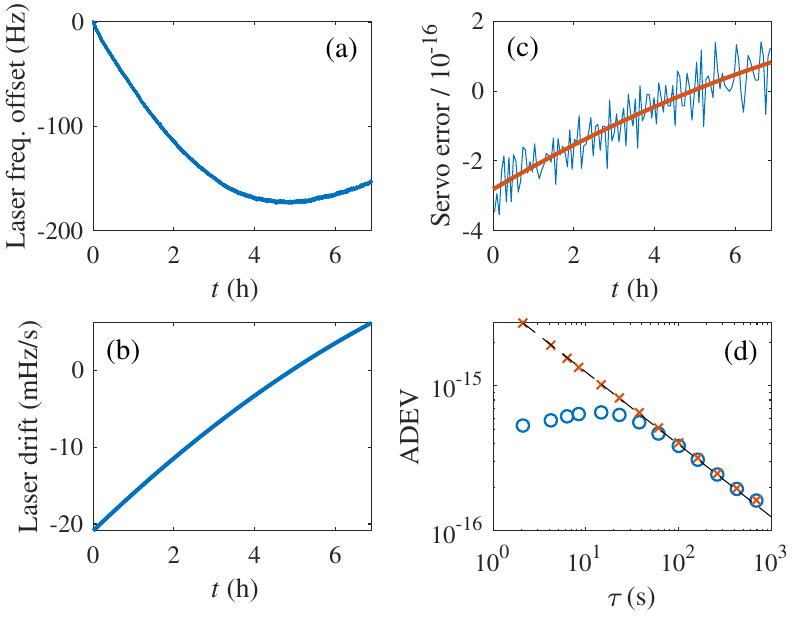}%
\caption{(a) Laser frequency offset measured by the ion, (b) laser drift evaluated from a third-order polynomial fit to (a) in order to avoid QPN, (c) fractional servo error evaluated using the gain-corrected line center (noisy blue curve; mean of 100 cycles to reduce QPN) and from Eq.~\eqref{eq:servo_error} (red curve), and (d) self-comparison using the normal clock data (blue circles) and the gain-corrected data (red crosses). The dashed line is a $\tau^{-1/2}$ fit.
\label{fig:servo_error}}
\end{figure}

\subsubsection{AOM frequency chirp}

A double-pass AOM is used to create the probe laser pulses. When the rf is turned on, the AOM crystal heats up, which changes its optical length via both thermal expansion and the temperature dependence of the refractive index, leading to a frequency chirp. 
The chirp was evaluated from the beat note between an unshifted reference beam and light passing through the double-pass AOM at different, high rf powers and different duty cycles. The shift is  proportional to the rf power and has a duty-cycle scaling of $1 - \tau_\mathrm{p}/(\tau_\mathrm{p} + \tau_\mathrm{d})$. 
We take the full estimated shift as the uncertainty, which, after averaging over the rf powers used for the three Zeeman pairs, gives an uncertainty  of \num{1.1e-19} for a probe time of 140\;ms. This could  be reduced by increasing the optical input power.

\subsubsection{Residual first-order Doppler shifts}

Residual first-order Doppler shifts can result from optical path length variations due to unstabilized fibers and free-space propagation, as well as from motion of the ion trap relative to the clock laser due to, e.g., thermal expansion or vibrations synchronized with the probe pulse. The only mechanical shutter on the optical table in use during clock operation (for the cooling light) is mounted to dampen vibrations on a separate breadboard, and the 5-ms delay before the probe pulse ensures that the mechanical motion has stopped. 

The 1-m fiber delivering the probe light from the breadboard to the physics package is unstabilized. The resulting Doppler shift is estimated using a temperature sensor with low thermal mass close to the fiber. An identical fiber was studied in a measurement similar to the AOM chirp measurement above. A phase-shift temperature coefficient of $1.1\times 10^{-5}\;\mathrm{K}^{-1}$ was found, close to typical literature values \cite{Lagakos1981a}. The ADEV of the measured noise is more than one order of magnitude below the clock instability at all time scales. 
The noise ADEV estimated using the temperature sensor is about a factor of two higher, possibly due to a shorter time constant, and is thus a conservative estimate. The mean shift estimated from the temperature tends to zero for long averaging times. The high uptime and unattended operation of our clock also means that there is negligible correlation between the uptime and the laboratory temperature.
However, the fiber measurements showed a persistent frequency shift of ${\sim}\num{1e-19}$. Although this was likely caused by some experimental nonideality, we take it as an upper limit for the uncertainty related to the 1-m fiber.

The fiber-coupled waveguide frequency doubler for the clock laser and its connection fibers (4\;m total length) are also not stabilized. As these are inside the temperature-stabilized thermal/acoustic enclosure around the cavity vacuum system, we expect similar shifts as above (or smaller), but as they have not been characterized, we take \num{5e-19} as a conservative estimate of the total uncertainty from unstabilized paths. The free-space paths have a negligible effect in comparison with the unstabilized fibers.

\subsection{Uncertainty budget}

An uncertainty budget for typical operational parameters is shown in Table~\ref{tab:u-bud}, with a total uncertainty of \num{7.9e-19}. 
We note that by reducing the first-order Doppler contribution to \num{1e-19} by stabilizing the fibers inside the cavity enclosure, the total uncertainty could be reduced to \num{6.1e-19}.

\begin{table}[tb]
\centering
  \caption{
    Typical uncertainty budget of the VTT MIKES $^{88}$Sr$^+$ optical clock in fractional units ($10^{-19}$) for a probe time of 140\;ms and $q_z = 0.422$.
   } \label{tab:u-bud}
  \begin{ruledtabular}
  \begin{tabular}{l S S} 
Contribution & {Shift} & {Uncertainty} \\
\midrule
BBR E1 shift, $T = 295.000(51)$\;K & 5256.9 &   \\
\quad BBR field  &   & 3.7  \\
\quad Diff.\ polarizability $\Delta\alpha_0$ & & 2.2 \\
\quad Dynamic correction $\eta$ & & 0.90 \\
BBR M1 shift & -0.102 & 0.002 \\
Collisional shift & 0 & 2.2 \\
Thermal motion, $T_\text{ion} = 0.8(2)$\;mK & -12.8 & 3.2 \\
EQS & 0 & 0.3 \\
Excess micromotion & 0 & 0.014 \\
Tensor Stark shift & 0 & 0.003 \\
\qty{674}{nm} E1 ac Stark shift& 0.034 & 0.034 \\
\qty{674}{nm} E2 ac Stark shift& 0 & 0.087 \\
Quadratic Zeeman shift, static field & 1.601 & 0.031 \\
AOM chirp & 0 & 1.1 \\
Servo errors & 0 & 1.0 \\
First-order Doppler shift & 0 & 5.0 \\[1ex]
Total & 5245.7 & 7.9 \\
  \end{tabular}
     \end{ruledtabular}
\end{table}

\section{Absolute frequency measurements} \label{sec:absfreq}

\subsection{2022 measurement against PTB CSF2}

A first absolute frequency measurement was performed against the PTB CSF2 cesium fountain clock \cite{Weyers2018a} during an optical clock comparison campaign \cite{Lindvall2025a} within the ROCIT project \cite{Margolis2024b} in March 2022. The Sr$^+$ clock was measured against a hydrogen maser (HM) at VTT, CSF2 was measured against the local UTC realization UTC(PTB), and the phase difference between the HM and UTC(PTB) was measured using an Integer Precise Point Positioning (IPPP) GNSS link \cite{Petit2015b}. 
The frequency chain is described by the following equation
\begin{eqnarray}
\frac{f_\mathrm{Sr^+}}{f_\mathrm{CSF2}} &=& 
\frac{f_{\mathrm{Sr}^+, T_1}}{f_{\mathrm{HM}, T_1}} \times 
\frac{f_{\mathrm{HM}, T_1}}{f_{\mathrm{HM}, T_\mathrm{tot}}} \times
\frac{f_{\mathrm{HM}, T_\mathrm{tot}}}{f_{\mathrm{UTC(PTB)}, T_\mathrm{tot}}} \nonumber \\
&&\times \frac{f_{\mathrm{UTC(PTB)}, T_\mathrm{tot}}}{f_{\mathrm{UTC(PTB)}, T_2}} \times 
\frac{f_{\mathrm{UTC(PTB)}, T_2}}{f_{\mathrm{CSF2},T_2}}, \label{eq:rocit}
\end{eqnarray}
where $T_1$ denotes the uptime of the Sr$^+$ clock, $T_2$ the uptime of CSF2, and $T_\mathrm{tot}$ is the total 12-day period of the IPPP comparison (MJD 59660.36--59672.36). As in \cite{Lindvall2025a}, a multiple of 1~day was chosen to allow diurnal effects in the link to average down. The fractional uptime of Sr$^+$ was 92.8\% and that of CSF2 98.0\%. 
The second and fourth ratios on the right-hand side of Eq.~\eqref{eq:rocit} introduce extrapolation uncertainty, although for CSF2 it is negligible compared to its statistical uncertainty. 

As customary, the analysis is carried out using fractional frequency-ratio corrections, $y = r/r_0 -1$, where $r = f_i/f_j$ is a frequency ratio and $r_0$ a reference value, typically chosen such that $|y|<10^{-14}$. This allows linearization of the calculations so that $y_{\mathrm{Sr}^+/\mathrm{CSF2}}$ is the sum of the $y$ values corresponding to each ratio on the right-hand side of Eq.~\eqref{eq:rocit}.

During this campaign, the systematic uncertainty of the Sr$^+$ clock was dominated by AOM chirp (\num{1.0e-17}) due to a higher rf power being used. The servo error uncertainty was also higher at \num{2.2e-18}, with the remaining contributions similar to those in Table~\ref{tab:u-bud}. A full uncertainty budget for the Sr$^+$ clock is available in \footnote{T.~Lindvall and A.~E.~Wallin, Zenodo, 2025, \href{http://doi.org/10.5281/zenodo.xxxxxxx}{http://doi.org/10.5281/zenodo.xxxxxxx}.}.
The uncertainties related to the comb and rf synthesis and distribution are described in Appendix~\ref{sec:comb_rf}. 
The relativistic red shift (RRS) of the Sr$^+$ clock was for this campaign determined using the GNSS/geoid method \cite{Denker2018a}, see Appendix~\ref{sec:RRS} for details. 
The analysis of the Sr$^+$/CSF2 ratio closely follows that of the GNSS optical-clock frequency ratios in \cite{Lindvall2025a}. An uncertainty budget for this measurement is shown in Table~\ref{tab:rocit} and the final frequency value is \num{444779044095485.49(15)}\;Hz.

\begin{table}[tb]
\centering
\caption{
Uncertainty budget for the the 2022 absolute frequency measurement against CSF2.
\label{tab:rocit}}
\begin{ruledtabular}
\begin{tabular}{l S } 
Contribution & {Uncertainty ($10^{-16}$)} \\
\midrule
Sr$^+$ statistical &  0.078  \\
Sr$^+$ systematic & 0.10   \\
Sr$^+$ relativistic red shift & 0.024 \\
HM extrapolation & 0.24 \\
Comb, statistical & 0.72 \\
Comb and rf distribution, systematic & 0.10 \\
IPPP link & 0.83 \\
CSFS statistical & 2.6 \\
CSF2 systematic (incl.\ RRS) & 1.7 \\
Total & 3.3  \\
\end{tabular}
\end{ruledtabular}
\end{table}

\subsection{2024--2025 measurement against TAI} \label{sec:TAI}

During the past ten years, absolute frequency measurements against TAI have become a popular alternative to using a local Cs fountain, see, e.g., \cite{Hachisu2017b,McGrew2019a,Pizzocaro2020a,Jian2023a}. In addition to not requiring operation of a local fountain, the ensemble of primary or secondary frequency standards (PSFSs) contributing to TAI has a lower systematic uncertainty than any single standard, which becomes relevant if the statistical uncertainty is averaged down in a long measurement.

A second absolute frequency measurement of our Sr$^+$ clock was carried out against TAI from June 2024 to March 2025 (MJD 60459--60764, Circular~T 438--447 \footnote{Circular T is a monthly publication by the Bureau International des Poids et Mesures (BIPM), available at \href{https://www.bipm.org/en/time-ftp/circular-t}{https://www.bipm.org/en/time-ftp/circular-t}.}). Between mid-August and mid-October, the clock was operated in `measurement campaign mode' with an uptime of 97\%. 
Outside this period, it was also used for other measurements, such as evaluation of systematics. 

During this ten-month period, only three optical clocks contributed to TAI with five submissions in total. This justifies determining the absolute frequency against TAI instead of using the more complex method of separately analyzing the frequency against the contributing cesium fountain clocks \cite{Nemitz2021a,Kobayashi2025a}.

As mentioned in Sec.~\ref{sec:thermal}, the ion heating rate and its uncertainty were larger during this campaign, resulting in a mean thermal-motion uncertainty of \num{8.4e-19}. For characterization purposes, two measurements with high EQS were carried out, which resulted in a higher EQS uncertainty of ${\approx} (3\ldots4)\times 10^{-18}$ for three months. The other uncertainty contributions were similar to those in Table~\ref{tab:u-bud}. Monthly uncertainty budgets for the Sr$^+$ clock are available in \cite{Note2}.

\begin{table*}[tb]
\centering
\caption{
Uncertainty budget for the absolute frequency measurement against TAI from June 2024 to March 2025 (Circular T 438--447). The last three rows show the fractional optical-clock uptimes, the monthly weights, and the reduced frequency, $f - \qty{444 779 044 095 485}{Hz}$. The uncertainty notation used in Circular T is given in parenthesis. The TAI calibration uncertainty, as it would appear in Circular T, is the square sum of the first six uncertainty contributions.
\label{tab:tai}}
\begin{ruledtabular}
\begin{tabular}{l S S S S S S S S S S S} 
  & {June 2024} & {July 2024} & {Aug 2024} & {Sep 2024} & {Oct 2024} &
{Nov 2024} & {Dec 2024} & {Jan 2025} & {Feb 2025} & {March 2025} & {Total}\\
Contribution & \multicolumn{10}{c}{Uncertainty ($10^{-16}$)} \\
\midrule
Sr$^+$ $u_\mathrm{A}$ & 0.017& 0.019& 0.019& 0.016& 0.024& 0.030& 0.041& 0.029& 0.023& 0.021& 0.008\\
Sr$^+$ $u_\mathrm{B}$ & 0.012 & 0.011 & 0.011 & 0.012 & 0.039 & 0.012 & 0.012 & 0.040 & 0.053 & 0.011 & 0.020\\
RRS (incl.\ in $u_\mathrm{B}$) & 0.024 & 0.024 & 0.024 & 0.024 & 0.024 & 0.024 & 0.024 & 0.024 & 0.024 & 0.024 & 0.024 \\
HM extrap.\ ($u_\mathrm{A/Lab}$) & 1.2& 2.4& 2.4& 0.051& 0.95& 1.5& 2.1& 0.96& 0.82& 0.49& 0.32\\
rf  ($u_\mathrm{B/Lab}$) & 0.20 & 0.20 & 0.20 & 0.20 & 0.20 & 0.20 & 0.20 & 0.20 & 0.20 & 0.20 & 0.20 \\
UTC link ($u_\mathrm{l/Tai}$) & 1.3 & 1.5 & 1.3 & 1.3 & 1.7 & 1.3 & 1.3 & 1.3 & 1.3 & 1.3 & 0.28\\
TAI extrapolation & 0& 1.3& 0& 0& 0& 0& 0& 0& 0& 0& 0.032\\
TAI/SI $u_\mathrm{A,TAI/SI}$ & 1.1& 0.93& 0.84& 0.91& 0.95& 1.0& 1.1& 0.85& 0.79& 0.80& 0.31\\
TAI/SI $u_\mathrm{B,TAI/SI}$ & 0.92& 0.94& 0.91& 0.87& 0.93& 0.98& 1.1& 0.85& 0.86& 0.87& 0.81\\
\midrule
Total & 2.3& 3.4& 3.0& 1.8& 2.4& 2.5& 2.9& 2.0& 1.9& 1.8& 0.98\\
TAI calibration & 1.8& 2.9& 2.7& 1.3& 2.0& 2.0& 2.4& 1.6& 1.6& 1.4& {---}\\
\midrule
Sr$^+$ uptime & 0.811& 0.679& 0.774& 0.989& 0.886& 0.795& 0.789& 0.868& 0.858& 0.917& 0.837\\
Weight & 0.070 & 0.025 & 0.044 & 0.161 & 0.106 & 0.083 & 0.041 & 0.119 & 0.170 & 0.179 & 1\\
Red. frequency (Hz) & 0.341 & 0.381 & 0.392 & 0.422 & 0.439 & 0.449 & 0.495 & 0.344 & 0.280 & 0.344 & 0.373\\
\end{tabular}
\end{ruledtabular}
\end{table*}

For each month, the absolute frequency is determined according to the following frequency chain
\begin{eqnarray}
\frac{f_{\mathrm{Sr}^+}}{f_\mathrm{SI}} &=& 
\frac{f_{\mathrm{Sr}^+, T_1}}{f_{\mathrm{HM}, T_1}} \times 
\frac{f_{\mathrm{HM}, T_1}}{f_{\mathrm{HM}, T_2}} \times
\frac{f_{\mathrm{HM}, T_2}}{f_{\text{UTC(MIKE)}, T_2}} \nonumber \\
&&\times \frac{f_{\text{UTC(MIKE)}, T_2}}{f_{\mathrm{TAI}, T_2}} \times 
\frac{f_{\mathrm{TAI}, T_2}}{f_{\mathrm{TAI},T_3}} \times
\frac{f_{\mathrm{TAI}, T_3}}{f_{\mathrm{SI},T_3}}. \label{eq:tai}
\end{eqnarray}
The Sr$^+$ clock is measured against the HM with an uptime $T_1$. Maser extrapolation is used to extend the result to the total measurement time $T_2$, which is a multiple of 5\;d as dictated by the Circular T format. 
The frequency of the HM relative to the local time scale UTC(MIKE) is obtained from a local phase measurement, and the frequency of UTC(MIKE) relative to TAI (UTC) is obtained from Circular T. 
For July 2024, the analysis excludes the last 5\;d period of the month and the second to last ratio in Eq.~\eqref{eq:tai} introduces an extrapolation uncertainty. For the other nine months, the total measurement time is the full duration of the Circular T month, $T_2 = T_3$.
The fractional deviation of TAI from the SI second is based on data from PSFSs submitted to the Bureau International des Poids et Mesures (BIPM) and is reported in Circular T as the TAI scale interval $d = -y_\text{TAI/SI}$, where $y_\text{TAI/SI} = f_{\mathrm{TAI}, T_3}/f_{\mathrm{SI},T_3} - 1$.

All extrapolation uncertainties are evaluated using the Fourier-transform method \cite{Dawkins2007a}. 
The noise model for the hydrogen maser is given in Appendix~\ref{sec:maser} and that of TAI (or EAL, \'Echelle Atomique Libre) is reported in the BIPM \emph{etoile} files \footnote{Files reporting the fractional frequency of EAL per Circular T month. Also list $u_\mathrm{A}$, $u_\mathrm{B}$, and weights of all contributing PSFS. Available at \href{https://webtai.bipm.org/ftp/pub/tai/other-products/etoile/}{https://webtai.bipm.org/ftp/pub/tai/other-products/etoile/}}. The uncertainty of the UTC link is evaluated using the recommended formula \cite{Panfilo2010a}, 
\begin{equation} \label{eq:UTClink}
    u_\mathrm{l}(\tau,u_{\mathrm{A,l,s}},u_{\mathrm{A,l,e}}) = \frac{\sqrt{u_{\mathrm{A,l,s}}^2 + u_{\mathrm{A,l,e}}^2}}{5\;\mathrm{d}} \left( \frac{\tau}{5\;\mathrm{d}}\right)^{-0.9},
\end{equation}
where $u_{\mathrm{A,l,s}}$ ($u_{\mathrm{A,l,e}}$) is the statistical uncertainty of $\mathrm{UTC}-\mathrm{UTC}(\text{MIKE})$ reported in Circular~T at the start (end) of the measurement interval and $\tau$ is the measurement time. For this campaign, the statistical uncertainty was \qty{0.2}{ns} for all months except for Circular T 442 (October 2024), when it was \qty{0.3}{ns}.

To allow the statistical uncertainty of $y_\text{TAI/SI}$, $u_\mathrm{A,TAI/SI}$, to average down when averaging multiple months, the published total uncertainty of $y_\text{TAI/SI}$ is divided into its systematic and statistical parts as follows. All measurements used to determine $y_\mathrm{TAI/SI}$ for a given month are published in the corresponding \emph{etoile} file \cite{Note4}, including the systematic uncertainty and weight of each measurement. This analysis includes also measurements taken 360 days before the month in question, whose statistical uncertainty includes an extrapolation uncertainty. For each PSFS included, we calculate the total weight and weighted-mean systematic uncertainty.
The total systematic uncertainty, $u_\mathrm{B,TAI/SI}$, is then obtained by assuming that the systematic uncertainties of the contributing PSFSs are uncorrelated. This is generally considered a good approximation for the uncertainty contributions of cesium fountain clocks. In the case of secondary representations of the second (SRSs), the uncertainty of the recommended frequency value ($u_\text{Srep}$ in Circular~T) and common atomic parameters, e.g., the differential polarizability used to evaluate the BBR shift, cause correlations between different clocks based on the same transition. Only during the last month of our measurement did two SRSs based on Yb \cite{Pizzocaro2020a,Kobayashi2022a} contribute to TAI, and the systematic uncertainty of both is dominated by the BBR temperature rather than the polarizability. The correlated $u_\text{Srep}$, on the other hand, was taken into account, which increased $u_\mathrm{B,TAI/SI}$ from \num{8.1e-17} to \num{8.7e-17} for this month (a third Yb clock contributed via historical data only and had a negligible weight). The PSFSs contributing to our absolute frequency value are listed in Table~\ref{tab:weights} in Appendix~\ref{sec:corr}. Having obtained the total $u_\mathrm{B,TAI/SI}$ of a month, we calculate  $u_\mathrm{A,TAI/SI}$ from the total uncertainty, using the value with three-digit precision in the \emph{etoile} file to avoid rounding errors. 

For the NIM5 cesium fountain clock, the TAI submissions for Circular~T 438--441 were corrected by $\Delta d_\text{NIM5} = \num{0.35e-15}$ in Circular~T 443. We apply this correction using the weights of NIM5 in the \emph{etoile} files, which increases the TAI scale interval $d$ by \num{0.01e-15} for these months.

Each month is analyzed similarly to how PSFS data are analyzed for TAI steering. We evaluate the mean of $y_{\mathrm{Sr^+/HM}}$ and correct the value to the midpoint  of the analysis interval using the measured maser drift (see Appendix~\ref{sec:maser}) and the offset between the midpoint and the barycenter of the data. The uncertainty from the maser drift is very small (${\lesssim} \num{2e-19}$) and is included in the maser extrapolation uncertainty. An uncertainty budget for each month is shown in Table~\ref{tab:tai}. The table also shows the TAI calibration uncertainty for each month, as it would appear in Circular T, which for seven out of ten months is ${\lesssim}\num{2e-16}$, the target of the roadmap criterion I.4 \cite{Dimarcq2024a}.

In order to account for correlations and to put more weight on months with low extrapolation uncertainty, a weighted mean of the monthly results was calculated using the Gauss-Markov theorem (generalized least-squares) \cite{Cox2006a,Pizzocaro2020a,Nemitz2021a}.
The systematic uncertainties of the RRS and rf synthesis and distribution were taken as fully correlated from month to month. For the varying systematic uncertainty of the Sr$^+$ clock, the smaller value was taken as the correlated fraction for each pair of months. The correlation coefficients between the monthly $u_\mathrm{B,TAI/SI}$ values were calculated using the weights from the \emph{etoile} files by assuming that the systematic uncertainty of each individual PSFS was fully correlated from month to month (and adding the small contribution from the correlated $u_\text{Srep}$ for the Yb clocks). The resulting correlation coefficients ranged from 0.55 to 0.99, where a large value means that the same PSFSs contributed with similar weights for the two months in question.

The correlations from maser extrapolation were evaluated using a generalization of the Fourier transform method as described in the Supplementary Material of \cite{Lindvall2025a}.
Here, only neighboring months were considered correlated, as the uncertainty of the maser-noise correlation function is larger than its absolute value for lags beyond 30\;d. The correlation from the laser drift uncertainty was also included but is negligible.

Equation~\eqref{eq:UTClink} for the link uncertainty is valid only for an unweighted mean of 5\;d periods. The mean (fractional) frequency of two intervals with duration $T_i$ and mean frequency $y_i$ is thus $y_{1+2} = (T_1 y_1 + T_2 y_2)/(T_1+T_2)$. 
The link-related covariance between adjacent intervals can be evaluated by requiring the uncertainty of $y_{1+2}$, evaluated using the law of propagation of uncertainty for correlated quantities, to fulfill 
Eq.~\eqref{eq:UTClink} for $\tau = T_1 +T_2$ \cite{Pizzocaro2020a}, which gives 
\begin{eqnarray}
    \sigma^2_{12} = (2 T_1 T_2)^{-1} \big[ 
    (T_1+T_2)^2 u_\mathrm{l}^2(T_1+T_2,u_\mathrm{A,l,1},u_\mathrm{A,l,2}) \nonumber \\
    - T_1^2 u_\mathrm{l}^2(T_1,u_\mathrm{A,l,1},u_\mathrm{A,l,1})  
    - T_2^2 u_\mathrm{l}^2(T_2,u_\mathrm{A,l,2},u_\mathrm{A,l,2}) \big].
\end{eqnarray}
For intervals with $T_1 = T_2 = T$ and $u_\mathrm{A,l,1} = u_\mathrm{A,l,2} = u_\mathrm{A,l}$, this leads to a correlation coefficient of $r_{12} = \sigma_{12}^2/u_\mathrm{l}^2(T,u_\mathrm{A,l},u_\mathrm{A,l}) = 2^{-0.8}-1 \approx -0.43$ due to the $(\tau/5\;\mathrm{d})^{-0.9}$ scaling, as opposed to $-0.5$ that would be obtained from $\tau^{-1}$ scaling.

The dominant contributions to the covariance matrix are of order $+10^{-32}$ between all months from  $u_\mathrm{B,TAI/SI}$ and $-10^{-32}$ between adjacent intervals from the UTC link. The contributions from maser extrapolation are one order of magnitude smaller, and the remaining ones even smaller.

\begin{figure}[t]
\includegraphics[width=1\columnwidth]{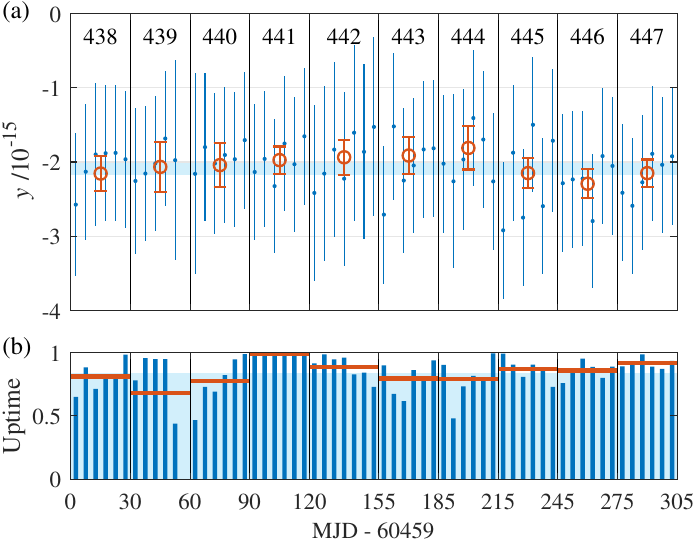}%
\caption{(a) Fractional frequency deviation from the 2021 CIPM recommended frequency value \qty{444 779 044 095 486.3}{Hz} \cite{Margolis2024a} for the 5\;d periods (dots with error bars), months (circles with error bars), and full 10 months (shaded area denoting $\pm1\sigma).$ Vertical lines separate the Circular T periods, labeled above the data. (b) Fractional uptime of the Sr$^+$ clock for the 5\;d periods (bars), months (horizontal lines), and full 10 months (shaded area).
\label{fig:yTAI}}
\end{figure}

The Gauss-Markov theorem then gives the weights and total uncertainties shown in Table~\ref{tab:tai}. The corresponding absolute frequency value is \num{444779044095485.373(44)}\;Hz. With a fractional uncertainty of \num{9.8e-17}, this is, to our knowledge, the most precise frequency measurement reported to date.
The monthly frequency values are shown in Fig.~\ref{fig:yTAI}(a). Even though they are not explicitly used in the analysis, also the 5\;d frequency values, which include significant TAI extrapolation uncertainties, are shown in order to demonstrate the statistical consistency. The reduced $\chi^2$ values of 0.12 and 0.49 for the 5\;d and monthly values, respectively, indicate that the uncertainties of the UTC link and $y_\mathrm{TAI/SI}$ as well as the instability of EAL are quite conservatively estimated by the BIPM, which was also concluded in \cite{Nemitz2021a}. Fig.~\ref{fig:yTAI}(b) shows the fractional uptime of the clock over the 5\;d, 1\;month, and 10\;month periods.

\subsection{Comparison with previous measurements}

In Fig.~\ref{fig:abs_freq}, the two absolute frequency values from this work are  compared to previously published values \cite{Margolis2004a,Dube2005a,Madej2012b,Barwood2014a,Dube2017a,Jian2023a,Steinel2023a,Marceau2025a} and to the 2021 CIPM recommended frequency value \cite{Margolis2024a}. All measurements agree within $2\sigma$, except \cite{Barwood2014a}, which in the light of recent high-accuracy values from PTB, NRC, and this work seems like a clear outlier. A weighted mean of these three values indicates that the 2021 CIPM recommended frequency value is too large by 1.6 times its uncertainty.

Our value against TAI is in excellent agreement with the previously most accurate value from NRC \cite{Marceau2025a}, while the value from PTB \cite{Steinel2023a} is lower than these by $1.3\sigma$ and $1.7\sigma$, respectively. The value in \cite{Steinel2023a} is derived from an optical $^{88}$Sr$^+$/$^{171}$Yb$^+$(E3) frequency ratio and the $^{171}$Yb$^+$(E3) absolute frequency value from \cite{Lange2021a}. For comparison, we have therefore plotted also the absolute frequency obtained from the optical ratio of our $^{88}$Sr$^+$ clock against the same $^{171}$Yb$^+$(E3) clock at PTB \cite{Lindvall2025a} combined with \cite{Lange2021a}. Even though the measurements against CSF2 and $^{171}$Yb$^+$(E3) consist of the same $^{88}$Sr$^+$ and IPPP link data to ${\approx}98.4\%$, this gives a lower frequency close to the one from \cite{Steinel2023a}, indicating that the lower PTB value likely is due to the $^{171}$Yb$^+$(E3) absolute frequency in \cite{Lange2021a}. There, the authors observed a difference of \num{2.6(2.7)e-16} between the frequencies measured against two Cs fountains and took a weighted mean.
This demonstrates that these absolute frequency measurements are at the very limit of what the current Cs-based definition of the second can achieve and highlights the importance of optical ratio measurements for validating the consistency of optical clocks independently of Cs.

\begin{figure}[t]
\includegraphics[width=1\columnwidth]{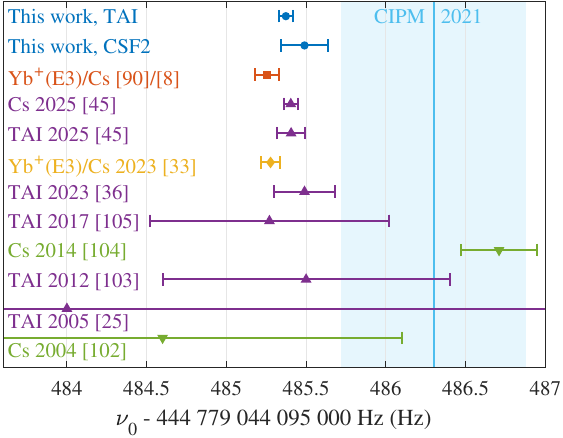}%
\caption{Absolute frequency values from this work ($\bullet$) compared to published values from NRC ($\blacktriangle$), PTB ($\blacklozenge$), 
and NPL ($\blacktriangledown$).  
Labels indicate means of measurement (against a Cs fountain, TAI, or via another optical transtion), publication year, and reference. 
Also shown is a value ($\blacksquare$) obtained from the $^{88}$Sr$^+$/$^{171}$Yb$^+$(E3) ratio in \cite{Lindvall2025a} and the $^{171}$Yb$^+$(E3) absolute frequency in \cite{Lange2021a}, which gives a similarly low value to \cite{Steinel2023a}. The vertical line and shaded area is the 2021 CIPM recommended frequency and its uncertainty.
\label{fig:abs_freq}}
\end{figure}

Note that the values in this work use the recent differential polarizability from \cite{Lindvall2025b}, 
while previous values from 2017 on (including our optical ratios in \cite{Lindvall2025a}) use the value from \cite{Dube2014a}. However, the difference, about \num{4e-18}, is relatively small compared to the uncertainties of the absolute frequency values.

Correlation coefficients between the two frequency measurements and with previous measurements are evaluated in Appendix~\ref{sec:corr}.

\section{Conclusions} \label{sec:conclusions}

We have presented a detailed uncertainty evaluation for a \Sr\ optical clock with a total estimated systematic uncertainty of \num{7.9e-19}. This  uncertainty, the third lowest reported to date, is enabled by low rf losses, the lowest blackbody-radiation temperature uncertainty reported for an ion clock, our recent measurement of the differential polarizability, and a low vacuum pressure. By stabilizing the remaining unstabilized optical paths, the uncertainty could be further reduced to about \num{6e-19}. The rf magnetic field measured in the ion trap is among the lowest reported thanks to the symmetric trap design. We have also demonstrated clock instabilities down to $\num{2e-15} \tau^{-1/2}$ with no added instability due to magnetic-field noise. A better clock laser should thus allow reaching $\num{\approx 1.5e-15} \tau^{-1/2}$.

Furthermore, we have presented absolute frequency measurements of the \Srtrans\ clock transition against a remote Cs fountain and against International Atomic Time with total uncertainties of \num{3.3e-16} and \num{9.8e-17}, respectively. The latter is the most accurate optical frequency measurement to date. It spanned ten months with monthly uptimes between 68\% and 99\%, which demonstrates the \Sr\ clock's potential for future TAI calibration. 
The uptime is limited mainly by ion loss and could thus be improved further by implementing automated ion reloading.

\section*{Acknowledgments}

We thank Heiner Denker for determining the relativistic red shift, G\'erard Petit for providing the IPPP solutions for the 2022 campaign, Hannu Koivula and Ulla Kallio for the height difference measurement, Pasi H\"akli and Mirjam Bilker-Koivula for advice on geodesy, Gianna Panfilo for helpful discussions on $d$ values,  Marco Pizzocaro for fruitful discussions on correlations, Pierre Dub\'e for numerous discussions about \Sr, Teemu Tomberg for building electronics, and Mikko Merimaa for initiating the optical clock research at MIKES.
This work has been partly supported by the projects 15SIB03 OC18, 17FUN07 CC4C, 18SIB05 ROCIT, and 20FUN01 TSCAC, which have received funding from the EMPIR programme co-financed by the Participating States and from the European Union’s Horizon 2020 research and innovation programme, and the projects 22IEM01 TOCK and 23FUN03 HIOC, which have received funding from the European Partnership on Metrology, co-financed from the European Union’s Horizon Europe Research and Innovation Programme and by the Participating States. 
The work has also been supported by the Research Council of Finland (decisions 339821 and 328786) and is part of the Research Council of Finland Flagship Programme `Photonics Research and Innovation' (PREIN, decision 320168).

\section*{Data availability}

The data that support the findings of this article are openly available \cite{Note2}.

\appendix

\section{Clock self-comparison}  \label{sec:self}

The clock measures the line-center frequencies $\nu_{1/2}$, $\nu_{3/2}$, and $\nu_{5/2}$ of three Zeeman pairs, where the subscripts refer to the absolute value of the involved $D_{5/2}$ sublevels. The tensor-shift-free clock frequency is $\nu_0 = (\nu_{1/2}+\nu_{3/2}+\nu_{5/2})/3$, but in the absence of another clock, we can only compare the line centers to each other. If they are uncorrelated, as is the case for quantum projection noise (QPN) or other white frequency noise processes, the Allan variance (AVAR) of the clock frequency is $\sigma_0^2 = (\sigma_{1/2-3/2}^2 + \sigma_{1/2-5/2}^2 + \sigma_{3/2-5/2}^2)/18$, where $\sigma_{i-j}^2$ denotes the AVAR of $\nu_i - \nu_j$.  This formula is exact even if the instabilities of the pairs differ, e.g., due to different pulse areas (which is not the case if Allan deviations are averaged as in \cite{Dube2015a}). 

The above formula only works if the tensor shifts are constant.
However, we can construct the linear combination $ \Delta_\mathrm{EQS} = (3\nu_{3/2} - 2\nu_{1/2} - \nu_{5/2})/42$, which consists of three line-center differences, has an identically zero tensor shift, and has the same QPN as the clock frequency if all pairs have the same instability. Its ADEV can be used as an EQS-free self-comparison that averages down as long as the change in tensor shift is negligible over a single servo cycle. Due to the different prefactors, it will, however, not be exact if the pairs have different instabilities.

\section{Differential dynamic polarizability \label{sec:polarizability}}

To evaluate the nonresonant  E1 ac Stark shifts, the differential dynamic polarizability of the clock transition is needed.
The dynamic polarizabilities of the \Soh\ and \Dfh\ states were calculated using the sum-over-states approach, including all transitions listed in \cite{UDportal,Jiang2009a} and adding the core and valence-core contributions (assumed to be static) from \cite{Jiang2009a}. This method gives an \Soh\ polarizability in good agreement with that available in \cite{UDportal}.

For the transitions $5s\,{}^2\!S_{1/2} \rightarrow 5p\,{}^2\!P_{1/2,\, 3/2},\;6p\,{}^2\!P_{1/2,\, 3/2}$ and $4d\,{}^2\!D_{5/2} \rightarrow 5p\,{}^2\!P_{3/2},\;  6p\,{}^2\!P_{3/2}$, low-uncertainty theoretical matrix elements from \cite{Roberts2023a} were then substituted.
The significant tail contribution to the \Dfh\ polarizability from $4d\,{}^2\!D_{5/2} \rightarrow nf\,{}^2\!F_{5/2,\, 7/2}$ transitions with $n>12$ is modeled by an effective transition at 140\;nm with a matrix element that is adjusted to make the differential static scalar polarizability agree with our experimental value \cite{Lindvall2025b}. 

These matrix elements were also used to calculate the dynamic correction \cite{Porsev2006a}: $\eta(295\;\mathrm{K}) =  -0.008\,95(17)$ and $\eta(300\;\mathrm{K}) =  -0.009\,27(17)$. The uncertainty is dominated by the uncertainty of the $4d\,{}^2\!D_{5/2} \rightarrow 5p\,{}^2\!P_{3/2}$ matrix element.

\section{Comb and rf synthesis and distribution uncertainty} \label{sec:comb_rf}

Systematic effects from signal distribution and optical-to-rf conversion have recently been studied for an intermittently operating clock, and diurnal effects were found to dominate, while the uncertainty related to `continuous errors' was constrained to \num{\approx5e-18} \cite{Nemitz2024a}. 
Due to our high uptimes and long measurement, diurnal effects are expected to be negligible. For example, a loop-back measurement of the longest (\qty{\approx 25}{m} one way) and least temperature-stable coaxial cable between our UTC and GNSS receiver laboratories averaged down to \num{-2.5(25)e-18} in 18~days.

All rf and DDS frequencies in our clock are phase locked to the HM.
The \qty{1348}{nm} laser frequency measured at the comb is $f_\text{IR} = f_0 + n f_\mathrm{r} + f_\mathrm{b} - f_\mathrm{fn}$, where $f_0$ is the carrier offset frequency, $n \approx \num{890000}$ is the comb mode number,  $f_\mathrm{b}$ is the beat note between between the comb and the \qty{1348}{nm} light, and $f_\mathrm{fn} = 75$\;MHz is the fiber-noise AOM frequency. The fourth harmonic of the \qty{250}{MHz} repetition rate $f_\mathrm{r}$ is locked or measured such that $f_\mathrm{r} = (\qty{980}{MHz} + f_\text{dmr})/4$, where $f_\text{dmr}$ stands for down-mixed repetition rate.

During the 2022 campaign, the repetition rate was locked to the HM by locking $f_\text{dmr}$ to a DDS. 
This arrangement multiplies the DDS quantization error by $n/2$ (at 674\;nm), causing a shift that was characterized to be \num{-1.71(5)e-16}. Allowing for undetected rf distribution errors, we rounded the systematic uncertainty related to the comb and rf distribution to \num{1.0e-17}. 
During this campaign, the software-defined radio (SDR) counter used to measure $f_\mathrm{b}$ was used in $\Omega$ counting mode. $\Pi$-averaging of $\Omega$ samples results in downsampling of high-frequency white phase noise and gives rise to a dead-time uncertainty averaging down as $\tau^{-1/2}$ (similar to what was found for nonoverlapping triangle weighting in \cite{Dawkins2007a} and to the Dick effect \cite{Dick1987a}). This is included as `comb, statistical' in Table~\ref{tab:rocit}. 
The frequency bias of $f_\mathrm{b}$ caused by the dead time resulted in a negligible clock shift of ${\lesssim}10^{-18}$.

For the 2024--2025 campaign, the comb repetition rate was locked to the clock laser and $f_\text{dmr}$ was measured against the HM.
Here, the accuracy of $f_\text{dmr}$ is critical. Only dead-time-free $\Lambda$ and $\Pi$ counter data were used. The SDR counter was also verified to be bias free by measuring a known DDS frequency and by comparing against K+K FXM50 and Keysight 53230A (in RCON mode) counters.

To evaluate possible errors related to rf synthesis and distribution, the frequency of a \qty{1378.3}{nm} laser relative to the HM was measured using two independent combs in different laboratories. The comparison comb is locked directly to the HM, while the \Sr\ clock comb repetition rate is compared to the HM as described above. The rf reference is delivered to the two combs via independent coaxial cables. After about 5 days, the frequency difference was \num{2(1)e-17}. The result is likely limited mainly by the unstabilized fiber between the laboratories, but we take \num{2e-17} as an upper limit for the systematic uncertainty related to rf synthesis and distribution.

\section{Relativistic red shift} \label{sec:RRS}

The relativistic red shift (RRS) was determined using the GNSS/geoid method \cite{Denker2018a}.
The coordinates of the MI05 GNSS antenna reference point (ARP) in the ITRF2014 realization of the International Terrestrial Reference System (at a certain epoch) were obtained from Precise Point Positioning (PPP) processing of daily RINEX (Receiver Independent Exchange Format) files. To average the annual variations of the PPP height, the height was determined for 20 days per year from epoch 2019.95 to 2022.85 and was corrected to the epoch 2022.2 of the ROCIT March 2022 campaign using the absolute land uplift of 4.2\;mm/y, obtained from the NKG2016LU land uplift model \cite{Vestol2019a}. The antenna height was then determined as the average of these 59 values with a standard deviation of 7\;mm.
To validate the result, the same procedure was repeated for the MI04 antenna (with the sampling offset by 1 day to avoid common outliers). Accounting for the known height difference between the ARPs, the results agree within 3\;mm. 

The height difference between the ARP and the ion was determined by surveying by the Finnish Geospatial Research Institute. To determine the difference between elliptical and normal height over the 48\;m horizontal displacement within the building, the ARP coordinates were transformed to the local European Terrestrial Reference Frame (ETRF) realization, EUREF-FIN, using the NKG2020 transformation \cite{Hakli2023a}, and the FIN2005N00 geoid model was used, which gave a 1\;mm correction.

The gravity potential at the position of the ion relative to the reference potential $W_0 = \qty{62636856.00}{m^2/s^2}$ was then determined as in \cite{Denker2018a}, yielding a relativistic red shift of $8.020(24)\times 10^{-16}$ at epoch 2022.2. Using the expression for the fractional red shift, $\Delta\nu_\mathrm{RRS}/\nu_0 = g H/c^2$, where $g = 9.819\,09(8)\;\mathrm{m/s^2}$ is the gravitational acceleration measured in the laboratory, this corresponds to a height of $H = 7.341(22)$\;m. The uncertainty contributions are the geoid model (19\;mm), the GNSS height (10\;mm), and the height difference between the antenna and the ion (1.7\;mm).
The height $H$ and the RRS can be corrected to other epochs using the levelled land uplift of $3.8\;\mathrm{mm/y}$ from NKG2016LU \cite{Vestol2019a}.

\begin{figure}[tb]
\includegraphics[width=1\columnwidth]{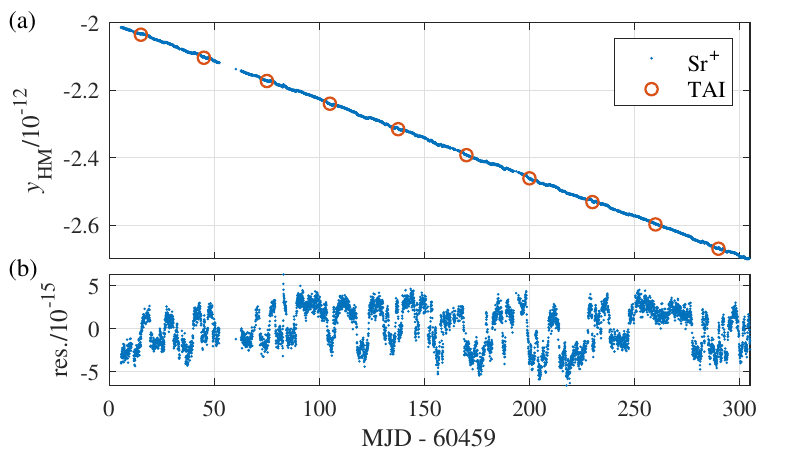}\\
\includegraphics[width=1\columnwidth]{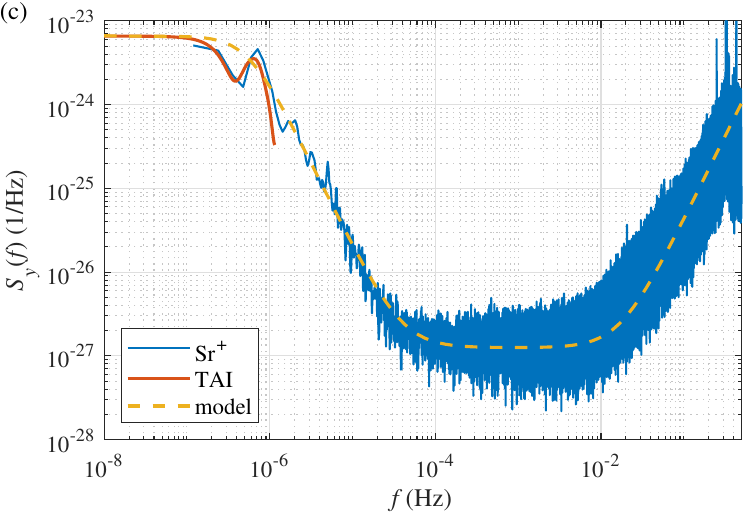}\\
\includegraphics[width=1\columnwidth]{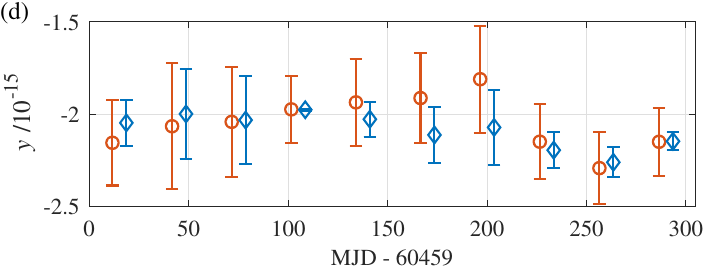}%
\caption{(a) Maser fractional frequency as measured against Sr$^+$ (3600\;s bins) and TAI (monthly values), (b) linear fit residuals of the Sr$^+$ data, and (c) fractional-frequency power spectral density $S_y$ showing measured (Sr$^+$ and 5\;d TAI) and modeled maser noise. (d) Monthly \Sr\ $y$ values from Fig.~\ref{fig:yTAI}(a) with total uncertainties as error bars (circles) compared to interpolated values with extrapolation uncertainties as error bars (diamonds). The circles and diamonds are offset horizontally for clarity.
\label{fig:maser}}
\end{figure}

\section{Maser noise model} \label{sec:maser}

Figure~\ref{fig:maser}(a) shows the fractional frequency of the maser as measured against the optical clock and against TAI. The maser drift is very linear, but the linear fit residuals in Fig.~\ref{fig:maser}(b) reveal that, in addition to the drift, the maser frequency undergoes random frequency jumps between two `levels'. This causes a bump in the ADEV and power spectral density (PSD) that cannot be modeled using polynomial-law noise types. As in \cite{Lindvall2025a}, this is modeled by a Lorentzian peak in the one-sided fractional-frequency PSD, 
\begin{equation}  
S_y(f) = h_2 f^2 + h_0 + h_{-1}/f + A/(1 + (f-f_{0})^2/\delta f^2),
\end{equation}
see Fig.~\ref{fig:maser}(c) and Table~\ref{tab:maser}. 

\begin{table}[tb]
\centering
\caption{
Maser noise coefficients. The coefficients for the polynomial-law noise types $h_i$ are given as ADEV at 1\;s and converted to fractional-frequency PSD $S_y$.
\label{tab:maser}}
\begin{ruledtabular}
\begin{tabular}
{l l l l} 
White phase noise & $h_2$ & $(\num{4.0e-13})^2/(0.076 / 2)$ & Hz$^{-3}$ \\
White frequency noise & $h_0$ & $2 (\num{2.5e-14})^2 $ & Hz$^{-1}$ \\
Flicker frequency noise & $h_{-1}$ & $(\num{0.5e-16})^2/(2 \ln{2})$ & 1 \\
Lorentzian peak & $A$ & \num{6.5e-24} & Hz$^{-1}$ \\
 & $f_0$ & \num{5e-8} & Hz \\
 & $\delta f$ & \num{0.55e-6} & Hz \\
\end{tabular}
\end{ruledtabular}
\end{table}

Note that during the 2022 ROCIT campaign, the maser frequency did not jump and thus a different  noise model was used for this particular maser in \cite{Lindvall2025a} and for the measurement against CSF2 in this work.

The random frequency jumps dominate the extrapolation uncertainty. To validate the noise model, we bin 3600 valid 1\;S/s values and interpolate the result onto a continuous 1\;h wall-clock grid. This assumes that no large frequency jumps occurred during the downtime of the Sr$^+$ clock. Figure~\ref{fig:maser}(d) shows the monthly means of the interpolated data compared to the values from Fig.~\ref{fig:yTAI}(a). The difference is ${\lesssim}1.3u_\text{extr}$ (extrapolation uncertainty) for each month, and the difference in weighted mean of all 10 months is $\num{-2.8e-17} = -0.87 u_\text{extr} = -0.28 u_\text{tot}$ (total uncertainty). This confirms that the maser noise model treats the frequency jumps adequately. 
The reduced $\chi^2$ of the interpolated values is 0.24 compared to 0.49 for the normal values.

\section{Correlations} \label{sec:corr}

Correlation coefficients between the absolute frequency measurement against PTB CSF2 ($q_1$) and the optical ratios measured during the ROCIT March 2022 optical clock comparison campaign were evaluated as described in the supplementary material of \cite{Lindvall2025a} and are listed in Table~\ref{tab:rocit-corr}. As expected, there is significant correlation with the optical ratios involving the Sr$^+$ clock (ratios 18, 19, 25, 26, 33--35), but also smaller ones with ratios involving the optical clocks at PTB due to the shared GNSS receiver and IPPP link.

Table~\ref{tab:weights} lists primary and secondary frequency standards (PSFS) with weight $\num{\geq0.001}$ in our absolute frequency value against TAI together with weights, systematic uncertainties weighted as described in Sec.~\ref{sec:TAI}, and references. The monthly weights of all PSFSs are given in \cite{Note2}.

\begin{table}[b]
\centering
\caption{
Nonzero correlation coefficients between the Sr$^+$/PTB CSF2 absolute frequency in this work ($q_1$) and the GNSS optical-clock frequency ratios in \cite{Lindvall2025a}, numbered following Table~2 in \cite{Lindvall2025a}. 
\label{tab:rocit-corr}}
\begin{ruledtabular}
\begin{tabular}
{l@{\ $=$\ }S[table-format=+1.4] l@{\ $=$\ }S[table-format=+1.4] l@{\ $=$\ }S[table-format=+1.3]} 
$r(q_1,13)$ &  0.013 & $r(q_1,16)$ & -0.007 & $r(q_1,17)$ & -0.001 \\
$r(q_1,18)$ & -0.069 & $r(q_1,19)$ & -0.236 & $r(q_1,20)$ &  0.002 \\
$r(q_1,22)$ & -0.007 & $r(q_1,23)$ & -0.002 & $r(q_1,25)$ & -0.186 \\
$r(q_1,26)$ & -0.081 & $r(q_1,28)$ &  0.002 & $r(q_1,33)$ &  0.094 \\
$r(q_1,34)$ &  0.112 & $r(q_1,35)$ &  0.148 & $r(q_1,36)$ &  0.004 \\
$r(q_1,38)$ & -0.001 \\
\end{tabular}
\end{ruledtabular}
\end{table}

Concerning correlations between the 2024--2025 absolute frequency ($q_2$) and measurements from March 2022, we note that the correlated systematic uncertainty of the Sr$^+$ clock is small (\num{\approx 1e-18}) compared to the total uncertainties, so its contribution can be neglected. 
The correlation coefficient between the two absolute frequency measurements is significant due to the high TAI weight of CSF2 (see Table~\ref{tab:weights}), whose systematic uncertainty can be considered fully correlated between 2022 and 2025: $r(q_2,q_1) = 0.238$.
Among the optical clocks contributing to both the ROCIT campaign \cite{Lindvall2025a} and TAI, the correlation coefficient between $q_2$ and the NMIJ-Yb1/VTT Sr$^+$ ratio is  $r(q_2,26) = -0.023$, where the $u_\mathrm{B}$ of the Yb clock was considered fully correlated, but not the rf distribution uncertainty.
The correlation coefficient with the ratio involving IT-Yb1 is small, $r(q_2,25) = -0.001$, and the one with the ratio involving NPL-Sr1 is negligible.

\begin{table}[h!]
\centering
\caption{
Primary and secondary frequency standards (PSFS) with weight $\num{\geq0.001}$ in our absolute frequency value against TAI, weights, and weighted systematic uncertainties. 
\label{tab:weights}}
\begin{ruledtabular}
\begin{tabular}
{l S[table-format=1.4] S c} 
PSFS        & {Weight} & {$u_{\mathrm{B,mean}}$} & Reference \\
\midrule
PTB-CSF2    & 0.267 & \num{1.7e-16} & \cite{Weyers2018a}\\
SU-CsFO2     & 0.163 & \num{2.2e-16} & \cite{Blinov2017a}\\
NRC-FCs2    & 0.138 & \num{2.1e-16} & \cite{Beattie2020a}\\
LNEOP-FO2      & 0.082 & \num{2.4e-16} & \cite{Guena2012a}\\
PTB-CSF1    & 0.078 & \num{2.9e-16} & \cite{Weyers2018a}\\
LNEOP-FORb\footnotemark[1]  & 0.051 & \num{4.2e-16} & \cite{Guena2014a}\\
NMIJ-Yb1\footnotemark[1]  & 0.049 & \num{2.4e-16} & \cite{Kobayashi2022a}\\
NTSC-CsF2    & 0.041 & \num{4.0e-16} & \cite{Wang2023a}\\
IT-Yb1\footnotemark[1]   & 0.035 & \num{1.9e-16} & \cite{Pizzocaro2020a}\\
IT-CsF2     & 0.034 & \num{2.3e-16} & \cite{Levi2014a}\\
NIM5    & 0.026 & \num{6.8e-16} & \cite{Fang2015a}\\
LNEOP-FO1      & 0.022 & \num{3.7e-16} & \cite{Guena2012a}\\
NPL-Sr1\footnotemark[1]  & 0.009 & \num{2.1e-16} & \cite{Hobson2020b}\\
METAS-FOC2     & 0.003 & \num{1.8e-15} & \cite{Jallageas2018a}\\
\end{tabular}
\end{ruledtabular}
\footnotetext[1]{Secondary representation of the second (SRS). $u_{\mathrm{B,mean}}$ includes $u_\text{Srep}$.}
\end{table}


\begin{thebibliography}{119}%
\makeatletter
\providecommand \@ifxundefined [1]{%
 \@ifx{#1\undefined}
}%
\providecommand \@ifnum [1]{%
 \ifnum #1\expandafter \@firstoftwo
 \else \expandafter \@secondoftwo
 \fi
}%
\providecommand \@ifx [1]{%
 \ifx #1\expandafter \@firstoftwo
 \else \expandafter \@secondoftwo
 \fi
}%
\providecommand \natexlab [1]{#1}%
\providecommand \enquote  [1]{``#1''}%
\providecommand \bibnamefont  [1]{#1}%
\providecommand \bibfnamefont [1]{#1}%
\providecommand \citenamefont [1]{#1}%
\providecommand \href@noop [0]{\@secondoftwo}%
\providecommand \href [0]{\begingroup \@sanitize@url \@href}%
\providecommand \@href[1]{\@@startlink{#1}\@@href}%
\providecommand \@@href[1]{\endgroup#1\@@endlink}%
\providecommand \@sanitize@url [0]{\catcode `\\12\catcode `\$12\catcode `\&12\catcode `\#12\catcode `\^12\catcode `\_12\catcode `\%12\relax}%
\providecommand \@@startlink[1]{}%
\providecommand \@@endlink[0]{}%
\providecommand \url  [0]{\begingroup\@sanitize@url \@url }%
\providecommand \@url [1]{\endgroup\@href {#1}{\urlprefix }}%
\providecommand \urlprefix  [0]{URL }%
\providecommand \Eprint [0]{\href }%
\providecommand \doibase [0]{https://doi.org/}%
\providecommand \selectlanguage [0]{\@gobble}%
\providecommand \bibinfo  [0]{\@secondoftwo}%
\providecommand \bibfield  [0]{\@secondoftwo}%
\providecommand \translation [1]{[#1]}%
\providecommand \BibitemOpen [0]{}%
\providecommand \bibitemStop [0]{}%
\providecommand \bibitemNoStop [0]{.\EOS\space}%
\providecommand \EOS [0]{\spacefactor3000\relax}%
\providecommand \BibitemShut  [1]{\csname bibitem#1\endcsname}%
\let\auto@bib@innerbib\@empty
\bibitem [{\citenamefont {Ludlow}\ \emph {et~al.}(2015)\citenamefont {Ludlow}, \citenamefont {Boyd}, \citenamefont {Ye}, \citenamefont {Peik},\ and\ \citenamefont {Schmidt}}]{Ludlow2015b}%
  \BibitemOpen
  \bibfield  {author} {\bibinfo {author} {\bibfnamefont {A.~D.}\ \bibnamefont {Ludlow}}, \bibinfo {author} {\bibfnamefont {M.~M.}\ \bibnamefont {Boyd}}, \bibinfo {author} {\bibfnamefont {J.}~\bibnamefont {Ye}}, \bibinfo {author} {\bibfnamefont {E.}~\bibnamefont {Peik}},\ and\ \bibinfo {author} {\bibfnamefont {P.~O.}\ \bibnamefont {Schmidt}},\ }\bibfield  {title} {\bibinfo {title} {{O}ptical atomic clocks},\ }\href {https://doi.org/10.1103/RevModPhys.87.637} {\bibfield  {journal} {\bibinfo  {journal} {Rev. Mod. Phys.}\ }\textbf {\bibinfo {volume} {87}},\ \bibinfo {pages} {637} (\bibinfo {year} {2015})}\BibitemShut {NoStop}%
\bibitem [{\citenamefont {Brewer}\ \emph {et~al.}(2019)\citenamefont {Brewer}, \citenamefont {Chen}, \citenamefont {Hankin}, \citenamefont {Clements}, \citenamefont {Chou}, \citenamefont {Wineland}, \citenamefont {Hume},\ and\ \citenamefont {Leibrandt}}]{Brewer2019a}%
  \BibitemOpen
  \bibfield  {author} {\bibinfo {author} {\bibfnamefont {S.~M.}\ \bibnamefont {Brewer}}, \bibinfo {author} {\bibfnamefont {J.-S.}\ \bibnamefont {Chen}}, \bibinfo {author} {\bibfnamefont {A.~M.}\ \bibnamefont {Hankin}}, \bibinfo {author} {\bibfnamefont {E.~R.}\ \bibnamefont {Clements}}, \bibinfo {author} {\bibfnamefont {C.~W.}\ \bibnamefont {Chou}}, \bibinfo {author} {\bibfnamefont {D.~J.}\ \bibnamefont {Wineland}}, \bibinfo {author} {\bibfnamefont {D.~B.}\ \bibnamefont {Hume}},\ and\ \bibinfo {author} {\bibfnamefont {D.~R.}\ \bibnamefont {Leibrandt}},\ }\bibfield  {title} {\bibinfo {title} {$^{27}${Al}$^{+}$ quantum-logic clock with a systematic uncertainty below ${10}^{-18}$},\ }\href {https://doi.org/10.1103/PhysRevLett.123.033201} {\bibfield  {journal} {\bibinfo  {journal} {Phys. Rev. Lett.}\ }\textbf {\bibinfo {volume} {123}},\ \bibinfo {pages} {033201} (\bibinfo {year} {2019})}\BibitemShut {NoStop}%
\bibitem [{\citenamefont {Aeppli}\ \emph {et~al.}(2024)\citenamefont {Aeppli}, \citenamefont {Kim}, \citenamefont {Warfield}, \citenamefont {Safronova},\ and\ \citenamefont {Ye}}]{Aeppli2024a}%
  \BibitemOpen
  \bibfield  {author} {\bibinfo {author} {\bibfnamefont {A.}~\bibnamefont {Aeppli}}, \bibinfo {author} {\bibfnamefont {K.}~\bibnamefont {Kim}}, \bibinfo {author} {\bibfnamefont {W.}~\bibnamefont {Warfield}}, \bibinfo {author} {\bibfnamefont {M.~S.}\ \bibnamefont {Safronova}},\ and\ \bibinfo {author} {\bibfnamefont {J.}~\bibnamefont {Ye}},\ }\bibfield  {title} {\bibinfo {title} {Clock with $8 \times 10^{-19}$ systematic uncertainty},\ }\href {https://doi.org/10.1103/physrevlett.133.023401} {\bibfield  {journal} {\bibinfo  {journal} {Phys. Rev. Lett.}\ }\textbf {\bibinfo {volume} {133}},\ \bibinfo {pages} {023401} (\bibinfo {year} {2024})} \BibitemShut {NoStop}%
\bibitem [{\citenamefont {Marshall}\ \emph {et~al.}(2025)\citenamefont {Marshall}, \citenamefont {Castillo}, \citenamefont {Arthur-Dworschack}, \citenamefont {Aeppli}, \citenamefont {Kim}, \citenamefont {Lee}, \citenamefont {Warfield}, \citenamefont {Hinrichs}, \citenamefont {Nardelli}, \citenamefont {Fortier}, \citenamefont {Ye}, \citenamefont {Leibrandt},\ and\ \citenamefont {Hume}}]{Marshall2025a}%
  \BibitemOpen
  \bibfield  {author} {\bibinfo {author} {\bibfnamefont {M.~C.}\ \bibnamefont {Marshall}}, \bibinfo {author} {\bibfnamefont {D.~A.~R.}\ \bibnamefont {Castillo}}, \bibinfo {author} {\bibfnamefont {W.~J.}\ \bibnamefont {Arthur-Dworschack}}, \bibinfo {author} {\bibfnamefont {A.}~\bibnamefont {Aeppli}}, \bibinfo {author} {\bibfnamefont {K.}~\bibnamefont {Kim}}, \bibinfo {author} {\bibfnamefont {D.}~\bibnamefont {Lee}}, \bibinfo {author} {\bibfnamefont {W.}~\bibnamefont {Warfield}}, \bibinfo {author} {\bibfnamefont {J.}~\bibnamefont {Hinrichs}}, \bibinfo {author} {\bibfnamefont {N.~V.}\ \bibnamefont {Nardelli}}, \bibinfo {author} {\bibfnamefont {T.~M.}\ \bibnamefont {Fortier}}, \bibinfo {author} {\bibfnamefont {J.}~\bibnamefont {Ye}}, \bibinfo {author} {\bibfnamefont {D.~R.}\ \bibnamefont {Leibrandt}},\ and\ \bibinfo {author} {\bibfnamefont {D.~B.}\ \bibnamefont {Hume}},\ }\bibfield  {title} {\bibinfo {title} {High-stability single-ion clock with $5.5\times10^{-19}$ systematic uncertainty},\ }\href
  {https://link.aps.org/doi/10.1103/hb3c-dk28} {\bibfield  {journal} {\bibinfo  {journal} {Phys. Rev. Lett.}\ }\textbf {\bibinfo {volume} {135}},\ \bibinfo {pages} {033201} (\bibinfo {year} {2025})} \BibitemShut {NoStop}%
\bibitem [{\citenamefont {Zhang}\ \emph {et~al.}(2025)\citenamefont {Zhang}, \citenamefont {Ma}, \citenamefont {Huang}, \citenamefont {Han}, \citenamefont {Hu}, \citenamefont {Wang}, \citenamefont {Zhang}, \citenamefont {Tang}, \citenamefont {Shi}, \citenamefont {Guan},\ and\ \citenamefont {Gao}}]{Zhang2025b}%
  \BibitemOpen
  \bibfield  {author} {\bibinfo {author} {\bibfnamefont {B.}~\bibnamefont {Zhang}}, \bibinfo {author} {\bibfnamefont {Z.}~\bibnamefont {Ma}}, \bibinfo {author} {\bibfnamefont {Y.}~\bibnamefont {Huang}}, \bibinfo {author} {\bibfnamefont {H.}~\bibnamefont {Han}}, \bibinfo {author} {\bibfnamefont {R.}~\bibnamefont {Hu}}, \bibinfo {author} {\bibfnamefont {Y.}~\bibnamefont {Wang}}, \bibinfo {author} {\bibfnamefont {H.}~\bibnamefont {Zhang}}, \bibinfo {author} {\bibfnamefont {L.}~\bibnamefont {Tang}}, \bibinfo {author} {\bibfnamefont {T.}~\bibnamefont {Shi}}, \bibinfo {author} {\bibfnamefont {H.}~\bibnamefont {Guan}},\ and\ \bibinfo {author} {\bibfnamefont {K.}~\bibnamefont {Gao}},\ }\bibfield  {title} {\bibinfo {title} {A liquid-nitrogen-cooled {Ca}$^+$ ion optical clock with a systematic uncertainty of $4.6\times 10^{-19}$}\ } (\bibinfo {year} {2025}),\ \Eprint {https://arxiv.org/abs/2506.17423} {arXiv:2506.17423 [physics.atom-ph]}
  \BibitemShut {NoStop}%
\bibitem [{\citenamefont {Delva}\ \emph {et~al.}(2017)\citenamefont {Delva}, \citenamefont {Lodewyck}, \citenamefont {Bilicki}, \citenamefont {Bookjans}, \citenamefont {Vallet}, \citenamefont {Le~Targat}, \citenamefont {Pottie}, \citenamefont {Guerlin}, \citenamefont {Meynadier}, \citenamefont {Le~Poncin-Lafitte}, \citenamefont {Lopez}, \citenamefont {Amy-Klein}, \citenamefont {Lee}, \citenamefont {Quintin}, \citenamefont {Lisdat}, \citenamefont {Al-Masoudi}, \citenamefont {D\"orscher}, \citenamefont {Grebing}, \citenamefont {Grosche}, \citenamefont {Kuhl}, \citenamefont {Raupach}, \citenamefont {Sterr}, \citenamefont {Hill}, \citenamefont {Hobson}, \citenamefont {Bowden}, \citenamefont {Kronj\"ager}, \citenamefont {Marra}, \citenamefont {Rolland}, \citenamefont {Baynes}, \citenamefont {Margolis},\ and\ \citenamefont {Gill}}]{Delva2017b}%
  \BibitemOpen
  \bibfield  {author} {\bibinfo {author} {\bibfnamefont {P.}~\bibnamefont {Delva}}, \bibinfo {author} {\bibfnamefont {J.}~\bibnamefont {Lodewyck}}, \bibinfo {author} {\bibfnamefont {S.}~\bibnamefont {Bilicki}}, \bibinfo {author} {\bibfnamefont {E.}~\bibnamefont {Bookjans}}, \bibinfo {author} {\bibfnamefont {G.}~\bibnamefont {Vallet}}, \bibinfo {author} {\bibfnamefont {R.}~\bibnamefont {Le~Targat}}, \bibinfo {author} {\bibfnamefont {P.-E.}\ \bibnamefont {Pottie}}, \bibinfo {author} {\bibfnamefont {C.}~\bibnamefont {Guerlin}}, \bibinfo {author} {\bibfnamefont {F.}~\bibnamefont {Meynadier}}, \bibinfo {author} {\bibfnamefont {C.}~\bibnamefont {Le~Poncin-Lafitte}  \emph {et~al.}},\ }\bibfield  {title} {\bibinfo
  {title} {{T}est of {S}pecial {R}elativity {U}sing a {F}iber {N}etwork of {O}ptical {C}locks},\ }\href {https://doi.org/10.1103/PhysRevLett.118.221102} {\bibfield  {journal} {\bibinfo  {journal} {Phys. Rev. Lett.}\ }\textbf {\bibinfo {volume} {118}},\ \bibinfo {pages} {221102} (\bibinfo {year} {2017})}\BibitemShut {NoStop}%
\bibitem [{\citenamefont {Sanner}\ \emph {et~al.}(2019)\citenamefont {Sanner}, \citenamefont {Huntemann}, \citenamefont {Lange}, \citenamefont {Tamm}, \citenamefont {Peik}, \citenamefont {Safronova},\ and\ \citenamefont {Porsev}}]{Sanner2019a}%
  \BibitemOpen
  \bibfield  {author} {\bibinfo {author} {\bibfnamefont {C.}~\bibnamefont {Sanner}}, \bibinfo {author} {\bibfnamefont {N.}~\bibnamefont {Huntemann}}, \bibinfo {author} {\bibfnamefont {R.}~\bibnamefont {Lange}}, \bibinfo {author} {\bibfnamefont {C.}~\bibnamefont {Tamm}}, \bibinfo {author} {\bibfnamefont {E.}~\bibnamefont {Peik}}, \bibinfo {author} {\bibfnamefont {M.~S.}\ \bibnamefont {Safronova}},\ and\ \bibinfo {author} {\bibfnamefont {S.~G.}\ \bibnamefont {Porsev}},\ }\bibfield  {title} {\bibinfo {title} {Optical clock comparison for {L}orentz symmetry testing},\ }\href {https://doi.org/10.1038/s41586-019-0972-2} {\bibfield  {journal} {\bibinfo  {journal} {Nature}\ }\textbf {\bibinfo {volume} {567}},\ \bibinfo {pages} {204} (\bibinfo {year} {2019})}\BibitemShut {NoStop}%
\bibitem [{\citenamefont {Lange}\ \emph {et~al.}(2021)\citenamefont {Lange}, \citenamefont {Huntemann}, \citenamefont {Rahm}, \citenamefont {Sanner}, \citenamefont {Shao}, \citenamefont {Lipphardt}, \citenamefont {Tamm}, \citenamefont {Weyers},\ and\ \citenamefont {Peik}}]{Lange2021a}%
  \BibitemOpen
  \bibfield  {author} {\bibinfo {author} {\bibfnamefont {R.}~\bibnamefont {Lange}}, \bibinfo {author} {\bibfnamefont {N.}~\bibnamefont {Huntemann}}, \bibinfo {author} {\bibfnamefont {J.}~\bibnamefont {Rahm}}, \bibinfo {author} {\bibfnamefont {C.}~\bibnamefont {Sanner}}, \bibinfo {author} {\bibfnamefont {H.}~\bibnamefont {Shao}}, \bibinfo {author} {\bibfnamefont {B.}~\bibnamefont {Lipphardt}}, \bibinfo {author} {\bibfnamefont {C.}~\bibnamefont {Tamm}}, \bibinfo {author} {\bibfnamefont {S.}~\bibnamefont {Weyers}},\ and\ \bibinfo {author} {\bibfnamefont {E.}~\bibnamefont {Peik}},\ }\bibfield  {title} {\bibinfo {title} {Improved limits for violations of local position invariance from atomic clock comparisons},\ }\href {https://doi.org/10.1103/physrevlett.126.011102} {\bibfield  {journal} {\bibinfo  {journal} {Phys. Rev. Lett.}\ }\textbf {\bibinfo {volume} {126}},\ \bibinfo {pages} {011102} (\bibinfo {year} {2021})}\BibitemShut {NoStop}%
\bibitem [{\citenamefont {Sherrill}\ \emph {et~al.}(2023)\citenamefont {Sherrill}, \citenamefont {Parsons}, \citenamefont {Baynham}, \citenamefont {Bowden}, \citenamefont {Curtis}, \citenamefont {Hendricks}, \citenamefont {Hill}, \citenamefont {Hobson}, \citenamefont {Margolis}, \citenamefont {Robertson}, \citenamefont {Schioppo}, \citenamefont {Szymaniec}, \citenamefont {Tofful}, \citenamefont {Tunesi}, \citenamefont {Godun},\ and\ \citenamefont {Calmet}}]{Sherrill2023a}%
  \BibitemOpen
  \bibfield  {author} {\bibinfo {author} {\bibfnamefont {N.}~\bibnamefont {Sherrill}}, \bibinfo {author} {\bibfnamefont {A.~O.}\ \bibnamefont {Parsons}}, \bibinfo {author} {\bibfnamefont {C.~F.~A.}\ \bibnamefont {Baynham}}, \bibinfo {author} {\bibfnamefont {W.}~\bibnamefont {Bowden}}, \bibinfo {author} {\bibfnamefont {E.~A.}\ \bibnamefont {Curtis}}, \bibinfo {author} {\bibfnamefont {R.}~\bibnamefont {Hendricks}}, \bibinfo {author} {\bibfnamefont {I.~R.}\ \bibnamefont {Hill}}, \bibinfo {author} {\bibfnamefont {R.}~\bibnamefont {Hobson}}, \bibinfo {author} {\bibfnamefont {H.~S.}\ \bibnamefont {Margolis}}, \bibinfo {author} {\bibfnamefont {B.~I.}\ \bibnamefont {Robertson}}, \bibinfo {author} {\bibfnamefont {M.}~\bibnamefont {Schioppo}}, \bibinfo {author} {\bibfnamefont {K.}~\bibnamefont {Szymaniec}}, \bibinfo {author} {\bibfnamefont {A.}~\bibnamefont {Tofful}}, \bibinfo {author} {\bibfnamefont {J.}~\bibnamefont {Tunesi}}, \bibinfo {author} {\bibfnamefont {R.~M.}\ \bibnamefont {Godun}},\ and\ \bibinfo {author}
  {\bibfnamefont {X.}~\bibnamefont {Calmet}},\ }\bibfield  {title} {\bibinfo {title} {Analysis of atomic-clock data to constrain variations of fundamental constants},\ }\href {https://doi.org/10.1088/1367-2630/aceff6} {\bibfield  {journal} {\bibinfo  {journal} {New J. Phys.}\ }\textbf {\bibinfo {volume} {25}},\ \bibinfo {pages} {093012} (\bibinfo {year} {2023})} \BibitemShut {NoStop}%
\bibitem [{\citenamefont {Roberts}\ \emph {et~al.}(2020)\citenamefont {Roberts}, \citenamefont {Delva}, \citenamefont {Al-Masoudi}, \citenamefont {Amy-Klein}, \citenamefont {B{\ae}rentsen}, \citenamefont {Baynham}, \citenamefont {Benkler}, \citenamefont {Bilicki}, \citenamefont {Bize}, \citenamefont {Bowden}, \citenamefont {Calvert}, \citenamefont {Cambier}, \citenamefont {Cantin}, \citenamefont {Curtis}, \citenamefont {Dörscher}, \citenamefont {Favier}, \citenamefont {Frank}, \citenamefont {Gill}, \citenamefont {Godun}, \citenamefont {Grosche}, \citenamefont {Guo}, \citenamefont {Hees}, \citenamefont {Hill}, \citenamefont {Hobson}, \citenamefont {Huntemann}, \citenamefont {Kronjäger}, \citenamefont {Koke}, \citenamefont {Kuhl}, \citenamefont {Lange}, \citenamefont {Legero}, \citenamefont {Lipphardt}, \citenamefont {Lisdat}, \citenamefont {Lodewyck}, \citenamefont {Lopez}, \citenamefont {Margolis}, \citenamefont {{\'{A}}lvarez-Mart{\'{\i}}nez}, \citenamefont {Meynadier}, \citenamefont {Ozimek},
  \citenamefont {Peik}, \citenamefont {Pottie}, \citenamefont {Quintin}, \citenamefont {Sanner}, \citenamefont {Sarlo}, \citenamefont {Schioppo}, \citenamefont {Schwarz}, \citenamefont {Silva}, \citenamefont {Sterr}, \citenamefont {Tamm}, \citenamefont {Targat}, \citenamefont {Tuckey}, \citenamefont {Vallet}, \citenamefont {Waterholter}, \citenamefont {Xu},\ and\ \citenamefont {Wolf}}]{Roberts2020a}%
  \BibitemOpen
  \bibfield  {author} {\bibinfo {author} {\bibfnamefont {B.~M.}\ \bibnamefont {Roberts}}, \bibinfo {author} {\bibfnamefont {P.}~\bibnamefont {Delva}}, \bibinfo {author} {\bibfnamefont {A.}~\bibnamefont {Al-Masoudi}}, \bibinfo {author} {\bibfnamefont {A.}~\bibnamefont {Amy-Klein}}, \bibinfo {author} {\bibfnamefont {C.}~\bibnamefont {B{\ae}rentsen}}, \bibinfo {author} {\bibfnamefont {C.~F.~A.}\ \bibnamefont {Baynham}}, \bibinfo {author} {\bibfnamefont {E.}~\bibnamefont {Benkler}}, \bibinfo {author} {\bibfnamefont {S.}~\bibnamefont {Bilicki}}, \bibinfo {author} {\bibfnamefont {S.}~\bibnamefont {Bize}}, \bibinfo {author} {\bibfnamefont {W.}~\bibnamefont {Bowden} \emph {et~al.}},\ }\bibfield  {title} {\bibinfo {title} {Search for transient variations of the fine structure constant and dark matter using fiber-linked optical atomic clocks},\ }\href {https://doi.org/10.1088/1367-2630/abaace} {\bibfield  {journal} {\bibinfo  {journal} {New J. Phys.}\ }\textbf {\bibinfo {volume} {22}},\ \bibinfo {pages} {093010} (\bibinfo {year} {2020})}\BibitemShut {NoStop}%
\bibitem [{\citenamefont {Kennedy}\ \emph {et~al.}(2020)\citenamefont {Kennedy}, \citenamefont {Oelker}, \citenamefont {Robinson}, \citenamefont {Bothwell}, \citenamefont {Kedar}, \citenamefont {Milner}, \citenamefont {Marti}, \citenamefont {Derevianko},\ and\ \citenamefont {Ye}}]{Kennedy2020a}%
  \BibitemOpen
  \bibfield  {author} {\bibinfo {author} {\bibfnamefont {C.~J.}\ \bibnamefont {Kennedy}}, \bibinfo {author} {\bibfnamefont {E.}~\bibnamefont {Oelker}}, \bibinfo {author} {\bibfnamefont {J.~M.}\ \bibnamefont {Robinson}}, \bibinfo {author} {\bibfnamefont {T.}~\bibnamefont {Bothwell}}, \bibinfo {author} {\bibfnamefont {D.}~\bibnamefont {Kedar}}, \bibinfo {author} {\bibfnamefont {W.~R.}\ \bibnamefont {Milner}}, \bibinfo {author} {\bibfnamefont {G.~E.}\ \bibnamefont {Marti}}, \bibinfo {author} {\bibfnamefont {A.}~\bibnamefont {Derevianko}},\ and\ \bibinfo {author} {\bibfnamefont {J.}~\bibnamefont {Ye}},\ }\bibfield  {title} {\bibinfo {title} {Precision metrology meets cosmology: Improved constraints on ultralight dark matter from atom-cavity frequency comparisons},\ }\href {https://doi.org/10.1103/physrevlett.125.201302} {\bibfield  {journal} {\bibinfo  {journal} {Phys. Rev. Lett.}\ }\textbf {\bibinfo {volume} {125}},\ \bibinfo {pages} {201302} (\bibinfo {year} {2020})}\BibitemShut {NoStop}%
\bibitem [{\citenamefont {Beloy}\ \emph {et~al.}(2021)\citenamefont {Beloy}, \citenamefont {Bodine}, \citenamefont {Bothwell}, \citenamefont {Brewer}, \citenamefont {Bromley}, \citenamefont {Chen}, \citenamefont {Deschênes}, \citenamefont {Diddams}, \citenamefont {Fasano}, \citenamefont {Fortier}, \citenamefont {Hassan}, \citenamefont {Hume}, \citenamefont {Kedar}, \citenamefont {Kennedy}, \citenamefont {Khader}, \citenamefont {Koepke}, \citenamefont {Leibrandt}, \citenamefont {Leopardi}, \citenamefont {Ludlow}, \citenamefont {McGrew}, \citenamefont {Milner}, \citenamefont {Newbury}, \citenamefont {Nicolodi}, \citenamefont {Oelker}, \citenamefont {Parker}, \citenamefont {Robinson}, \citenamefont {Romisch}, \citenamefont {Schäffer}, \citenamefont {Sherman}, \citenamefont {Sinclair}, \citenamefont {Sonderhouse}, \citenamefont {Swann}, \citenamefont {Yao}, \citenamefont {Ye},\ and\ \citenamefont {Zhang}}]{Beloy2021a}%
  \BibitemOpen
  \bibfield  {author} {\bibinfo {author} {\bibfnamefont {K.}~\bibnamefont {Beloy}}, \bibinfo {author} {\bibfnamefont {M.~I.}\ \bibnamefont {Bodine}}, \bibinfo {author} {\bibfnamefont {T.}~\bibnamefont {Bothwell}}, \bibinfo {author} {\bibfnamefont {S.~M.}\ \bibnamefont {Brewer}}, \bibinfo {author} {\bibfnamefont {S.~L.}\ \bibnamefont {Bromley}}, \bibinfo {author} {\bibfnamefont {J.-S.}\ \bibnamefont {Chen}}, \bibinfo {author} {\bibfnamefont {J.-D.}\ \bibnamefont {Deschênes}}, \bibinfo {author} {\bibfnamefont {S.~A.}\ \bibnamefont {Diddams}}, \bibinfo {author} {\bibfnamefont {R.~J.}\ \bibnamefont {Fasano}}, \bibinfo {author} {\bibfnamefont {T.~M.}\ \bibnamefont {Fortier} \emph {et~al.} (BACON Collaboration)},\ }\bibfield  {title} {\bibinfo {title} {Frequency ratio measurements at 18-digit accuracy using an optical clock network},\ }\href {https://doi.org/10.1038/s41586-021-03253-4} {\bibfield  {journal} {\bibinfo  {journal} {Nature}\ }\textbf {\bibinfo {volume} {591}},\ \bibinfo {pages} {564} (\bibinfo {year} {2021})}\BibitemShut {NoStop}%
\bibitem [{\citenamefont {Kobayashi}\ \emph {et~al.}(2022)\citenamefont {Kobayashi}, \citenamefont {Takamizawa}, \citenamefont {Akamatsu}, \citenamefont {Kawasaki}, \citenamefont {Nishiyama}, \citenamefont {Hosaka}, \citenamefont {Hisai}, \citenamefont {Wada}, \citenamefont {Inaba}, \citenamefont {Tanabe},\ and\ \citenamefont {Yasuda}}]{Kobayashi2022a}%
  \BibitemOpen
  \bibfield  {author} {\bibinfo {author} {\bibfnamefont {T.}~\bibnamefont {Kobayashi}}, \bibinfo {author} {\bibfnamefont {A.}~\bibnamefont {Takamizawa}}, \bibinfo {author} {\bibfnamefont {D.}~\bibnamefont {Akamatsu}}, \bibinfo {author} {\bibfnamefont {A.}~\bibnamefont {Kawasaki}}, \bibinfo {author} {\bibfnamefont {A.}~\bibnamefont {Nishiyama}}, \bibinfo {author} {\bibfnamefont {K.}~\bibnamefont {Hosaka}}, \bibinfo {author} {\bibfnamefont {Y.}~\bibnamefont {Hisai}}, \bibinfo {author} {\bibfnamefont {M.}~\bibnamefont {Wada}}, \bibinfo {author} {\bibfnamefont {H.}~\bibnamefont {Inaba}}, \bibinfo {author} {\bibfnamefont {T.}~\bibnamefont {Tanabe}},\ and\ \bibinfo {author} {\bibfnamefont {M.}~\bibnamefont {Yasuda}},\ }\bibfield  {title} {\bibinfo {title} {Search for ultralight dark matter from long-term frequency comparisons of optical and microwave atomic clocks},\ }\href {https://doi.org/10.1103/physrevlett.129.241301} {\bibfield  {journal} {\bibinfo  {journal} {Phys. Rev. Lett.}\ }\textbf {\bibinfo {volume}
  {129}},\ \bibinfo {pages} {241301} (\bibinfo {year} {2022})} \BibitemShut {NoStop}%
\bibitem [{\citenamefont {Filzinger}\ \emph {et~al.}(2023)\citenamefont {Filzinger}, \citenamefont {Dörscher}, \citenamefont {Lange}, \citenamefont {Klose}, \citenamefont {Steinel}, \citenamefont {Benkler}, \citenamefont {Peik}, \citenamefont {Lisdat},\ and\ \citenamefont {Huntemann}}]{Filzinger2023a}%
  \BibitemOpen
  \bibfield  {author} {\bibinfo {author} {\bibfnamefont {M.}~\bibnamefont {Filzinger}}, \bibinfo {author} {\bibfnamefont {S.}~\bibnamefont {Dörscher}}, \bibinfo {author} {\bibfnamefont {R.}~\bibnamefont {Lange}}, \bibinfo {author} {\bibfnamefont {J.}~\bibnamefont {Klose}}, \bibinfo {author} {\bibfnamefont {M.}~\bibnamefont {Steinel}}, \bibinfo {author} {\bibfnamefont {E.}~\bibnamefont {Benkler}}, \bibinfo {author} {\bibfnamefont {E.}~\bibnamefont {Peik}}, \bibinfo {author} {\bibfnamefont {C.}~\bibnamefont {Lisdat}},\ and\ \bibinfo {author} {\bibfnamefont {N.}~\bibnamefont {Huntemann}},\ }\bibfield  {title} {\bibinfo {title} {Improved limits on the coupling of ultralight bosonic dark matter to photons from optical atomic clock comparisons},\ }\href {https://doi.org/10.1103/physrevlett.130.253001} {\bibfield  {journal} {\bibinfo  {journal} {Phys. Rev. Lett.}\ }\textbf {\bibinfo {volume} {130}},\ \bibinfo {pages} {253001} (\bibinfo {year} {2023})} \BibitemShut {NoStop}%
\bibitem [{\citenamefont {Chou}\ \emph {et~al.}(2010)\citenamefont {Chou}, \citenamefont {Hume}, \citenamefont {Koelemeij}, \citenamefont {Wineland},\ and\ \citenamefont {Rosenband}}]{Chou2010a}%
  \BibitemOpen
  \bibfield  {author} {\bibinfo {author} {\bibfnamefont {C.~W.}\ \bibnamefont {Chou}}, \bibinfo {author} {\bibfnamefont {D.~B.}\ \bibnamefont {Hume}}, \bibinfo {author} {\bibfnamefont {J.~C.~J.}\ \bibnamefont {Koelemeij}}, \bibinfo {author} {\bibfnamefont {D.~J.}\ \bibnamefont {Wineland}},\ and\ \bibinfo {author} {\bibfnamefont {T.}~\bibnamefont {Rosenband}},\ }\bibfield  {title} {\bibinfo {title} {{F}requency {C}omparison of {T}wo {H}igh-{A}ccuracy {A}l$^+$ {O}ptical {C}locks},\ }\href {https://doi.org/10.1103/PhysRevLett.104.070802} {\bibfield  {journal} {\bibinfo  {journal} {Phys. Rev. Lett.}\ }\textbf {\bibinfo {volume} {104}},\ \bibinfo {pages} {070802} (\bibinfo {year} {2010})}\BibitemShut {NoStop}%
\bibitem [{\citenamefont {Grotti}\ \emph {et~al.}(2018)\citenamefont {Grotti}, \citenamefont {Koller}, \citenamefont {Vogt}, \citenamefont {H\"afner}, \citenamefont {Sterr}, \citenamefont {Lisdat}, \citenamefont {Denker}, \citenamefont {Voigt}, \citenamefont {Timmen}, \citenamefont {Rolland}, \citenamefont {Baynes}, \citenamefont {Margolis}, \citenamefont {Zampaolo}, \citenamefont {Thoumany}, \citenamefont {Pizzocaro}, \citenamefont {Rauf}, \citenamefont {Bregolin}, \citenamefont {Tampellini}, \citenamefont {Barbieri}, \citenamefont {Zucco}, \citenamefont {Costanzo}, \citenamefont {Clivati}, \citenamefont {Levi},\ and\ \citenamefont {Calonico}}]{Grotti2018a}%
  \BibitemOpen
  \bibfield  {author} {\bibinfo {author} {\bibfnamefont {J.}~\bibnamefont {Grotti}}, \bibinfo {author} {\bibfnamefont {S.}~\bibnamefont {Koller}}, \bibinfo {author} {\bibfnamefont {S.}~\bibnamefont {Vogt}}, \bibinfo {author} {\bibfnamefont {S.}~\bibnamefont {H\"afner}}, \bibinfo {author} {\bibfnamefont {U.}~\bibnamefont {Sterr}}, \bibinfo {author} {\bibfnamefont {C.}~\bibnamefont {Lisdat}}, \bibinfo {author} {\bibfnamefont {H.}~\bibnamefont {Denker}}, \bibinfo {author} {\bibfnamefont {C.}~\bibnamefont {Voigt}}, \bibinfo {author} {\bibfnamefont {L.}~\bibnamefont {Timmen}}, \bibinfo {author} {\bibfnamefont {A.}~\bibnamefont {Rolland} \emph{et~al.}},\ }\bibfield  {title} {\bibinfo {title} {{G}eodesy and metrology with a transportable optical clock},\ }\href {https://doi.org/10.1038/s41567-017-0042-3} {\bibfield  {journal} {\bibinfo  {journal} {Nat. Physics}\ }\textbf {\bibinfo {volume} {14}},\ \bibinfo {pages} {437} (\bibinfo {year} {2018})}\BibitemShut {NoStop}%
\bibitem [{\citenamefont {Takamoto}\ \emph {et~al.}(2020)\citenamefont {Takamoto}, \citenamefont {Ushijima}, \citenamefont {Ohmae}, \citenamefont {Yahagi}, \citenamefont {Kokado}, \citenamefont {Shinkai},\ and\ \citenamefont {Katori}}]{Takamoto2020a}%
  \BibitemOpen
  \bibfield  {author} {\bibinfo {author} {\bibfnamefont {M.}~\bibnamefont {Takamoto}}, \bibinfo {author} {\bibfnamefont {I.}~\bibnamefont {Ushijima}}, \bibinfo {author} {\bibfnamefont {N.}~\bibnamefont {Ohmae}}, \bibinfo {author} {\bibfnamefont {T.}~\bibnamefont {Yahagi}}, \bibinfo {author} {\bibfnamefont {K.}~\bibnamefont {Kokado}}, \bibinfo {author} {\bibfnamefont {H.}~\bibnamefont {Shinkai}},\ and\ \bibinfo {author} {\bibfnamefont {H.}~\bibnamefont {Katori}},\ }\bibfield  {title} {\bibinfo {title} {Test of general relativity by a pair of transportable optical lattice clocks},\ }\href {https://doi.org/10.1038/s41566-020-0619-8} {\bibfield  {journal} {\bibinfo  {journal} {Nat. Photonics}\ }\textbf {\bibinfo {volume} {14}},\ \bibinfo {pages} {411} (\bibinfo {year} {2020})}\BibitemShut {NoStop}%
\bibitem [{\citenamefont {Grotti}\ \emph {et~al.}(2024)\citenamefont {Grotti}, \citenamefont {Nosske}, \citenamefont {Koller}, \citenamefont {Herbers}, \citenamefont {Denker}, \citenamefont {Timmen}, \citenamefont {Vishnyakova}, \citenamefont {Grosche}, \citenamefont {Waterholter}, \citenamefont {Kuhl}, \citenamefont {Koke}, \citenamefont {Benkler}, \citenamefont {Giunta}, \citenamefont {Maisenbacher}, \citenamefont {Matveev}, \citenamefont {Dörscher}, \citenamefont {Schwarz}, \citenamefont {Al-Masoudi}, \citenamefont {Hänsch}, \citenamefont {Udem}, \citenamefont {Holzwarth},\ and\ \citenamefont {Lisdat}}]{GRotti2024a}%
  \BibitemOpen
  \bibfield  {author} {\bibinfo {author} {\bibfnamefont {J.}~\bibnamefont {Grotti}}, \bibinfo {author} {\bibfnamefont {I.}~\bibnamefont {Nosske}}, \bibinfo {author} {\bibfnamefont {S.~B.}\ \bibnamefont {Koller}}, \bibinfo {author} {\bibfnamefont {S.}~\bibnamefont {Herbers}}, \bibinfo {author} {\bibfnamefont {H.}~\bibnamefont {Denker}}, \bibinfo {author} {\bibfnamefont {L.}~\bibnamefont {Timmen}}, \bibinfo {author} {\bibfnamefont {G.}~\bibnamefont {Vishnyakova}}, \bibinfo {author} {\bibfnamefont {G.}~\bibnamefont {Grosche}}, \bibinfo {author} {\bibfnamefont {T.}~\bibnamefont {Waterholter}}, \bibinfo {author} {\bibfnamefont {A.}~\bibnamefont {Kuhl} \emph{et~al.}},\ }\bibfield  {title} {\bibinfo {title} {Long-distance chronometric leveling with a portable optical clock},\ }\href {https://doi.org/10.1103/physrevapplied.21.l061001} {\bibfield  {journal} {\bibinfo  {journal} {Phys. Rev. Applied}\ }\textbf {\bibinfo {volume} {21}},\ \bibinfo {pages} {L061001} (\bibinfo {year} {2024})} \BibitemShut {NoStop}%
\bibitem [{\citenamefont {Dimarcq}\ \emph {et~al.}(2024)\citenamefont {Dimarcq}, \citenamefont {Gertsvolf}, \citenamefont {Mileti}, \citenamefont {Bize}, \citenamefont {Oates}, \citenamefont {Peik}, \citenamefont {Calonico}, \citenamefont {Ido}, \citenamefont {Tavella}, \citenamefont {Meynadier}, \citenamefont {Petit}, \citenamefont {Panfilo}, \citenamefont {Bartholomew}, \citenamefont {Defraigne}, \citenamefont {Donley}, \citenamefont {Hedekvist}, \citenamefont {Sesia}, \citenamefont {Wouters}, \citenamefont {Dube}, \citenamefont {Fang}, \citenamefont {Levi}, \citenamefont {Lodewyck}, \citenamefont {Margolis}, \citenamefont {Newell}, \citenamefont {Slyusarev}, \citenamefont {Weyers}, \citenamefont {Uzan}, \citenamefont {Yasuda}, \citenamefont {Yu}, \citenamefont {Rieck}, \citenamefont {Schnatz}, \citenamefont {Hanado}, \citenamefont {Fujieda}, \citenamefont {Pottie}, \citenamefont {Hanssen}, \citenamefont {Malimon},\ and\ \citenamefont {Ashby}}]{Dimarcq2024a}%
  \BibitemOpen
  \bibfield  {author} {\bibinfo {author} {\bibfnamefont {N.}~\bibnamefont {Dimarcq}}, \bibinfo {author} {\bibfnamefont {M.}~\bibnamefont {Gertsvolf}}, \bibinfo {author} {\bibfnamefont {G.}~\bibnamefont {Mileti}}, \bibinfo {author} {\bibfnamefont {S.}~\bibnamefont {Bize}}, \bibinfo {author} {\bibfnamefont {C.~W.}\ \bibnamefont {Oates}}, \bibinfo {author} {\bibfnamefont {E.}~\bibnamefont {Peik}}, \bibinfo {author} {\bibfnamefont {D.}~\bibnamefont {Calonico}}, \bibinfo {author} {\bibfnamefont {T.}~\bibnamefont {Ido}}, \bibinfo {author} {\bibfnamefont {P.}~\bibnamefont {Tavella}}, \bibinfo {author} {\bibfnamefont {F.}~\bibnamefont {Meynadier} \emph{et~al.}},\ }\bibfield  {title} {\bibinfo {title} {Roadmap towards the redefinition of the second},\ }\href {https://doi.org/10.1088/1681-7575/ad17d2} {\bibfield  {journal} {\bibinfo  {journal} {Metrologia}\ }\textbf {\bibinfo {volume} {61}},\ \bibinfo {pages} {012001} (\bibinfo {year} {2024})} \BibitemShut {NoStop}%
\bibitem [{\citenamefont {Tofful}\ \emph {et~al.}(2024)\citenamefont {Tofful}, \citenamefont {Baynham}, \citenamefont {Curtis}, \citenamefont {Parsons}, \citenamefont {Robertson}, \citenamefont {Schioppo}, \citenamefont {Tunesi}, \citenamefont {Margolis}, \citenamefont {Hendricks}, \citenamefont {Whale}, \citenamefont {Thompson},\ and\ \citenamefont {Godun}}]{Tofful2024a}%
  \BibitemOpen
  \bibfield  {author} {\bibinfo {author} {\bibfnamefont {A.}~\bibnamefont {Tofful}}, \bibinfo {author} {\bibfnamefont {C.~F.~A.}\ \bibnamefont {Baynham}}, \bibinfo {author} {\bibfnamefont {E.~A.}\ \bibnamefont {Curtis}}, \bibinfo {author} {\bibfnamefont {A.~O.}\ \bibnamefont {Parsons}}, \bibinfo {author} {\bibfnamefont {B.~I.}\ \bibnamefont {Robertson}}, \bibinfo {author} {\bibfnamefont {M.}~\bibnamefont {Schioppo}}, \bibinfo {author} {\bibfnamefont {J.}~\bibnamefont {Tunesi}}, \bibinfo {author} {\bibfnamefont {H.~S.}\ \bibnamefont {Margolis}}, \bibinfo {author} {\bibfnamefont {R.~J.}\ \bibnamefont {Hendricks}}, \bibinfo {author} {\bibfnamefont {J.}~\bibnamefont {Whale}}, \bibinfo {author} {\bibfnamefont {R.~C.}\ \bibnamefont {Thompson}},\ and\ \bibinfo {author} {\bibfnamefont {R.~M.}\ \bibnamefont {Godun}},\ }\bibfield  {title} {\bibinfo {title} {$^{171}${Yb}$^+$ optical clock with $2.2\times 10^{-18}$ systematic uncertainty and absolute frequency measurements},\ }\href
  {https://doi.org/10.1088/1681-7575/ad53cd} {\bibfield  {journal} {\bibinfo  {journal} {Metrologia}\ }\textbf {\bibinfo {volume} {61}},\ \bibinfo {pages} {045001} (\bibinfo {year} {2024})} \BibitemShut {NoStop}%
\bibitem [{\citenamefont {Hausser}\ \emph {et~al.}(2025)\citenamefont {Hausser}, \citenamefont {Keller}, \citenamefont {Nordmann}, \citenamefont {Bhatt}, \citenamefont {Kiethe}, \citenamefont {Liu}, \citenamefont {Richter}, \citenamefont {von Boehn}, \citenamefont {Rahm}, \citenamefont {Weyers}, \citenamefont {Benkler}, \citenamefont {Lipphardt}, \citenamefont {Dörscher}, \citenamefont {Stahl}, \citenamefont {Klose}, \citenamefont {Lisdat}, \citenamefont {Filzinger}, \citenamefont {Huntemann}, \citenamefont {Peik},\ and\ \citenamefont {Mehlstäubler}}]{Hausser2025a}%
  \BibitemOpen
  \bibfield  {author} {\bibinfo {author} {\bibfnamefont {H.}~\bibnamefont {Hausser}}, \bibinfo {author} {\bibfnamefont {J.}~\bibnamefont {Keller}}, \bibinfo {author} {\bibfnamefont {T.}~\bibnamefont {Nordmann}}, \bibinfo {author} {\bibfnamefont {N.}~\bibnamefont {Bhatt}}, \bibinfo {author} {\bibfnamefont {J.}~\bibnamefont {Kiethe}}, \bibinfo {author} {\bibfnamefont {H.}~\bibnamefont {Liu}}, \bibinfo {author} {\bibfnamefont {I.}~\bibnamefont {Richter}}, \bibinfo {author} {\bibfnamefont {M.}~\bibnamefont {von Boehn}}, \bibinfo {author} {\bibfnamefont {J.}~\bibnamefont {Rahm}}, \bibinfo {author} {\bibfnamefont {S.}~\bibnamefont {Weyers}}, \bibinfo {author} {\bibfnamefont {E.}~\bibnamefont {Benkler}}, \bibinfo {author} {\bibfnamefont {B.}~\bibnamefont {Lipphardt}}, \bibinfo {author} {\bibfnamefont {S.}~\bibnamefont {Dörscher}}, \bibinfo {author} {\bibfnamefont {K.}~\bibnamefont {Stahl}}, \bibinfo {author} {\bibfnamefont {J.}~\bibnamefont {Klose}}, \bibinfo {author} {\bibfnamefont {C.}~\bibnamefont {Lisdat}},
  \bibinfo {author} {\bibfnamefont {M.}~\bibnamefont {Filzinger}}, \bibinfo {author} {\bibfnamefont {N.}~\bibnamefont {Huntemann}}, \bibinfo {author} {\bibfnamefont {E.}~\bibnamefont {Peik}},\ and\ \bibinfo {author} {\bibfnamefont {T.}~\bibnamefont {Mehlstäubler}},\ }\bibfield  {title} {\bibinfo {title} {$^{115}${In}$^+$-$^{172}${Yb}$^+$ Coulomb crystal clock with $2.5\times10^{-18}$ systematic uncertainty},\ }\href {https://doi.org/10.1103/physrevlett.134.023201} {\bibfield  {journal} {\bibinfo  {journal} {Phys. Rev. Lett.}\ }\textbf {\bibinfo {volume} {134}},\ \bibinfo {pages} {023201} (\bibinfo {year} {2025})} \BibitemShut {NoStop}%
\bibitem [{\citenamefont {McGrew}\ \emph {et~al.}(2018)\citenamefont {McGrew}, \citenamefont {Zhang}, \citenamefont {Fasano}, \citenamefont {Sch\"affer}, \citenamefont {Beloy}, \citenamefont {Nicolodi}, \citenamefont {Brown}, \citenamefont {Hinkley}, \citenamefont {Milani}, \citenamefont {Schioppo}, \citenamefont {Yoon},\ and\ \citenamefont {Ludlow}}]{McGrew2018a}%
  \BibitemOpen
  \bibfield  {author} {\bibinfo {author} {\bibfnamefont {W.~F.}\ \bibnamefont {McGrew}}, \bibinfo {author} {\bibfnamefont {X.}~\bibnamefont {Zhang}}, \bibinfo {author} {\bibfnamefont {R.~J.}\ \bibnamefont {Fasano}}, \bibinfo {author} {\bibfnamefont {S.~A.}\ \bibnamefont {Sch\"affer}}, \bibinfo {author} {\bibfnamefont {K.}~\bibnamefont {Beloy}}, \bibinfo {author} {\bibfnamefont {D.}~\bibnamefont {Nicolodi}}, \bibinfo {author} {\bibfnamefont {R.~C.}\ \bibnamefont {Brown}}, \bibinfo {author} {\bibfnamefont {N.}~\bibnamefont {Hinkley}}, \bibinfo {author} {\bibfnamefont {G.}~\bibnamefont {Milani}}, \bibinfo {author} {\bibfnamefont {M.}~\bibnamefont {Schioppo}}, \bibinfo {author} {\bibfnamefont {T.~H.}\ \bibnamefont {Yoon}},\ and\ \bibinfo {author} {\bibfnamefont {A.~D.}\ \bibnamefont {Ludlow}},\ }\bibfield  {title} {\bibinfo {title} {Atomic clock performance enabling geodesy below the centimetre level},\ }\href {https://doi.org/10.1038/s41586-018-0738-2} {\bibfield  {journal} {\bibinfo  {journal} {Nature}\
  }\textbf {\bibinfo {volume} {564}},\ \bibinfo {pages} {87} (\bibinfo {year} {2018})}\BibitemShut {NoStop}%
\bibitem [{\citenamefont {Bothwell}\ \emph {et~al.}(2019)\citenamefont {Bothwell}, \citenamefont {Kedar}, \citenamefont {Oelker}, \citenamefont {Robinson}, \citenamefont {Bromley}, \citenamefont {Tew}, \citenamefont {Ye},\ and\ \citenamefont {Kennedy}}]{Bothwell2019a}%
  \BibitemOpen
  \bibfield  {author} {\bibinfo {author} {\bibfnamefont {T.}~\bibnamefont {Bothwell}}, \bibinfo {author} {\bibfnamefont {D.}~\bibnamefont {Kedar}}, \bibinfo {author} {\bibfnamefont {E.}~\bibnamefont {Oelker}}, \bibinfo {author} {\bibfnamefont {J.~M.}\ \bibnamefont {Robinson}}, \bibinfo {author} {\bibfnamefont {S.~L.}\ \bibnamefont {Bromley}}, \bibinfo {author} {\bibfnamefont {W.~L.}\ \bibnamefont {Tew}}, \bibinfo {author} {\bibfnamefont {J.}~\bibnamefont {Ye}},\ and\ \bibinfo {author} {\bibfnamefont {C.~J.}\ \bibnamefont {Kennedy}},\ }\bibfield  {title} {\bibinfo {title} {{JILA} {SrI} optical lattice clock with uncertainty of $2.0 \times 10^{-18}$},\ }\href {https://doi.org/10.1088/1681-7575/ab4089} {\bibfield  {journal} {\bibinfo  {journal} {Metrologia}\ }\textbf {\bibinfo {volume} {56}},\ \bibinfo {pages} {065004} (\bibinfo {year} {2019})}\BibitemShut {NoStop}%
\bibitem [{\citenamefont {Lu}\ \emph {et~al.}(2025)\citenamefont {Lu}, \citenamefont {Guo}, \citenamefont {Liu}, \citenamefont {Cao}, \citenamefont {Li}, \citenamefont {Xia}, \citenamefont {Xu}, \citenamefont {Lu}, \citenamefont {Wang},\ and\ \citenamefont {Chang}}]{Lu2025a}%
  \BibitemOpen
  \bibfield  {author} {\bibinfo {author} {\bibfnamefont {X.}~\bibnamefont {Lu}}, \bibinfo {author} {\bibfnamefont {F.}~\bibnamefont {Guo}}, \bibinfo {author} {\bibfnamefont {Y.}~\bibnamefont {Liu}}, \bibinfo {author} {\bibfnamefont {J.}~\bibnamefont {Cao}}, \bibinfo {author} {\bibfnamefont {J.}~\bibnamefont {Li}}, \bibinfo {author} {\bibfnamefont {J.}~\bibnamefont {Xia}}, \bibinfo {author} {\bibfnamefont {Q.}~\bibnamefont {Xu}}, \bibinfo {author} {\bibfnamefont {B.}~\bibnamefont {Lu}}, \bibinfo {author} {\bibfnamefont {Y.}~\bibnamefont {Wang}},\ and\ \bibinfo {author} {\bibfnamefont {H.}~\bibnamefont {Chang}},\ }\bibfield  {title} {\bibinfo {title} {{NTSC} {SrII} optical lattice clock with uncertainty of $2\times 10^{-18}$},\ }\href {https://doi.org/10.1088/1681-7575/addc77} {\bibfield  {journal} {\bibinfo  {journal} {Metrologia}\ }\textbf {\bibinfo {volume} {62}},\ \bibinfo {pages} {035007} (\bibinfo {year} {2025})}\BibitemShut {NoStop}%
\bibitem [{\citenamefont {Dub\'e}\ \emph {et~al.}(2005)\citenamefont {Dub\'e}, \citenamefont {Madej}, \citenamefont {Bernard}, \citenamefont {Marmet}, \citenamefont {Boulanger},\ and\ \citenamefont {Cundy}}]{Dube2005a}%
  \BibitemOpen
  \bibfield  {author} {\bibinfo {author} {\bibfnamefont {P.}~\bibnamefont {Dub\'e}}, \bibinfo {author} {\bibfnamefont {A.~A.}\ \bibnamefont {Madej}}, \bibinfo {author} {\bibfnamefont {J.~E.}\ \bibnamefont {Bernard}}, \bibinfo {author} {\bibfnamefont {L.}~\bibnamefont {Marmet}}, \bibinfo {author} {\bibfnamefont {J.-S.}\ \bibnamefont {Boulanger}},\ and\ \bibinfo {author} {\bibfnamefont {S.}~\bibnamefont {Cundy}},\ }\bibfield  {title} {\bibinfo {title} {{E}lectric {Q}uadrupole {S}hift {C}ancellation in {S}ingle-{I}on {O}ptical {F}requency {S}tandards},\ }\href {https://doi.org/10.1103/PhysRevLett.95.033001} {\bibfield  {journal} {\bibinfo  {journal} {Phys. Rev. Lett.}\ }\textbf {\bibinfo {volume} {95}},\ \bibinfo {pages} {033001} (\bibinfo {year} {2005})}\BibitemShut {NoStop}%
\bibitem [{\citenamefont {Lindvall}\ \emph {et~al.}(2022)\citenamefont {Lindvall}, \citenamefont {Hanhij\"arvi}, \citenamefont {Fordell},\ and\ \citenamefont {Wallin}}]{Lindvall2022a}%
  \BibitemOpen
  \bibfield  {author} {\bibinfo {author} {\bibfnamefont {T.}~\bibnamefont {Lindvall}}, \bibinfo {author} {\bibfnamefont {K.~J.}\ \bibnamefont {Hanhij\"arvi}}, \bibinfo {author} {\bibfnamefont {T.}~\bibnamefont {Fordell}},\ and\ \bibinfo {author} {\bibfnamefont {A.~E.}\ \bibnamefont {Wallin}},\ }\bibfield  {title} {\bibinfo {title} {High-accuracy determination of {P}aul-trap stability parameters for electric-quadrupole-shift prediction},\ }\href {https://doi.org/10.1063/5.0106633} {\bibfield  {journal} {\bibinfo  {journal} {J. Appl. Phys.}\ }\textbf {\bibinfo {volume} {132}},\ \bibinfo {pages} {124401} (\bibinfo {year} {2022})}\BibitemShut {NoStop}%
\bibitem [{\citenamefont {{Dubé}}\ \emph {et~al.}(2014)\citenamefont {{Dubé}}, \citenamefont {{Madej}}, \citenamefont {{Tibbo}},\ and\ \citenamefont {{Bernard}}}]{Dube2014b}%
  \BibitemOpen
  \bibfield  {author} {\bibinfo {author} {\bibfnamefont {P.}~\bibnamefont {{Dubé}}}, \bibinfo {author} {\bibfnamefont {A.~A.}\ \bibnamefont {{Madej}}}, \bibinfo {author} {\bibfnamefont {M.}~\bibnamefont {{Tibbo}}},\ and\ \bibinfo {author} {\bibfnamefont {J.~E.}\ \bibnamefont {{Bernard}}},\ }\bibfield  {title} {\bibinfo {title} {Measurement of the static scalar polarizability of the $^{88}${Sr}$^+$ clock transition},\ }in\ \href {https://doi.org/10.1109/EFTF.2014.7331531} {\emph {\bibinfo {booktitle} {2014 European Frequency and Time Forum (EFTF)}}}\ (\bibinfo {year} {2014})\ pp.\ \bibinfo {pages} {443--446}\BibitemShut {NoStop}%
\bibitem [{\citenamefont {Gan}\ \emph {et~al.}(2018)\citenamefont {Gan}, \citenamefont {Maslennikov}, \citenamefont {Tseng}, \citenamefont {Tan}, \citenamefont {Kaewuam}, \citenamefont {Arnold}, \citenamefont {Matsukevich},\ and\ \citenamefont {Barrett}}]{Gan2018a}%
  \BibitemOpen
  \bibfield  {author} {\bibinfo {author} {\bibfnamefont {H.~C.~J.}\ \bibnamefont {Gan}}, \bibinfo {author} {\bibfnamefont {G.}~\bibnamefont {Maslennikov}}, \bibinfo {author} {\bibfnamefont {K.-W.}\ \bibnamefont {Tseng}}, \bibinfo {author} {\bibfnamefont {T.~R.}\ \bibnamefont {Tan}}, \bibinfo {author} {\bibfnamefont {R.}~\bibnamefont {Kaewuam}}, \bibinfo {author} {\bibfnamefont {K.~J.}\ \bibnamefont {Arnold}}, \bibinfo {author} {\bibfnamefont {D.}~\bibnamefont {Matsukevich}},\ and\ \bibinfo {author} {\bibfnamefont {M.~D.}\ \bibnamefont {Barrett}},\ }\bibfield  {title} {\bibinfo {title} {Oscillating-magnetic-field effects in high-precision metrology},\ }\href {https://doi.org/10.1103/PhysRevA.98.032514} {\bibfield  {journal} {\bibinfo  {journal} {Phys. Rev. A}\ }\textbf {\bibinfo {volume} {98}},\ \bibinfo {pages} {032514} (\bibinfo {year} {2018})}\BibitemShut {NoStop}%
\bibitem [{\citenamefont {Lindvall}\ \emph {et~al.}(2025{\natexlab{a}})\citenamefont {Lindvall}, \citenamefont {Hanhij\"arvi}, \citenamefont {Fordell},\ and\ \citenamefont {Wallin}}]{Lindvall2025b}%
  \BibitemOpen
  \bibfield  {author} {\bibinfo {author} {\bibfnamefont {T.}~\bibnamefont {Lindvall}}, \bibinfo {author} {\bibfnamefont {K.~J.}\ \bibnamefont {Hanhij\"arvi}}, \bibinfo {author} {\bibfnamefont {T.}~\bibnamefont {Fordell}},\ and\ \bibinfo {author} {\bibfnamefont {A.~E.}\ \bibnamefont {Wallin}},\ }\bibfield  {title} {\bibinfo {title} {Measurement of the differential static scalar polarizability of the $^{88}${Sr}$^{+}$ clock transition},\ }\href {https://doi.org/10.1103/52by-28mr} {\bibfield  {journal} {\bibinfo  {journal} {Phys. Rev. Lett.}\ }\textbf {\bibinfo {volume} {135}},\ \bibinfo {pages} {043402} (\bibinfo {year} {2025})} \BibitemShut {NoStop}%
\bibitem [{\citenamefont {Dub\'e}\ \emph {et~al.}(2021)\citenamefont {Dub\'e}, \citenamefont {Kato}, \citenamefont {Bernard},\ and\ \citenamefont {Jian}}]{Dube2021a}%
  \BibitemOpen
  \bibfield  {author} {\bibinfo {author} {\bibfnamefont {P.}~\bibnamefont {Dub\'e}}, \bibinfo {author} {\bibfnamefont {K.}~\bibnamefont {Kato}}, \bibinfo {author} {\bibfnamefont {J.}~\bibnamefont {Bernard}},\ and\ \bibinfo {author} {\bibfnamefont {B.}~\bibnamefont {Jian}},\ }\bibfield  {title} {\bibinfo {title} {Progress towards a transportable and high-accuracy {Sr}$^+$ ion clock at {NRC}},\ }in\ \href {https://doi.org/10.1109/eftf/ifcs52194.2021.9604288} {\emph {\bibinfo {booktitle} {2021 Joint Conference of the European Frequency and Time Forum and {IEEE} International Frequency Control Symposium ({EFTF}/{IFCS})}}}\ (\bibinfo  {publisher} {{IEEE}},\ \bibinfo {year} {2021})\BibitemShut {NoStop}%
\bibitem [{\citenamefont {Spampinato}\ \emph {et~al.}(2024)\citenamefont {Spampinato}, \citenamefont {Stacey}, \citenamefont {Mulholland}, \citenamefont {Robertson}, \citenamefont {Klein}, \citenamefont {Huang}, \citenamefont {Barwood},\ and\ \citenamefont {Gill}}]{Spampinato2024a}%
  \BibitemOpen
  \bibfield  {author} {\bibinfo {author} {\bibfnamefont {A.}~\bibnamefont {Spampinato}}, \bibinfo {author} {\bibfnamefont {J.}~\bibnamefont {Stacey}}, \bibinfo {author} {\bibfnamefont {S.}~\bibnamefont {Mulholland}}, \bibinfo {author} {\bibfnamefont {B.~I.}\ \bibnamefont {Robertson}}, \bibinfo {author} {\bibfnamefont {H.~A.}\ \bibnamefont {Klein}}, \bibinfo {author} {\bibfnamefont {G.}~\bibnamefont {Huang}}, \bibinfo {author} {\bibfnamefont {G.~P.}\ \bibnamefont {Barwood}},\ and\ \bibinfo {author} {\bibfnamefont {P.}~\bibnamefont {Gill}},\ }\bibfield  {title} {\bibinfo {title} {An ion trap design for a space-deployable strontium-ion optical clock},\ }\href {https://doi.org/10.1098/rspa.2023.0593} {\bibfield  {journal} {\bibinfo  {journal} {Proc. R. Soc. A}\ }\textbf {\bibinfo {volume} {480}},\ \bibinfo {pages} {20230593} (\bibinfo {year} {2024})}\BibitemShut {NoStop}%
\bibitem [{\citenamefont {Fairbank}\ \emph {et~al.}(2025)\citenamefont {Fairbank}, \citenamefont {Lee}, \citenamefont {Notcutt},\ and\ \citenamefont {Davila-Rodriguez}}]{Fairbank2025a}%
  \BibitemOpen
  \bibfield  {author} {\bibinfo {author} {\bibfnamefont {D.}~\bibnamefont {Fairbank}}, \bibinfo {author} {\bibfnamefont {D.}~\bibnamefont {Lee}}, \bibinfo {author} {\bibfnamefont {M.}~\bibnamefont {Notcutt}},\ and\ \bibinfo {author} {\bibfnamefont {J.}~\bibnamefont {Davila-Rodriguez}},\ }\bibfield  {title} {\bibinfo {title} {Towards a commercial strontium-ion optical clock},\ }in\ \href {https://epapers2.org/ifcs-eftf2025/ESR/paper_details.php?paper_id=2128} {\emph {\bibinfo {booktitle} {IFCS-EFTF 2025}}}\ (\bibinfo {year} {2025})\BibitemShut {NoStop}%
\bibitem [{\citenamefont {Steinel}\ \emph {et~al.}(2023)\citenamefont {Steinel}, \citenamefont {Shao}, \citenamefont {Filzinger}, \citenamefont {Lipphardt}, \citenamefont {Brinkmann}, \citenamefont {Didier}, \citenamefont {Mehlstäubler}, \citenamefont {Lindvall}, \citenamefont {Peik},\ and\ \citenamefont {Huntemann}}]{Steinel2023a}%
  \BibitemOpen
  \bibfield  {author} {\bibinfo {author} {\bibfnamefont {M.}~\bibnamefont {Steinel}}, \bibinfo {author} {\bibfnamefont {H.}~\bibnamefont {Shao}}, \bibinfo {author} {\bibfnamefont {M.}~\bibnamefont {Filzinger}}, \bibinfo {author} {\bibfnamefont {B.}~\bibnamefont {Lipphardt}}, \bibinfo {author} {\bibfnamefont {M.}~\bibnamefont {Brinkmann}}, \bibinfo {author} {\bibfnamefont {A.}~\bibnamefont {Didier}}, \bibinfo {author} {\bibfnamefont {T.~E.}\ \bibnamefont {Mehlstäubler}}, \bibinfo {author} {\bibfnamefont {T.}~\bibnamefont {Lindvall}}, \bibinfo {author} {\bibfnamefont {E.}~\bibnamefont {Peik}},\ and\ \bibinfo {author} {\bibfnamefont {N.}~\bibnamefont {Huntemann}},\ }\bibfield  {title} {\bibinfo {title} {Evaluation of a $^{88}${Sr}$^+$ optical clock with a direct measurement of the blackbody radiation shift and determination of the clock frequency},\ }\href {https://doi.org/10.1103/physrevlett.131.083002} {\bibfield  {journal} {\bibinfo  {journal} {Phys. Rev. Lett.}\ }\textbf {\bibinfo {volume} {131}},\ \bibinfo
  {pages} {083002} (\bibinfo {year} {2023})} \BibitemShut {NoStop}%
\bibitem [{\citenamefont {Akerman}\ and\ \citenamefont {Ozeri}(2025)}]{Akerman2025a}%
  \BibitemOpen
  \bibfield  {author} {\bibinfo {author} {\bibfnamefont {N.}~\bibnamefont {Akerman}}\ and\ \bibinfo {author} {\bibfnamefont {R.}~\bibnamefont {Ozeri}},\ }\bibfield  {title} {\bibinfo {title} {Operating a multi-ion clock with dynamical decoupling},\ }\href {https://doi.org/10.1103/physrevlett.134.013201} {\bibfield  {journal} {\bibinfo  {journal} {Phys. Rev. Lett.}\ }\textbf {\bibinfo {volume} {134}},\ \bibinfo {pages} {013201} (\bibinfo {year} {2025})} \BibitemShut {NoStop}%
\bibitem [{\citenamefont {Loh}\ \emph {et~al.}(2025)\citenamefont {Loh}, \citenamefont {Reens}, \citenamefont {Kharas}, \citenamefont {Sumant}, \citenamefont {Belanger}, \citenamefont {Maxson}, \citenamefont {Medeiros}, \citenamefont {Setzer}, \citenamefont {Gray}, \citenamefont {DeBry}, \citenamefont {Bruzewicz}, \citenamefont {Plant}, \citenamefont {Liddell}, \citenamefont {West}, \citenamefont {Doshi}, \citenamefont {Roychowdhury}, \citenamefont {Kim}, \citenamefont {Braje}, \citenamefont {Juodawlkis}, \citenamefont {Chiaverini},\ and\ \citenamefont {McConnell}}]{Loh2025a}%
  \BibitemOpen
  \bibfield  {author} {\bibinfo {author} {\bibfnamefont {W.}~\bibnamefont {Loh}}, \bibinfo {author} {\bibfnamefont {D.}~\bibnamefont {Reens}}, \bibinfo {author} {\bibfnamefont {D.}~\bibnamefont {Kharas}}, \bibinfo {author} {\bibfnamefont {A.}~\bibnamefont {Sumant}}, \bibinfo {author} {\bibfnamefont {C.}~\bibnamefont {Belanger}}, \bibinfo {author} {\bibfnamefont {R.~T.}\ \bibnamefont {Maxson}}, \bibinfo {author} {\bibfnamefont {A.}~\bibnamefont {Medeiros}}, \bibinfo {author} {\bibfnamefont {W.}~\bibnamefont {Setzer}}, \bibinfo {author} {\bibfnamefont {D.}~\bibnamefont {Gray}}, \bibinfo {author} {\bibfnamefont {K.}~\bibnamefont {DeBry} \emph{et~al.}},\ }\bibfield  {title} {\bibinfo {title} {Optical atomic clock interrogation using an integrated spiral cavity laser},\ }\href {https://doi.org/10.1038/s41566-024-01588-8} {\bibfield  {journal} {\bibinfo  {journal} {Nat. Photonics}\ }\textbf {\bibinfo {volume} {19}},\ \bibinfo {pages} {277} (\bibinfo {year} {2025})} \BibitemShut {NoStop}%
\bibitem [{\citenamefont {Jian}\ \emph {et~al.}(2023)\citenamefont {Jian}, \citenamefont {Bernard}, \citenamefont {Gertsvolf},\ and\ \citenamefont {Dub{\'{e}}}}]{Jian2023a}%
  \BibitemOpen
  \bibfield  {author} {\bibinfo {author} {\bibfnamefont {B.}~\bibnamefont {Jian}}, \bibinfo {author} {\bibfnamefont {J.}~\bibnamefont {Bernard}}, \bibinfo {author} {\bibfnamefont {M.}~\bibnamefont {Gertsvolf}},\ and\ \bibinfo {author} {\bibfnamefont {P.}~\bibnamefont {Dub{\'{e}}}},\ }\bibfield  {title} {\bibinfo {title} {Improved absolute frequency measurement of the strontium ion clock using a {GPS} link to the {SI} second},\ }\href {https://doi.org/10.1088/1681-7575/aca615} {\bibfield  {journal} {\bibinfo  {journal} {Metrologia}\ }\textbf {\bibinfo {volume} {60}},\ \bibinfo {pages} {015007} (\bibinfo {year} {2023})}\BibitemShut {NoStop}%
\bibitem [{\citenamefont {Barwood}\ \emph {et~al.}(2015)\citenamefont {Barwood}, \citenamefont {Huang}, \citenamefont {King}, \citenamefont {Klein},\ and\ \citenamefont {Gill}}]{Barwood2015a}%
  \BibitemOpen
  \bibfield  {author} {\bibinfo {author} {\bibfnamefont {G.~P.}\ \bibnamefont {Barwood}}, \bibinfo {author} {\bibfnamefont {G.}~\bibnamefont {Huang}}, \bibinfo {author} {\bibfnamefont {S.~A.}\ \bibnamefont {King}}, \bibinfo {author} {\bibfnamefont {H.~A.}\ \bibnamefont {Klein}},\ and\ \bibinfo {author} {\bibfnamefont {P.}~\bibnamefont {Gill}},\ }\bibfield  {title} {\bibinfo {title} {{F}requency noise processes in a strontium ion optical clock},\ }\href {https://doi.org/10.1088/0953-4075/48/3/035401} {\bibfield  {journal} {\bibinfo  {journal} {J. Phys. B: At. Mol. Opt. Phys.}\ }\textbf {\bibinfo {volume} {48}},\ \bibinfo {pages} {035401} (\bibinfo {year} {2015})}\BibitemShut {NoStop}%
\bibitem [{\citenamefont {Zeng}\ \emph {et~al.}(2023)\citenamefont {Zeng}, \citenamefont {Ma}, \citenamefont {Hu}, \citenamefont {Zhang}, \citenamefont {Hao}, \citenamefont {Zhang}, \citenamefont {Huang}, \citenamefont {Guan},\ and\ \citenamefont {Gao}}]{Zeng2023c}%
  \BibitemOpen
  \bibfield  {author} {\bibinfo {author} {\bibfnamefont {M.}~\bibnamefont {Zeng}}, \bibinfo {author} {\bibfnamefont {Z.}~\bibnamefont {Ma}}, \bibinfo {author} {\bibfnamefont {R.}~\bibnamefont {Hu}}, \bibinfo {author} {\bibfnamefont {B.}~\bibnamefont {Zhang}}, \bibinfo {author} {\bibfnamefont {Y.}~\bibnamefont {Hao}}, \bibinfo {author} {\bibfnamefont {H.}~\bibnamefont {Zhang}}, \bibinfo {author} {\bibfnamefont {Y.}~\bibnamefont {Huang}}, \bibinfo {author} {\bibfnamefont {H.}~\bibnamefont {Guan}},\ and\ \bibinfo {author} {\bibfnamefont {K.}~\bibnamefont {Gao}},\ }\bibfield  {title} {\bibinfo {title} {A combined magnetic field stabilization system for improving the stability of $^{40}${Ca}$^+$ optical clock},\ }\href {https://doi.org/10.1088/1674-1056/acf5d5} {\bibfield  {journal} {\bibinfo  {journal} {Chinese Phys. B}\ }\textbf {\bibinfo {volume} {32}},\ \bibinfo {pages} {110704} (\bibinfo {year} {2023})}\BibitemShut {NoStop}%
\bibitem [{\citenamefont {Weyers}\ \emph {et~al.}(2018)\citenamefont {Weyers}, \citenamefont {Gerginov}, \citenamefont {Kazda}, \citenamefont {Rahm}, \citenamefont {Lipphardt}, \citenamefont {Dobrev},\ and\ \citenamefont {Gibble}}]{Weyers2018a}%
  \BibitemOpen
  \bibfield  {author} {\bibinfo {author} {\bibfnamefont {S.}~\bibnamefont {Weyers}}, \bibinfo {author} {\bibfnamefont {V.}~\bibnamefont {Gerginov}}, \bibinfo {author} {\bibfnamefont {M.}~\bibnamefont {Kazda}}, \bibinfo {author} {\bibfnamefont {J.}~\bibnamefont {Rahm}}, \bibinfo {author} {\bibfnamefont {B.}~\bibnamefont {Lipphardt}}, \bibinfo {author} {\bibfnamefont {G.}~\bibnamefont {Dobrev}},\ and\ \bibinfo {author} {\bibfnamefont {K.}~\bibnamefont {Gibble}},\ }\bibfield  {title} {\bibinfo {title} {Advances in the accuracy, stability, and reliability of the {PTB} primary fountain clocks},\ }\href {https://doi.org/10.1088/1681-7575/aae008} {\bibfield  {journal} {\bibinfo  {journal} {Metrologia}\ }\textbf {\bibinfo {volume} {55}},\ \bibinfo {pages} {789} (\bibinfo {year} {2018})}\BibitemShut {NoStop}%
\bibitem [{\citenamefont {Petit}\ \emph {et~al.}(2015)\citenamefont {Petit}, \citenamefont {Kanj}, \citenamefont {Loyer}, \citenamefont {Delporte}, \citenamefont {Mercier},\ and\ \citenamefont {Perosanz}}]{Petit2015b}%
  \BibitemOpen
  \bibfield  {author} {\bibinfo {author} {\bibfnamefont {G.}~\bibnamefont {Petit}}, \bibinfo {author} {\bibfnamefont {A.}~\bibnamefont {Kanj}}, \bibinfo {author} {\bibfnamefont {S.}~\bibnamefont {Loyer}}, \bibinfo {author} {\bibfnamefont {J.}~\bibnamefont {Delporte}}, \bibinfo {author} {\bibfnamefont {F.}~\bibnamefont {Mercier}},\ and\ \bibinfo {author} {\bibfnamefont {F.}~\bibnamefont {Perosanz}},\ }\bibfield  {title} {\bibinfo {title} {$1 \times 10^{-16}$ frequency transfer by {GPS} {PPP} with integer ambiguity resolution},\ }\href {https://doi.org/10.1088/0026-1394/52/2/301} {\bibfield  {journal} {\bibinfo  {journal} {Metrologia}\ }\textbf {\bibinfo {volume} {52}},\ \bibinfo {pages} {301} (\bibinfo {year} {2015})}\BibitemShut {NoStop}%
\bibitem [{\citenamefont {McGrew}\ \emph {et~al.}(2019)\citenamefont {McGrew}, \citenamefont {Zhang}, \citenamefont {Leopardi}, \citenamefont {Fasano}, \citenamefont {Nicolodi}, \citenamefont {Beloy}, \citenamefont {Yao}, \citenamefont {Sherman}, \citenamefont {Sch\"{a}ffer}, \citenamefont {Savory}, \citenamefont {Brown}, \citenamefont {R\"{o}misch}, \citenamefont {Oates}, \citenamefont {Parker}, \citenamefont {Fortier},\ and\ \citenamefont {Ludlow}}]{McGrew2019a}%
  \BibitemOpen
  \bibfield  {author} {\bibinfo {author} {\bibfnamefont {W.~F.}\ \bibnamefont {McGrew}}, \bibinfo {author} {\bibfnamefont {X.}~\bibnamefont {Zhang}}, \bibinfo {author} {\bibfnamefont {H.}~\bibnamefont {Leopardi}}, \bibinfo {author} {\bibfnamefont {R.~J.}\ \bibnamefont {Fasano}}, \bibinfo {author} {\bibfnamefont {D.}~\bibnamefont {Nicolodi}}, \bibinfo {author} {\bibfnamefont {K.}~\bibnamefont {Beloy}}, \bibinfo {author} {\bibfnamefont {J.}~\bibnamefont {Yao}}, \bibinfo {author} {\bibfnamefont {J.~A.}\ \bibnamefont {Sherman}}, \bibinfo {author} {\bibfnamefont {S.~A.}\ \bibnamefont {Sch\"{a}ffer}}, \bibinfo {author} {\bibfnamefont {J.}~\bibnamefont {Savory}}, \bibinfo {author} {\bibfnamefont {R.~C.}\ \bibnamefont {Brown}}, \bibinfo {author} {\bibfnamefont {S.}~\bibnamefont {R\"{o}misch}}, \bibinfo {author} {\bibfnamefont {C.~W.}\ \bibnamefont {Oates}}, \bibinfo {author} {\bibfnamefont {T.~E.}\ \bibnamefont {Parker}}, \bibinfo {author} {\bibfnamefont {T.~M.}\ \bibnamefont {Fortier}},\ and\ \bibinfo {author}
  {\bibfnamefont {A.~D.}\ \bibnamefont {Ludlow}},\ }\bibfield  {title} {\bibinfo {title} {Towards the optical second: verifying optical clocks at the {SI} limit},\ }\href {https://doi.org/10.1364/OPTICA.6.000448} {\bibfield  {journal} {\bibinfo  {journal} {Optica}\ }\textbf {\bibinfo {volume} {6}},\ \bibinfo {pages} {448} (\bibinfo {year} {2019})}\BibitemShut {NoStop}%
\bibitem [{\citenamefont {Schwarz}\ \emph {et~al.}(2020)\citenamefont {Schwarz}, \citenamefont {D\"orscher}, \citenamefont {Al-Masoudi}, \citenamefont {Benkler}, \citenamefont {Legero}, \citenamefont {Sterr}, \citenamefont {Weyers}, \citenamefont {Rahm}, \citenamefont {Lipphardt},\ and\ \citenamefont {Lisdat}}]{Schwarz2020a}%
  \BibitemOpen
  \bibfield  {author} {\bibinfo {author} {\bibfnamefont {R.}~\bibnamefont {Schwarz}}, \bibinfo {author} {\bibfnamefont {S.}~\bibnamefont {D\"orscher}}, \bibinfo {author} {\bibfnamefont {A.}~\bibnamefont {Al-Masoudi}}, \bibinfo {author} {\bibfnamefont {E.}~\bibnamefont {Benkler}}, \bibinfo {author} {\bibfnamefont {T.}~\bibnamefont {Legero}}, \bibinfo {author} {\bibfnamefont {U.}~\bibnamefont {Sterr}}, \bibinfo {author} {\bibfnamefont {S.}~\bibnamefont {Weyers}}, \bibinfo {author} {\bibfnamefont {J.}~\bibnamefont {Rahm}}, \bibinfo {author} {\bibfnamefont {B.}~\bibnamefont {Lipphardt}},\ and\ \bibinfo {author} {\bibfnamefont {C.}~\bibnamefont {Lisdat}},\ }\bibfield  {title} {\bibinfo {title} {Long term measurement of the $^{87}\mathrm{Sr}$ clock frequency at the limit of primary {Cs} clocks},\ }\href {https://doi.org/10.1103/PhysRevResearch.2.033242} {\bibfield  {journal} {\bibinfo  {journal} {Phys. Rev. Res.}\ }\textbf {\bibinfo {volume} {2}},\ \bibinfo {pages} {033242} (\bibinfo {year} {2020})}\BibitemShut
  {NoStop}%
\bibitem [{\citenamefont {Nemitz}\ \emph {et~al.}(2021)\citenamefont {Nemitz}, \citenamefont {Gotoh}, \citenamefont {Nakagawa}, \citenamefont {Ito}, \citenamefont {Hanado}, \citenamefont {Ido},\ and\ \citenamefont {Hachisu}}]{Nemitz2021a}%
  \BibitemOpen
  \bibfield  {author} {\bibinfo {author} {\bibfnamefont {N.}~\bibnamefont {Nemitz}}, \bibinfo {author} {\bibfnamefont {T.}~\bibnamefont {Gotoh}}, \bibinfo {author} {\bibfnamefont {F.}~\bibnamefont {Nakagawa}}, \bibinfo {author} {\bibfnamefont {H.}~\bibnamefont {Ito}}, \bibinfo {author} {\bibfnamefont {Y.}~\bibnamefont {Hanado}}, \bibinfo {author} {\bibfnamefont {T.}~\bibnamefont {Ido}},\ and\ \bibinfo {author} {\bibfnamefont {H.}~\bibnamefont {Hachisu}},\ }\bibfield  {title} {\bibinfo {title} {Absolute frequency of $^{87}${Sr} at $1.8 \times 10^{-16}$ uncertainty by reference to remote primary frequency standards},\ }\href {https://doi.org/10.1088/1681-7575/abc232} {\bibfield  {journal} {\bibinfo  {journal} {Metrologia}\ }\textbf {\bibinfo {volume} {58}},\ \bibinfo {pages} {025006} (\bibinfo {year} {2021})}\BibitemShut {NoStop}%
\bibitem [{\citenamefont {Kobayashi}\ \emph {et~al.}(2025)\citenamefont {Kobayashi}, \citenamefont {Nishiyama}, \citenamefont {Hosaka}, \citenamefont {Akamatsu}, \citenamefont {Kawasaki}, \citenamefont {Wada}, \citenamefont {Inaba}, \citenamefont {Tanabe},\ and\ \citenamefont {Yasuda}}]{Kobayashi2025a}%
  \BibitemOpen
  \bibfield  {author} {\bibinfo {author} {\bibfnamefont {T.}~\bibnamefont {Kobayashi}}, \bibinfo {author} {\bibfnamefont {A.}~\bibnamefont {Nishiyama}}, \bibinfo {author} {\bibfnamefont {K.}~\bibnamefont {Hosaka}}, \bibinfo {author} {\bibfnamefont {D.}~\bibnamefont {Akamatsu}}, \bibinfo {author} {\bibfnamefont {A.}~\bibnamefont {Kawasaki}}, \bibinfo {author} {\bibfnamefont {M.}~\bibnamefont {Wada}}, \bibinfo {author} {\bibfnamefont {H.}~\bibnamefont {Inaba}}, \bibinfo {author} {\bibfnamefont {T.}~\bibnamefont {Tanabe}},\ and\ \bibinfo {author} {\bibfnamefont {M.}~\bibnamefont {Yasuda}},\ }\bibfield  {title} {\bibinfo {title} {Improved absolute frequency measurement of $^{171}${Yb} at {NMIJ} with uncertainty below $2\times 10^{-16}$},\ }\href {https://doi.org/10.1088/1681-7575/adb754} {\bibfield  {journal} {\bibinfo  {journal} {Metrologia}\ }\textbf {\bibinfo {volume} {62}},\ \bibinfo {pages} {025006} (\bibinfo {year} {2025})}\BibitemShut {NoStop}%
\bibitem [{\citenamefont {Marceau}\ \emph {et~al.}(2025)\citenamefont {Marceau}, \citenamefont {Beattie}, \citenamefont {Kato}, \citenamefont {Jian}, \citenamefont {Gertsvolf},\ and\ \citenamefont {Dubé}}]{Marceau2025a}%
  \BibitemOpen
  \bibfield  {author} {\bibinfo {author} {\bibfnamefont {C.}~\bibnamefont {Marceau}}, \bibinfo {author} {\bibfnamefont {S.}~\bibnamefont {Beattie}}, \bibinfo {author} {\bibfnamefont {K.}~\bibnamefont {Kato}}, \bibinfo {author} {\bibfnamefont {B.}~\bibnamefont {Jian}}, \bibinfo {author} {\bibfnamefont {M.}~\bibnamefont {Gertsvolf}},\ and\ \bibinfo {author} {\bibfnamefont {P.}~\bibnamefont {Dubé}},\ }\bibfield  {title} {\bibinfo {title} {Absolute frequency measurement of a $^{88}${Sr}$^+$ clock by direct comparison to a primary frequency standard},\ }\href {https://doi.org/10.1088/1681-7575/ade4d1} {\bibfield  {journal} {\bibinfo  {journal} {Metrologia}\ }\textbf {\bibinfo {volume} {62}},\ \bibinfo {pages} {045001} (\bibinfo {year} {2025})}\BibitemShut {NoStop}%
\bibitem [{\citenamefont {Brownnutt}\ \emph {et~al.}(2007)\citenamefont {Brownnutt}, \citenamefont {Letchumanan}, \citenamefont {Wilpers}, \citenamefont {Thompson}, \citenamefont {Gill},\ and\ \citenamefont {Sinclair}}]{Brownnutt2007a}%
  \BibitemOpen
  \bibfield  {author} {\bibinfo {author} {\bibfnamefont {M.}~\bibnamefont {Brownnutt}}, \bibinfo {author} {\bibfnamefont {V.}~\bibnamefont {Letchumanan}}, \bibinfo {author} {\bibfnamefont {G.}~\bibnamefont {Wilpers}}, \bibinfo {author} {\bibfnamefont {R.~C.}\ \bibnamefont {Thompson}}, \bibinfo {author} {\bibfnamefont {P.}~\bibnamefont {Gill}},\ and\ \bibinfo {author} {\bibfnamefont {A.~G.}\ \bibnamefont {Sinclair}},\ }\bibfield  {title} {\bibinfo {title} {{C}ontrolled photoionization loading of $^{88}${S}r$^+$ for precision ion-trap experiments},\ }\href {https://doi.org/10.1007/s00340-007-2624-8} {\bibfield  {journal} {\bibinfo  {journal} {Appl. Phys. B}\ }\textbf {\bibinfo {volume} {87}},\ \bibinfo {pages} {411} (\bibinfo {year} {2007})}\BibitemShut {NoStop}%
\bibitem [{\citenamefont {Fordell}\ and\ \citenamefont {Lindvall}(2019)}]{Fordell2019a}%
  \BibitemOpen
  \bibfield  {author} {\bibinfo {author} {\bibfnamefont {T.}~\bibnamefont {Fordell}}\ and\ \bibinfo {author} {\bibfnamefont {T.}~\bibnamefont {Lindvall}},\ }\bibfield  {title} {\bibinfo {title} {Broadband lasers for photo-ionization and repumping of trapped ions},\ }\href {https://doi.org/10.1364/JOSAB.36.000415} {\bibfield  {journal} {\bibinfo  {journal} {J. Opt. Soc. Am. B}\ }\textbf {\bibinfo {volume} {36}},\ \bibinfo {pages} {415} (\bibinfo {year} {2019})} \BibitemShut {NoStop}%
\bibitem [{\citenamefont {Haq}\ \emph {et~al.}(2006)\citenamefont {Haq}, \citenamefont {Mahmood}, \citenamefont {Amin}, \citenamefont {Jamil}, \citenamefont {Ali},\ and\ \citenamefont {Baig}}]{Haq2006a}%
  \BibitemOpen
  \bibfield  {author} {\bibinfo {author} {\bibfnamefont {S.-U.}\ \bibnamefont {Haq}}, \bibinfo {author} {\bibfnamefont {S.}~\bibnamefont {Mahmood}}, \bibinfo {author} {\bibfnamefont {N.}~\bibnamefont {Amin}}, \bibinfo {author} {\bibfnamefont {Y.}~\bibnamefont {Jamil}}, \bibinfo {author} {\bibfnamefont {R.}~\bibnamefont {Ali}},\ and\ \bibinfo {author} {\bibfnamefont {M.~A.}\ \bibnamefont {Baig}},\ }\bibfield  {title} {\bibinfo {title} {{M}easurements of photoionization cross sections from the 5s5p $^1{P}_1$ and 5s6s $^1{S}_0$ excited states of strontium},\ }\href {https://doi.org/10.1088/0953-4075/39/7/003} {\bibfield  {journal} {\bibinfo  {journal} {J. Phys. B: At. Mol. Opt. Phys.}\ }\textbf {\bibinfo {volume} {39}},\ \bibinfo {pages} {1587} (\bibinfo {year} {2006})}\BibitemShut {NoStop}%
\bibitem [{\citenamefont {Shiner}\ \emph {et~al.}(2007)\citenamefont {Shiner}, \citenamefont {Madej}, \citenamefont {Dub\'e},\ and\ \citenamefont {Bernard}}]{Shiner2007a}%
  \BibitemOpen
  \bibfield  {author} {\bibinfo {author} {\bibfnamefont {A.~D.}\ \bibnamefont {Shiner}}, \bibinfo {author} {\bibfnamefont {A.~A.}\ \bibnamefont {Madej}}, \bibinfo {author} {\bibfnamefont {P.}~\bibnamefont {Dub\'e}},\ and\ \bibinfo {author} {\bibfnamefont {J.~E.}\ \bibnamefont {Bernard}},\ }\bibfield  {title} {\bibinfo {title} {{A}bsolute optical frequency measurement of saturated absorption lines in {R}b near 422 nm},\ }\href {https://doi.org/10.1007/s00340-007-2836-y} {\bibfield  {journal} {\bibinfo  {journal} {Appl. Phys. B}\ }\textbf {\bibinfo {volume} {89}},\ \bibinfo {pages} {595} (\bibinfo {year} {2007})}\BibitemShut {NoStop}%
\bibitem [{\citenamefont {Lindvall}\ \emph {et~al.}(2013)\citenamefont {Lindvall}, \citenamefont {Fordell}, \citenamefont {Tittonen},\ and\ \citenamefont {Merimaa}}]{Lindvall2013a}%
  \BibitemOpen
  \bibfield  {author} {\bibinfo {author} {\bibfnamefont {T.}~\bibnamefont {Lindvall}}, \bibinfo {author} {\bibfnamefont {T.}~\bibnamefont {Fordell}}, \bibinfo {author} {\bibfnamefont {I.}~\bibnamefont {Tittonen}},\ and\ \bibinfo {author} {\bibfnamefont {M.}~\bibnamefont {Merimaa}},\ }\bibfield  {title} {\bibinfo {title} {{U}npolarized, incoherent repumping light for prevention of dark states in a trapped and laser-cooled single ion},\ }\href {https://doi.org/10.1103/PhysRevA.87.013439} {\bibfield  {journal} {\bibinfo  {journal} {Phys. Rev. A}\ }\textbf {\bibinfo {volume} {87}},\ \bibinfo {pages} {013439} (\bibinfo {year} {2013})}\BibitemShut {NoStop}%
\bibitem [{\citenamefont {Fordell}\ \emph {et~al.}(2015)\citenamefont {Fordell}, \citenamefont {Lindvall}, \citenamefont {Dub\'{e}}, \citenamefont {Madej}, \citenamefont {Wallin},\ and\ \citenamefont {Merimaa}}]{Fordell2015a}%
  \BibitemOpen
  \bibfield  {author} {\bibinfo {author} {\bibfnamefont {T.}~\bibnamefont {Fordell}}, \bibinfo {author} {\bibfnamefont {T.}~\bibnamefont {Lindvall}}, \bibinfo {author} {\bibfnamefont {P.}~\bibnamefont {Dub\'{e}}}, \bibinfo {author} {\bibfnamefont {A.~A.}\ \bibnamefont {Madej}}, \bibinfo {author} {\bibfnamefont {A.~E.}\ \bibnamefont {Wallin}},\ and\ \bibinfo {author} {\bibfnamefont {M.}~\bibnamefont {Merimaa}},\ }\bibfield  {title} {\bibinfo {title} {{B}roadband, unpolarized repumping and clearout light sources for {S}r$^+$ single-ion clocks},\ }\href {https://doi.org/10.1364/OL.40.001822} {\bibfield  {journal} {\bibinfo  {journal} {Opt. Lett.}\ }\textbf {\bibinfo {volume} {40}},\ \bibinfo {pages} {1822} (\bibinfo {year} {2015})}\BibitemShut {NoStop}%
\bibitem [{\citenamefont {Lindvall}\ \emph {et~al.}(2012)\citenamefont {Lindvall}, \citenamefont {Merimaa}, \citenamefont {Tittonen},\ and\ \citenamefont {Madej}}]{Lindvall2012a}%
  \BibitemOpen
  \bibfield  {author} {\bibinfo {author} {\bibfnamefont {T.}~\bibnamefont {Lindvall}}, \bibinfo {author} {\bibfnamefont {M.}~\bibnamefont {Merimaa}}, \bibinfo {author} {\bibfnamefont {I.}~\bibnamefont {Tittonen}},\ and\ \bibinfo {author} {\bibfnamefont {A.~A.}\ \bibnamefont {Madej}},\ }\bibfield  {title} {\bibinfo {title} {{D}ark-state suppression and optimization of laser cooling and fluorescence in a trapped alkaline-earth-metal single ion},\ }\href {https://doi.org/10.1103/PhysRevA.86.033403} {\bibfield  {journal} {\bibinfo  {journal} {Phys. Rev. A}\ }\textbf {\bibinfo {volume} {86}},\ \bibinfo {pages} {033403} (\bibinfo {year} {2012})}\BibitemShut {NoStop}%
\bibitem [{\citenamefont {H\"afner}\ \emph {et~al.}(2015)\citenamefont {H\"afner}, \citenamefont {Falke}, \citenamefont {Grebing}, \citenamefont {Vogt}, \citenamefont {Legero}, \citenamefont {Merimaa}, \citenamefont {Lisdat},\ and\ \citenamefont {Sterr}}]{Hafner2015a}%
  \BibitemOpen
  \bibfield  {author} {\bibinfo {author} {\bibfnamefont {S.}~\bibnamefont {H\"afner}}, \bibinfo {author} {\bibfnamefont {S.}~\bibnamefont {Falke}}, \bibinfo {author} {\bibfnamefont {C.}~\bibnamefont {Grebing}}, \bibinfo {author} {\bibfnamefont {S.}~\bibnamefont {Vogt}}, \bibinfo {author} {\bibfnamefont {T.}~\bibnamefont {Legero}}, \bibinfo {author} {\bibfnamefont {M.}~\bibnamefont {Merimaa}}, \bibinfo {author} {\bibfnamefont {C.}~\bibnamefont {Lisdat}},\ and\ \bibinfo {author} {\bibfnamefont {U.}~\bibnamefont {Sterr}},\ }\bibfield  {title} {\bibinfo {title} {$8 \times 10^{-17}$ fractional laser frequency instability with a long room-temperature cavity},\ }\href {https://doi.org/10.1364/OL.40.002112} {\bibfield  {journal} {\bibinfo  {journal} {Opt. Lett.}\ }\textbf {\bibinfo {volume} {40}},\ \bibinfo {pages} {2112} (\bibinfo {year} {2015})}\BibitemShut {NoStop}%
\bibitem [{Note1()}]{Note1}%
  \BibitemOpen
  \bibinfo {note} {Martin Steinel (private communication).}\BibitemShut {Stop}%
\bibitem [{\citenamefont {Lindvall}\ \emph {et~al.}(2023)\citenamefont {Lindvall}, \citenamefont {Wallin}, \citenamefont {Hanhij\"arvi},\ and\ \citenamefont {Fordell}}]{Lindvall2023a}%
  \BibitemOpen
  \bibfield  {author} {\bibinfo {author} {\bibfnamefont {T.}~\bibnamefont {Lindvall}}, \bibinfo {author} {\bibfnamefont {A.~E.}\ \bibnamefont {Wallin}}, \bibinfo {author} {\bibfnamefont {K.~J.}\ \bibnamefont {Hanhij\"arvi}},\ and\ \bibinfo {author} {\bibfnamefont {T.}~\bibnamefont {Fordell}},\ }\bibfield  {title} {\bibinfo {title} {Noise-induced servo errors in optical clocks utilizing {R}abi interrogation},\ }\href {https://doi.org/10.1088/1681-7575/acdfd4} {\bibfield  {journal} {\bibinfo  {journal} {Metrologia}\ }\textbf {\bibinfo {volume} {60}},\ \bibinfo {pages} {045008} (\bibinfo {year} {2023})}\BibitemShut {NoStop}%
\bibitem [{\citenamefont {Bourdeauducq}\ \emph {et~al.}(2016)\citenamefont {Bourdeauducq}, \citenamefont {J\"ordens}, \citenamefont {Zotov}, \citenamefont {Britton}, \citenamefont {Slichter}, \citenamefont {Leibrandt}, \citenamefont {Allcock}, \citenamefont {Hankin}, \citenamefont {Kermarrec}, \citenamefont {Sionneau}, \citenamefont {Srinivas}, \citenamefont {Tan},\ and\ \citenamefont {Bohnet}}]{Bourdeauducq2016a}%
  \BibitemOpen
  \bibfield  {author} {\bibinfo {author} {\bibfnamefont {S.}~\bibnamefont {Bourdeauducq}}, \bibinfo {author} {\bibfnamefont {R.}~\bibnamefont {J\"ordens}}, \bibinfo {author} {\bibfnamefont {P.}~\bibnamefont {Zotov}}, \bibinfo {author} {\bibfnamefont {J.}~\bibnamefont {Britton}}, \bibinfo {author} {\bibfnamefont {D.}~\bibnamefont {Slichter}}, \bibinfo {author} {\bibfnamefont {D.}~\bibnamefont {Leibrandt}}, \bibinfo {author} {\bibfnamefont {D.}~\bibnamefont {Allcock}}, \bibinfo {author} {\bibfnamefont {A.}~\bibnamefont {Hankin}}, \bibinfo {author} {\bibfnamefont {F.}~\bibnamefont {Kermarrec}}, \bibinfo {author} {\bibfnamefont {Y.}~\bibnamefont {Sionneau}}, \bibinfo {author} {\bibfnamefont {R.}~\bibnamefont {Srinivas}}, \bibinfo {author} {\bibfnamefont {T.~R.}\ \bibnamefont {Tan}},\ and\ \bibinfo {author} {\bibfnamefont {J.}~\bibnamefont {Bohnet}},\ }\href {https://doi.org/10.5281/zenodo.51303} {\bibinfo {title} {{ARTIQ} 1.0}} (\bibinfo {year} {2016})\BibitemShut {NoStop}%
\bibitem [{\citenamefont {Wu}\ \emph {et~al.}(2021)\citenamefont {Wu}, \citenamefont {Mills}, \citenamefont {West}, \citenamefont {Heaven},\ and\ \citenamefont {Hudson}}]{Wu2021b}%
  \BibitemOpen
  \bibfield  {author} {\bibinfo {author} {\bibfnamefont {H.}~\bibnamefont {Wu}}, \bibinfo {author} {\bibfnamefont {M.}~\bibnamefont {Mills}}, \bibinfo {author} {\bibfnamefont {E.}~\bibnamefont {West}}, \bibinfo {author} {\bibfnamefont {M.~C.}\ \bibnamefont {Heaven}},\ and\ \bibinfo {author} {\bibfnamefont {E.~R.}\ \bibnamefont {Hudson}},\ }\bibfield  {title} {\bibinfo {title} {Increase of the barium ion-trap lifetime via photodissociation},\ }\href {https://doi.org/10.1103/physreva.104.063103} {\bibfield  {journal} {\bibinfo  {journal} {Phys. Rev. A}\ }\textbf {\bibinfo {volume} {104}},\ \bibinfo {pages} {063103} (\bibinfo {year} {2021})} \BibitemShut {NoStop}%
\bibitem [{\citenamefont {Dub\'e}\ \emph {et~al.}(2015)\citenamefont {Dub\'e}, \citenamefont {Madej}, \citenamefont {Shiner},\ and\ \citenamefont {Jian}}]{Dube2015a}%
  \BibitemOpen
  \bibfield  {author} {\bibinfo {author} {\bibfnamefont {P.}~\bibnamefont {Dub\'e}}, \bibinfo {author} {\bibfnamefont {A.~A.}\ \bibnamefont {Madej}}, \bibinfo {author} {\bibfnamefont {A.}~\bibnamefont {Shiner}},\ and\ \bibinfo {author} {\bibfnamefont {B.}~\bibnamefont {Jian}},\ }\bibfield  {title} {\bibinfo {title} {$^{88}\mathrm{Sr}^{+}$ single-ion optical clock with a stability approaching the quantum projection noise limit},\ }\href {https://doi.org/10.1103/PhysRevA.92.042119} {\bibfield  {journal} {\bibinfo  {journal} {Phys. Rev. A}\ }\textbf {\bibinfo {volume} {92}},\ \bibinfo {pages} {042119} (\bibinfo {year} {2015})}\BibitemShut {NoStop}%
\bibitem [{\citenamefont {Dub\'e}\ \emph {et~al.}(2013)\citenamefont {Dub\'e}, \citenamefont {Madej}, \citenamefont {Zhou},\ and\ \citenamefont {Bernard}}]{Dube2013a}%
  \BibitemOpen
  \bibfield  {author} {\bibinfo {author} {\bibfnamefont {P.}~\bibnamefont {Dub\'e}}, \bibinfo {author} {\bibfnamefont {A.~A.}\ \bibnamefont {Madej}}, \bibinfo {author} {\bibfnamefont {Z.}~\bibnamefont {Zhou}},\ and\ \bibinfo {author} {\bibfnamefont {J.~E.}\ \bibnamefont {Bernard}},\ }\bibfield  {title} {\bibinfo {title} {{E}valuation of systematic shifts of the $^{88}${S}r$^+$ single-ion optical frequency standard at the $10^{-17}$ level},\ }\href {https://doi.org/10.1103/PhysRevA.87.023806} {\bibfield  {journal} {\bibinfo  {journal} {Phys. Rev. A}\ }\textbf {\bibinfo {volume} {87}},\ \bibinfo {pages} {023806} (\bibinfo {year} {2013})}\BibitemShut {NoStop}%
\bibitem [{\citenamefont {Berkeland}\ \emph {et~al.}(1998)\citenamefont {Berkeland}, \citenamefont {Miller}, \citenamefont {Bergquist}, \citenamefont {Itano},\ and\ \citenamefont {Wineland}}]{Berkeland1998a}%
  \BibitemOpen
  \bibfield  {author} {\bibinfo {author} {\bibfnamefont {D.~J.}\ \bibnamefont {Berkeland}}, \bibinfo {author} {\bibfnamefont {J.~D.}\ \bibnamefont {Miller}}, \bibinfo {author} {\bibfnamefont {J.~C.}\ \bibnamefont {Bergquist}}, \bibinfo {author} {\bibfnamefont {W.~M.}\ \bibnamefont {Itano}},\ and\ \bibinfo {author} {\bibfnamefont {D.~J.}\ \bibnamefont {Wineland}},\ }\bibfield  {title} {\bibinfo {title} {{M}inimization of ion micromotion in a {P}aul trap},\ }\href {https://doi.org/10.1063/1.367318} {\bibfield  {journal} {\bibinfo  {journal} {J. Appl. Phys.}\ }\textbf {\bibinfo {volume} {83}},\ \bibinfo {pages} {5025} (\bibinfo {year} {1998})}\BibitemShut {NoStop}%
\bibitem [{\citenamefont {Keller}\ \emph {et~al.}(2015)\citenamefont {Keller}, \citenamefont {Partner}, \citenamefont {Burgermeister},\ and\ \citenamefont {Mehlst\"aubler}}]{Keller2015a}%
  \BibitemOpen
  \bibfield  {author} {\bibinfo {author} {\bibfnamefont {J.}~\bibnamefont {Keller}}, \bibinfo {author} {\bibfnamefont {H.~L.}\ \bibnamefont {Partner}}, \bibinfo {author} {\bibfnamefont {T.}~\bibnamefont {Burgermeister}},\ and\ \bibinfo {author} {\bibfnamefont {T.~E.}\ \bibnamefont {Mehlst\"aubler}},\ }\bibfield  {title} {\bibinfo {title} {{P}recise determination of micromotion for trapped-ion optical clocks},\ }\href {https://doi.org/10.1063/1.4930037} {\bibfield  {journal} {\bibinfo  {journal} {J. Appl. Phys.}\ }\textbf {\bibinfo {volume} {118}},\ \bibinfo {eid} {104501} (\bibinfo {year} {2015})}\BibitemShut {NoStop}%
\bibitem [{\citenamefont {Dub\'e}\ \emph {et~al.}(2014)\citenamefont {Dub\'e}, \citenamefont {Madej}, \citenamefont {Tibbo},\ and\ \citenamefont {Bernard}}]{Dube2014a}%
  \BibitemOpen
  \bibfield  {author} {\bibinfo {author} {\bibfnamefont {P.}~\bibnamefont {Dub\'e}}, \bibinfo {author} {\bibfnamefont {A.~A.}\ \bibnamefont {Madej}}, \bibinfo {author} {\bibfnamefont {M.}~\bibnamefont {Tibbo}},\ and\ \bibinfo {author} {\bibfnamefont {J.~E.}\ \bibnamefont {Bernard}},\ }\bibfield  {title} {\bibinfo {title} {{H}igh-{A}ccuracy {M}easurement of the {D}ifferential {S}calar {P}olarizability of a $^{88}${Sr}$^{+}$ {C}lock {U}sing the {T}ime-{D}ilation {E}ffect},\ }\href {https://doi.org/10.1103/PhysRevLett.112.173002} {\bibfield  {journal} {\bibinfo  {journal} {Phys. Rev. Lett.}\ }\textbf {\bibinfo {volume} {112}},\ \bibinfo {pages} {173002} (\bibinfo {year} {2014})}\BibitemShut {NoStop}%
\bibitem [{\citenamefont {Barakhshan}\ \emph {et~al.}()\citenamefont {Barakhshan}, \citenamefont {Marrs}, \citenamefont {Bhosale}, \citenamefont {Arora}, \citenamefont {Eigenmann},\ and\ \citenamefont {Safronova}}]{UDportal}%
  \BibitemOpen
  \bibfield  {author} {\bibinfo {author} {\bibfnamefont {P.}~\bibnamefont {Barakhshan}}, \bibinfo {author} {\bibfnamefont {A.}~\bibnamefont {Marrs}}, \bibinfo {author} {\bibfnamefont {A.}~\bibnamefont {Bhosale}}, \bibinfo {author} {\bibfnamefont {B.}~\bibnamefont {Arora}}, \bibinfo {author} {\bibfnamefont {R.}~\bibnamefont {Eigenmann}},\ and\ \bibinfo {author} {\bibfnamefont {M.~S.}\ \bibnamefont {Safronova}},\ }\href {https://www.udel.edu/atom} {}\bibinfo {howpublished} {{\textit{Portal for High-Precision Atomic Data and Computation}} (version 2.0). University of Delaware, Newark, DE, USA. URL: {\tt{https://www.udel.edu/atom}} [February 2022].}\BibitemShut {Stop}%
\bibitem [{\citenamefont {Brownnutt}\ \emph {et~al.}(2015)\citenamefont {Brownnutt}, \citenamefont {Kumph}, \citenamefont {Rabl},\ and\ \citenamefont {Blatt}}]{Brownnutt2015a}%
  \BibitemOpen
  \bibfield  {author} {\bibinfo {author} {\bibfnamefont {M.}~\bibnamefont {Brownnutt}}, \bibinfo {author} {\bibfnamefont {M.}~\bibnamefont {Kumph}}, \bibinfo {author} {\bibfnamefont {P.}~\bibnamefont {Rabl}},\ and\ \bibinfo {author} {\bibfnamefont {R.}~\bibnamefont {Blatt}},\ }\bibfield  {title} {\bibinfo {title} {{I}on-trap measurements of electric-field noise near surfaces},\ }\href {https://doi.org/10.1103/RevModPhys.87.1419} {\bibfield  {journal} {\bibinfo  {journal} {Rev. Mod. Phys.}\ }\textbf {\bibinfo {volume} {87}},\ \bibinfo {pages} {1419} (\bibinfo {year} {2015})}\BibitemShut {NoStop}%
\bibitem [{\citenamefont {Dole\v{z}al}\ \emph {et~al.}(2015)\citenamefont {Dole\v{z}al}, \citenamefont {Balling}, \citenamefont {Nisbet-Jones}, \citenamefont {King}, \citenamefont {Jones}, \citenamefont {Klein}, \citenamefont {Gill}, \citenamefont {Lindvall}, \citenamefont {Wallin}, \citenamefont {Merimaa}, \citenamefont {Tamm}, \citenamefont {Sanner}, \citenamefont {Huntemann}, \citenamefont {Scharnhorst}, \citenamefont {Leroux}, \citenamefont {Schmidt}, \citenamefont {Burgermeister}, \citenamefont {Mehlst\"aubler},\ and\ \citenamefont {Peik}}]{Dolezal2015a}%
  \BibitemOpen
  \bibfield  {author} {\bibinfo {author} {\bibfnamefont {M.}~\bibnamefont {Dole\v{z}al}}, \bibinfo {author} {\bibfnamefont {P.}~\bibnamefont {Balling}}, \bibinfo {author} {\bibfnamefont {P.~B.~R.}\ \bibnamefont {Nisbet-Jones}}, \bibinfo {author} {\bibfnamefont {S.~A.}\ \bibnamefont {King}}, \bibinfo {author} {\bibfnamefont {J.~M.}\ \bibnamefont {Jones}}, \bibinfo {author} {\bibfnamefont {H.~A.}\ \bibnamefont {Klein}}, \bibinfo {author} {\bibfnamefont {P.}~\bibnamefont {Gill}}, \bibinfo {author} {\bibfnamefont {T.}~\bibnamefont {Lindvall}}, \bibinfo {author} {\bibfnamefont {A.~E.}\ \bibnamefont {Wallin}}, \bibinfo {author} {\bibfnamefont {M.}~\bibnamefont {Merimaa}}, \bibinfo {author} {\bibfnamefont {C.}~\bibnamefont {Tamm}}, \bibinfo {author} {\bibfnamefont {C.}~\bibnamefont {Sanner}}, \bibinfo {author} {\bibfnamefont {N.}~\bibnamefont {Huntemann}}, \bibinfo {author} {\bibfnamefont {N.}~\bibnamefont {Scharnhorst}}, \bibinfo {author} {\bibfnamefont {I.~D.}\ \bibnamefont {Leroux}}, \bibinfo {author}
  {\bibfnamefont {P.~O.}\ \bibnamefont {Schmidt}}, \bibinfo {author} {\bibfnamefont {T.}~\bibnamefont {Burgermeister}}, \bibinfo {author} {\bibfnamefont {T.~E.}\ \bibnamefont {Mehlst\"aubler}},\ and\ \bibinfo {author} {\bibfnamefont {E.}~\bibnamefont {Peik}},\ }\bibfield  {title} {\bibinfo {title} {{A}nalysis of thermal radiation in ion traps for optical frequency standards},\ }\href {https://doi.org/10.1088/0026-1394/52/6/842} {\bibfield  {journal} {\bibinfo  {journal} {Metrologia}\ }\textbf {\bibinfo {volume} {52}},\ \bibinfo {pages} {842} (\bibinfo {year} {2015})}\BibitemShut {NoStop}%
\bibitem [{\citenamefont {Abdel-Hafiz}\ \emph {et~al.}(2019)\citenamefont {Abdel-Hafiz}, \citenamefont {Ablewski}, \citenamefont {Al-Masoudi}, \citenamefont {\'{A}lvarez Mart\'{i}nez}, \citenamefont {Balling}, \citenamefont {Barwood}, \citenamefont {Benkler}, \citenamefont {Bober}, \citenamefont {Borkowski}, \citenamefont {Bowden}, \citenamefont {Ciury{\l}o}, \citenamefont {Cybulski}, \citenamefont {Didier}, \citenamefont {Dole\v{z}al}, \citenamefont {D\"orscher}, \citenamefont {Falke}, \citenamefont {Godun}, \citenamefont {Hamid}, \citenamefont {Hill}, \citenamefont {Hobson}, \citenamefont {Huntemann}, \citenamefont {Coq}, \citenamefont {Le~Targat}, \citenamefont {Legero}, \citenamefont {Lindvall}, \citenamefont {Lisdat}, \citenamefont {Lodewyck}, \citenamefont {Margolis}, \citenamefont {Mehlst\"aubler}, \citenamefont {Peik}, \citenamefont {Pelzer}, \citenamefont {Pizzocaro}, \citenamefont {Rauf}, \citenamefont {Rolland}, \citenamefont {Scharnhorst}, \citenamefont {Schioppo}, \citenamefont {Schmidt},
  \citenamefont {Schwarz}, \citenamefont {\c{S}enel}, \citenamefont {Spethmann}, \citenamefont {Sterr}, \citenamefont {Tamm}, \citenamefont {Thomsen}, \citenamefont {Vianello},\ and\ \citenamefont {Zawada}}]{OC18-Guidelines}%
  \BibitemOpen
  \bibfield  {author} {\bibinfo {author} {\bibfnamefont {M.}~\bibnamefont {Abdel-Hafiz}}, \bibinfo {author} {\bibfnamefont {P.}~\bibnamefont {Ablewski}}, \bibinfo {author} {\bibfnamefont {A.}~\bibnamefont {Al-Masoudi}}, \bibinfo {author} {\bibfnamefont {H.}~\bibnamefont {\'{A}lvarez Mart\'{i}nez}}, \bibinfo {author} {\bibfnamefont {P.}~\bibnamefont {Balling}}, \bibinfo {author} {\bibfnamefont {G.}~\bibnamefont {Barwood}}, \bibinfo {author} {\bibfnamefont {E.}~\bibnamefont {Benkler}}, \bibinfo {author} {\bibfnamefont {M.}~\bibnamefont {Bober}}, \bibinfo {author} {\bibfnamefont {M.}~\bibnamefont {Borkowski}}, \bibinfo {author} {\bibfnamefont {W.}~\bibnamefont {Bowden} \emph{et~al.}},\ } {{\bibinfo {title} {Guidelines for developing optical clocks with $10^{-18}$
  fractional frequency uncertainty}}},\ edited by\ \bibinfo {editor} {\bibfnamefont {T.}~\bibnamefont {Lindvall}}\ (\bibinfo {year} {2019})\ \Eprint {https://arxiv.org/abs/1906.11495} {arXiv:1906.11495} \BibitemShut {NoStop}%
\bibitem [{\citenamefont {Nisbet-Jones}\ \emph {et~al.}(2016)\citenamefont {Nisbet-Jones}, \citenamefont {King}, \citenamefont {Jones}, \citenamefont {Godun}, \citenamefont {Baynham}, \citenamefont {Bongs}, \citenamefont {Dole{\v{z}}al}, \citenamefont {Balling},\ and\ \citenamefont {Gill}}]{Nisbet-Jones2016a}%
  \BibitemOpen
  \bibfield  {author} {\bibinfo {author} {\bibfnamefont {P.~B.~R.}\ \bibnamefont {Nisbet-Jones}}, \bibinfo {author} {\bibfnamefont {S.~A.}\ \bibnamefont {King}}, \bibinfo {author} {\bibfnamefont {J.~M.}\ \bibnamefont {Jones}}, \bibinfo {author} {\bibfnamefont {R.~M.}\ \bibnamefont {Godun}}, \bibinfo {author} {\bibfnamefont {C.~F.~A.}\ \bibnamefont {Baynham}}, \bibinfo {author} {\bibfnamefont {K.}~\bibnamefont {Bongs}}, \bibinfo {author} {\bibfnamefont {M.}~\bibnamefont {Dole{\v{z}}al}}, \bibinfo {author} {\bibfnamefont {P.}~\bibnamefont {Balling}},\ and\ \bibinfo {author} {\bibfnamefont {P.}~\bibnamefont {Gill}},\ }\bibfield  {title} {\bibinfo {title} {{A} single-ion trap with minimized ion--environment interactions},\ }\href {https://doi.org/10.1007/s00340-016-6327-x} {\bibfield  {journal} {\bibinfo  {journal} {Appl. Phys. B}\ }\textbf {\bibinfo {volume} {122}},\ \bibinfo {pages} {57} (\bibinfo {year} {2016})}\BibitemShut {NoStop}%
\bibitem [{\citenamefont {Nordmann}\ \emph {et~al.}(2020)\citenamefont {Nordmann}, \citenamefont {Didier}, \citenamefont {Dole{\v{z}}al}, \citenamefont {Balling}, \citenamefont {Burgermeister},\ and\ \citenamefont {Mehlst\"aubler}}]{Nordmann2020a}%
  \BibitemOpen
  \bibfield  {author} {\bibinfo {author} {\bibfnamefont {T.}~\bibnamefont {Nordmann}}, \bibinfo {author} {\bibfnamefont {A.}~\bibnamefont {Didier}}, \bibinfo {author} {\bibfnamefont {M.}~\bibnamefont {Dole{\v{z}}al}}, \bibinfo {author} {\bibfnamefont {P.}~\bibnamefont {Balling}}, \bibinfo {author} {\bibfnamefont {T.}~\bibnamefont {Burgermeister}},\ and\ \bibinfo {author} {\bibfnamefont {T.~E.}\ \bibnamefont {Mehlst\"aubler}},\ }\bibfield  {title} {\bibinfo {title} {Sub-kelvin temperature management in ion traps for optical clocks},\ }\href {https://doi.org/10.1063/5.0024693} {\bibfield  {journal} {\bibinfo  {journal} {Rev. Sci. Instrum.}\ }\textbf {\bibinfo {volume} {91}},\ \bibinfo {pages} {111301} (\bibinfo {year} {2020})}\BibitemShut {NoStop}%
\bibitem [{\citenamefont {Ablewski}\ \emph {et~al.}(2020)\citenamefont {Ablewski}, \citenamefont {Bober},\ and\ \citenamefont {Zawada}}]{Ablewski2020a}%
  \BibitemOpen
  \bibfield  {author} {\bibinfo {author} {\bibfnamefont {P.}~\bibnamefont {Ablewski}}, \bibinfo {author} {\bibfnamefont {M.}~\bibnamefont {Bober}},\ and\ \bibinfo {author} {\bibfnamefont {M.}~\bibnamefont {Zawada}},\ }\bibfield  {title} {\bibinfo {title} {Emissivities of vacuum compatible materials: towards minimising blackbody radiation shift uncertainty in optical atomic clocks at room temperatures},\ }\href {https://doi.org/10.1088/1681-7575/ab63ae} {\bibfield  {journal} {\bibinfo  {journal} {Metrologia}\ }\textbf {\bibinfo {volume} {57}},\ \bibinfo {pages} {035004} (\bibinfo {year} {2020})}\BibitemShut {NoStop}%
\bibitem [{\citenamefont {Gaiser}\ \emph {et~al.}(2022)\citenamefont {Gaiser}, \citenamefont {Fellmuth}, \citenamefont {Gavioso}, \citenamefont {Kalemci}, \citenamefont {Kytin}, \citenamefont {Nakano}, \citenamefont {Pokhodun}, \citenamefont {Rourke}, \citenamefont {Rusby}, \citenamefont {Sparasci}, \citenamefont {Steur}, \citenamefont {Tew}, \citenamefont {Underwood}, \citenamefont {White}, \citenamefont {Yang},\ and\ \citenamefont {Zhang}}]{Gaiser2022a}%
  \BibitemOpen
  \bibfield  {author} {\bibinfo {author} {\bibfnamefont {C.}~\bibnamefont {Gaiser}}, \bibinfo {author} {\bibfnamefont {B.}~\bibnamefont {Fellmuth}}, \bibinfo {author} {\bibfnamefont {R.~M.}\ \bibnamefont {Gavioso}}, \bibinfo {author} {\bibfnamefont {M.}~\bibnamefont {Kalemci}}, \bibinfo {author} {\bibfnamefont {V.}~\bibnamefont {Kytin}}, \bibinfo {author} {\bibfnamefont {T.}~\bibnamefont {Nakano}}, \bibinfo {author} {\bibfnamefont {A.}~\bibnamefont {Pokhodun}}, \bibinfo {author} {\bibfnamefont {P.~M.~C.}\ \bibnamefont {Rourke}}, \bibinfo {author} {\bibfnamefont {R.}~\bibnamefont {Rusby}}, \bibinfo {author} {\bibfnamefont {F.}~\bibnamefont {Sparasci}}, \bibinfo {author} {\bibfnamefont {P.~P.~M.}\ \bibnamefont {Steur}}, \bibinfo {author} {\bibfnamefont {W.~L.}\ \bibnamefont {Tew}}, \bibinfo {author} {\bibfnamefont {R.}~\bibnamefont {Underwood}}, \bibinfo {author} {\bibfnamefont {R.}~\bibnamefont {White}}, \bibinfo {author} {\bibfnamefont {I.}~\bibnamefont {Yang}},\ and\ \bibinfo {author} {\bibfnamefont
  {J.}~\bibnamefont {Zhang}},\ }\bibfield  {title} {\bibinfo {title} {2022 update for the differences between thermodynamic temperature and {ITS}-90 below 335 {K}},\ }\href {https://doi.org/10.1063/5.0131026} {\bibfield  {journal} {\bibinfo  {journal} {J. Phys. Chem. Ref. Data}\ }\textbf {\bibinfo {volume} {51}},\ \bibinfo {pages} {043105} (\bibinfo {year} {2022})}\BibitemShut {NoStop}%
\bibitem [{\citenamefont {Sansonetti}(2012)}]{Sansonetti2012a}%
  \BibitemOpen
  \bibfield  {author} {\bibinfo {author} {\bibfnamefont {J.~E.}\ \bibnamefont {Sansonetti}},\ }\bibfield  {title} {\bibinfo {title} {{W}avelengths, {T}ransition {P}robabilities, and {E}nergy {L}evels for the {S}pectra of {S}trontium {I}ons ({S}r {II} through {S}r {XXXVIII})},\ }\href {https://doi.org/10.1063/1.3659413} {\bibfield  {journal} {\bibinfo  {journal} {J. Phys. Chem. Ref. Data}\ }\textbf {\bibinfo {volume} {41}},\ \bibinfo {pages} {013102} (\bibinfo {year} {2012})}\BibitemShut {NoStop}%
\bibitem [{\citenamefont {Farley}\ and\ \citenamefont {Wing}(1981)}]{Farley1981a}%
  \BibitemOpen
  \bibfield  {author} {\bibinfo {author} {\bibfnamefont {J.~W.}\ \bibnamefont {Farley}}\ and\ \bibinfo {author} {\bibfnamefont {W.~H.}\ \bibnamefont {Wing}},\ }\bibfield  {title} {\bibinfo {title} {{A}ccurate calculation of dynamic {S}tark shifts and depopulation rates of {R}ydberg energy levels induced by blackbody radiation. {H}ydrogen, helium, and alkali-metal atoms},\ }\href {https://doi.org/10.1103/PhysRevA.23.2397} {\bibfield  {journal} {\bibinfo  {journal} {Phys. Rev. A}\ }\textbf {\bibinfo {volume} {23}},\ \bibinfo {pages} {2397} (\bibinfo {year} {1981})}\BibitemShut {NoStop}%
\bibitem [{\citenamefont {Porsev}\ and\ \citenamefont {Derevianko}(2006)}]{Porsev2006a}%
  \BibitemOpen
  \bibfield  {author} {\bibinfo {author} {\bibfnamefont {S.~G.}\ \bibnamefont {Porsev}}\ and\ \bibinfo {author} {\bibfnamefont {A.}~\bibnamefont {Derevianko}},\ }\bibfield  {title} {\bibinfo {title} {{M}ultipolar theory of blackbody radiation shift of atomic energy levels and its implications for optical lattice clocks},\ }\href {https://doi.org/10.1103/PhysRevA.74.020502} {\bibfield  {journal} {\bibinfo  {journal} {Phys. Rev. A}\ }\textbf {\bibinfo {volume} {74}},\ \bibinfo {pages} {020502} (\bibinfo {year} {2006})}\BibitemShut {NoStop}%
\bibitem [{\citenamefont {Arora}\ \emph {et~al.}(2012)\citenamefont {Arora}, \citenamefont {Nandy},\ and\ \citenamefont {Sahoo}}]{Arora2012a}%
  \BibitemOpen
  \bibfield  {author} {\bibinfo {author} {\bibfnamefont {B.}~\bibnamefont {Arora}}, \bibinfo {author} {\bibfnamefont {D.~K.}\ \bibnamefont {Nandy}},\ and\ \bibinfo {author} {\bibfnamefont {B.~K.}\ \bibnamefont {Sahoo}},\ }\bibfield  {title} {\bibinfo {title} {{M}ultipolar blackbody radiation shifts for single-ion clocks},\ }\href {https://doi.org/10.1103/PhysRevA.85.012506} {\bibfield  {journal} {\bibinfo  {journal} {Phys. Rev. A}\ }\textbf {\bibinfo {volume} {85}},\ \bibinfo {pages} {012506} (\bibinfo {year} {2012})}\BibitemShut {NoStop}%
\bibitem [{\citenamefont {Safronova}\ and\ \citenamefont {Safronova}(2011)}]{Safronova2011c}%
  \BibitemOpen
  \bibfield  {author} {\bibinfo {author} {\bibfnamefont {U.~I.}\ \bibnamefont {Safronova}}\ and\ \bibinfo {author} {\bibfnamefont {M.~S.}\ \bibnamefont {Safronova}},\ }\bibfield  {title} {\bibinfo {title} {Excitation energies, {E1}, {M1}, and {E2} transition rates, and lifetimes in {Ca}$^+$, {Sr}$^+$, {Cd}$^+$, {Ba}$^+$, and {Hg}$^+$},\ }\href {https://doi.org/10.1139/p11-004} {\bibfield  {journal} {\bibinfo  {journal} {Can. J. Phys.}\ }\textbf {\bibinfo {volume} {89}},\ \bibinfo {pages} {465} (\bibinfo {year} {2011})}\BibitemShut {NoStop}%
\bibitem [{\citenamefont {Tang}\ \emph {et~al.}(2024)\citenamefont {Tang}, \citenamefont {Wei}, \citenamefont {Sahoo}, \citenamefont {Li}, \citenamefont {Yang}, \citenamefont {Zou},\ and\ \citenamefont {Huang}}]{Tang2024a}%
  \BibitemOpen
  \bibfield  {author} {\bibinfo {author} {\bibfnamefont {Z.-M.}\ \bibnamefont {Tang}}, \bibinfo {author} {\bibfnamefont {Y.-F.}\ \bibnamefont {Wei}}, \bibinfo {author} {\bibfnamefont {B.~K.}\ \bibnamefont {Sahoo}}, \bibinfo {author} {\bibfnamefont {C.-B.}\ \bibnamefont {Li}}, \bibinfo {author} {\bibfnamefont {Y.}~\bibnamefont {Yang}}, \bibinfo {author} {\bibfnamefont {Y.}~\bibnamefont {Zou}},\ and\ \bibinfo {author} {\bibfnamefont {X.-R.}\ \bibnamefont {Huang}},\ }\bibfield  {title} {\bibinfo {title} {Blackbody-radiation {Z}eeman shifts in optical clocks: The role of fine-structure intramanifold resonances},\ }\href {https://doi.org/10.1103/physreva.110.043108} {\bibfield  {journal} {\bibinfo  {journal} {Phys. Rev. A}\ }\textbf {\bibinfo {volume} {110}},\ \bibinfo {pages} {043108} (\bibinfo {year} {2024})} \BibitemShut {NoStop}%
\bibitem [{\citenamefont {Arnold}\ \emph {et~al.}(2020)\citenamefont {Arnold}, \citenamefont {Kaewuam}, \citenamefont {Chanu}, \citenamefont {Tan}, \citenamefont {Zhang},\ and\ \citenamefont {Barrett}}]{Arnold2020a}%
  \BibitemOpen
  \bibfield  {author} {\bibinfo {author} {\bibfnamefont {K.~J.}\ \bibnamefont {Arnold}}, \bibinfo {author} {\bibfnamefont {R.}~\bibnamefont {Kaewuam}}, \bibinfo {author} {\bibfnamefont {S.~R.}\ \bibnamefont {Chanu}}, \bibinfo {author} {\bibfnamefont {T.~R.}\ \bibnamefont {Tan}}, \bibinfo {author} {\bibfnamefont {Z.}~\bibnamefont {Zhang}},\ and\ \bibinfo {author} {\bibfnamefont {M.~D.}\ \bibnamefont {Barrett}},\ }\bibfield  {title} {\bibinfo {title} {Precision measurements of the $^{138}${Ba}$^{+}$ $6s{^{2}S}_{1/2}\ensuremath{-}5d{^{2}D}_{5/2}$ clock transition},\ }\href {https://doi.org/10.1103/PhysRevLett.124.193001} {\bibfield  {journal} {\bibinfo  {journal} {Phys. Rev. Lett.}\ }\textbf {\bibinfo {volume} {124}},\ \bibinfo {pages} {193001} (\bibinfo {year} {2020})}\BibitemShut {NoStop}%
\bibitem [{\citenamefont {Curtis}\ \emph {et~al.}(2024)\citenamefont {Curtis}, \citenamefont {Tofful}, \citenamefont {Parsons}, \citenamefont {Robertson},\ and\ \citenamefont {Godun}}]{Curtis2024a}%
  \BibitemOpen
  \bibfield  {author} {\bibinfo {author} {\bibfnamefont {E.~A.}\ \bibnamefont {Curtis}}, \bibinfo {author} {\bibfnamefont {A.}~\bibnamefont {Tofful}}, \bibinfo {author} {\bibfnamefont {A.~O.}\ \bibnamefont {Parsons}}, \bibinfo {author} {\bibfnamefont {B.~I.}\ \bibnamefont {Robertson}},\ and\ \bibinfo {author} {\bibfnamefont {R.~M.}\ \bibnamefont {Godun}},\ }\bibfield  {title} {\bibinfo {title} {$^{171}${Yb}$^{+}$ ion optical clock at {NPL}: advances in automation and assessment of the trap-induced ac {Z}eeman shift in an endcap trap geometry},\ }\href {https://doi.org/10.1088/1742-6596/2889/1/012044} {\bibfield  {journal} {\bibinfo  {journal} {J. Phys. Conf. Ser.}\ }\textbf {\bibinfo {volume} {2889}},\ \bibinfo {pages} {012044} (\bibinfo {year} {2024})}\BibitemShut {NoStop}%
\bibitem [{\citenamefont {Ivory}\ \emph {et~al.}(2024)\citenamefont {Ivory}, \citenamefont {Nordquist}, \citenamefont {Young}, \citenamefont {Hogle}, \citenamefont {Clark},\ and\ \citenamefont {Revelle}}]{Ivory2024a}%
  \BibitemOpen
  \bibfield  {author} {\bibinfo {author} {\bibfnamefont {M.}~\bibnamefont {Ivory}}, \bibinfo {author} {\bibfnamefont {C.~D.}\ \bibnamefont {Nordquist}}, \bibinfo {author} {\bibfnamefont {K.}~\bibnamefont {Young}}, \bibinfo {author} {\bibfnamefont {C.~W.}\ \bibnamefont {Hogle}}, \bibinfo {author} {\bibfnamefont {S.~M.}\ \bibnamefont {Clark}},\ and\ \bibinfo {author} {\bibfnamefont {M.~C.}\ \bibnamefont {Revelle}},\ }\bibfield  {title} {\bibinfo {title} {{AC} {Z}eeman effect in microfabricated surface traps},\ }\href {https://doi.org/10.1063/5.0204413} {\bibfield  {journal} {\bibinfo  {journal} {Rev. Sci. Instrum.}\ }\textbf {\bibinfo {volume} {95}},\ \bibinfo {pages} {093202} (\bibinfo {year} {2024})}\BibitemShut {NoStop}%
\bibitem [{\citenamefont {James}(1998)}]{James1998a}%
  \BibitemOpen
  \bibfield  {author} {\bibinfo {author} {\bibfnamefont {D.~F.~V.}\ \bibnamefont {James}},\ }\bibfield  {title} {\bibinfo {title} {{Q}uantum dynamics of cold trapped ions with application to quantum computation},\ }\href {https://doi.org/10.1007/s003400050373} {\bibfield  {journal} {\bibinfo  {journal} {Appl. Phys. B}\ }\textbf {\bibinfo {volume} {66}},\ \bibinfo {pages} {181} (\bibinfo {year} {1998})}\BibitemShut {NoStop}%
\bibitem [{\citenamefont {Safronova}\ \emph {et~al.}(2017)\citenamefont {Safronova}, \citenamefont {Safronova},\ and\ \citenamefont {Johnson}}]{Safronova2017a}%
  \BibitemOpen
  \bibfield  {author} {\bibinfo {author} {\bibfnamefont {U.~I.}\ \bibnamefont {Safronova}}, \bibinfo {author} {\bibfnamefont {M.~S.}\ \bibnamefont {Safronova}},\ and\ \bibinfo {author} {\bibfnamefont {W.~R.}\ \bibnamefont {Johnson}},\ }\bibfield  {title} {\bibinfo {title} {{F}orbidden ${M}1$ and ${E}2$ transitions in monovalent atoms and ions},\ }\href {https://doi.org/10.1103/PhysRevA.95.042507} {\bibfield  {journal} {\bibinfo  {journal} {Phys. Rev. A}\ }\textbf {\bibinfo {volume} {95}},\ \bibinfo {pages} {042507} (\bibinfo {year} {2017})}\BibitemShut {NoStop}%
\bibitem [{\citenamefont {Yudin}\ \emph {et~al.}(2023)\citenamefont {Yudin}, \citenamefont {Taichenachev}, \citenamefont {Prudnikov}, \citenamefont {Basalaev}, \citenamefont {Pal'chikov}, \citenamefont {von Boehn}, \citenamefont {Mehlstäubler},\ and\ \citenamefont {Bagayev}}]{Yudin2023a}%
  \BibitemOpen
  \bibfield  {author} {\bibinfo {author} {\bibfnamefont {V.~I.}\ \bibnamefont {Yudin}}, \bibinfo {author} {\bibfnamefont {A.~V.}\ \bibnamefont {Taichenachev}}, \bibinfo {author} {\bibfnamefont {O.~N.}\ \bibnamefont {Prudnikov}}, \bibinfo {author} {\bibfnamefont {M.~Y.}\ \bibnamefont {Basalaev}}, \bibinfo {author} {\bibfnamefont {V.~G.}\ \bibnamefont {Pal'chikov}}, \bibinfo {author} {\bibfnamefont {M.}~\bibnamefont {von Boehn}}, \bibinfo {author} {\bibfnamefont {T.~E.}\ \bibnamefont {Mehlstäubler}},\ and\ \bibinfo {author} {\bibfnamefont {S.~N.}\ \bibnamefont {Bagayev}},\ }\bibfield  {title} {\bibinfo {title} {Probe-field-ellipticity-induced shift in an atomic clock},\ }\href {https://doi.org/10.1103/physrevapplied.19.014022} {\bibfield  {journal} {\bibinfo  {journal} {Phys. Rev. Appl.}\ }\textbf {\bibinfo {volume} {19}},\ \bibinfo {pages} {014022} (\bibinfo {year} {2023})} \BibitemShut {NoStop}%
\bibitem [{\citenamefont {Roberts}\ \emph {et~al.}(2023)\citenamefont {Roberts}, \citenamefont {Fairhall},\ and\ \citenamefont {Ginges}}]{Roberts2023a}%
  \BibitemOpen
  \bibfield  {author} {\bibinfo {author} {\bibfnamefont {B.~M.}\ \bibnamefont {Roberts}}, \bibinfo {author} {\bibfnamefont {C.~J.}\ \bibnamefont {Fairhall}},\ and\ \bibinfo {author} {\bibfnamefont {J.~S.~M.}\ \bibnamefont {Ginges}},\ }\bibfield  {title} {\bibinfo {title} {Electric-dipole transition amplitudes for atoms and ions with one valence electron},\ }\href {https://doi.org/10.1103/physreva.107.052812} {\bibfield  {journal} {\bibinfo  {journal} {Phys. Rev. A}\ }\textbf {\bibinfo {volume} {107}},\ \bibinfo {pages} {052812} (\bibinfo {year} {2023})} \BibitemShut {NoStop}%
\bibitem [{\citenamefont {Kersevan}\ and\ \citenamefont {Ady}(2019)}]{Kersevan2019a}%
  \BibitemOpen
  \bibfield  {author} {\bibinfo {author} {\bibfnamefont {R.}~\bibnamefont {Kersevan}}\ and\ \bibinfo {author} {\bibfnamefont {M.}~\bibnamefont {Ady}},\ }\bibfield  {title} {\bibinfo {title} {Recent developments of {M}onte-{C}arlo codes {M}olflow+ and {S}ynrad+},\ }\href {https://doi.org/10.18429/JACOW-IPAC2019-TUPMP037} {\bibfield  {journal} {\bibinfo  {journal} {Proceedings of the 10th Int. Particle Accelerator Conf.}\ }\textbf {\bibinfo {volume} {IPAC2019}},\ \bibinfo {pages} {Australia} (\bibinfo {year} {2019})}\BibitemShut {NoStop}%
\bibitem [{\citenamefont {Likforman}\ \emph {et~al.}(2016)\citenamefont {Likforman}, \citenamefont {Tugay\'e}, \citenamefont {Guibal},\ and\ \citenamefont {Guidoni}}]{Likforman2016a}%
  \BibitemOpen
  \bibfield  {author} {\bibinfo {author} {\bibfnamefont {J.-P.}\ \bibnamefont {Likforman}}, \bibinfo {author} {\bibfnamefont {V.}~\bibnamefont {Tugay\'e}}, \bibinfo {author} {\bibfnamefont {S.}~\bibnamefont {Guibal}},\ and\ \bibinfo {author} {\bibfnamefont {L.}~\bibnamefont {Guidoni}},\ }\bibfield  {title} {\bibinfo {title} {{P}recision measurement of the branching fractions of the $5p\,^2{P}_{1/2}$ state in $^{88}\mathrm{Sr}^+$ with a single ion in a microfabricated surface trap},\ }\href {https://doi.org/10.1103/PhysRevA.93.052507} {\bibfield  {journal} {\bibinfo  {journal} {Phys. Rev. A}\ }\textbf {\bibinfo {volume} {93}},\ \bibinfo {pages} {052507} (\bibinfo {year} {2016})}\BibitemShut {NoStop}%
\bibitem [{\citenamefont {Barton}\ \emph {et~al.}(2000)\citenamefont {Barton}, \citenamefont {Donald}, \citenamefont {Lucas}, \citenamefont {Stevens}, \citenamefont {Steane},\ and\ \citenamefont {Stacey}}]{Barton2000a}%
  \BibitemOpen
  \bibfield  {author} {\bibinfo {author} {\bibfnamefont {P.~A.}\ \bibnamefont {Barton}}, \bibinfo {author} {\bibfnamefont {C.~J.~S.}\ \bibnamefont {Donald}}, \bibinfo {author} {\bibfnamefont {D.~M.}\ \bibnamefont {Lucas}}, \bibinfo {author} {\bibfnamefont {D.~A.}\ \bibnamefont {Stevens}}, \bibinfo {author} {\bibfnamefont {A.~M.}\ \bibnamefont {Steane}},\ and\ \bibinfo {author} {\bibfnamefont {D.~N.}\ \bibnamefont {Stacey}},\ }\bibfield  {title} {\bibinfo {title} {Measurement of the lifetime of the $3d{}^{2}{D}_{5/2}$ state in ${}^{40}${Ca}${}^{+}$},\ }\href {https://doi.org/10.1103/PhysRevA.62.032503} {\bibfield  {journal} {\bibinfo  {journal} {Phys. Rev. A}\ }\textbf {\bibinfo {volume} {62}},\ \bibinfo {pages} {032503} (\bibinfo {year} {2000})}\BibitemShut {NoStop}%
\bibitem [{\citenamefont {Pagano}\ \emph {et~al.}(2019)\citenamefont {Pagano}, \citenamefont {Hess}, \citenamefont {Kaplan}, \citenamefont {Tan}, \citenamefont {Richerme}, \citenamefont {Becker}, \citenamefont {Kyprianidis}, \citenamefont {Zhang}, \citenamefont {Birckelbaw}, \citenamefont {Hernandez}, \citenamefont {Wu},\ and\ \citenamefont {Monroe}}]{Pagano2019a}%
  \BibitemOpen
  \bibfield  {author} {\bibinfo {author} {\bibfnamefont {G.}~\bibnamefont {Pagano}}, \bibinfo {author} {\bibfnamefont {P.~W.}\ \bibnamefont {Hess}}, \bibinfo {author} {\bibfnamefont {H.~B.}\ \bibnamefont {Kaplan}}, \bibinfo {author} {\bibfnamefont {W.~L.}\ \bibnamefont {Tan}}, \bibinfo {author} {\bibfnamefont {P.}~\bibnamefont {Richerme}}, \bibinfo {author} {\bibfnamefont {P.}~\bibnamefont {Becker}}, \bibinfo {author} {\bibfnamefont {A.}~\bibnamefont {Kyprianidis}}, \bibinfo {author} {\bibfnamefont {J.}~\bibnamefont {Zhang}}, \bibinfo {author} {\bibfnamefont {E.}~\bibnamefont {Birckelbaw}}, \bibinfo {author} {\bibfnamefont {M.~R.}\ \bibnamefont {Hernandez}}, \bibinfo {author} {\bibfnamefont {Y.}~\bibnamefont {Wu}},\ and\ \bibinfo {author} {\bibfnamefont {C.}~\bibnamefont {Monroe}},\ }\bibfield  {title} {\bibinfo {title} {Cryogenic trapped-ion system for large scale quantum simulation},\ }\href {https://doi.org/10.1088/2058-9565/aae0fe} {\bibfield  {journal} {\bibinfo  {journal} {Quantum Sci. Technol.}\
  }\textbf {\bibinfo {volume} {4}},\ \bibinfo {pages} {014004} (\bibinfo {year} {2019})}\BibitemShut {NoStop}%
\bibitem [{\citenamefont {Hankin}\ \emph {et~al.}(2019)\citenamefont {Hankin}, \citenamefont {Clements}, \citenamefont {Huang}, \citenamefont {Brewer}, \citenamefont {Chen}, \citenamefont {Chou}, \citenamefont {Hume},\ and\ \citenamefont {Leibrandt}}]{Hankin2019a}%
  \BibitemOpen
  \bibfield  {author} {\bibinfo {author} {\bibfnamefont {A.~M.}\ \bibnamefont {Hankin}}, \bibinfo {author} {\bibfnamefont {E.~R.}\ \bibnamefont {Clements}}, \bibinfo {author} {\bibfnamefont {Y.}~\bibnamefont {Huang}}, \bibinfo {author} {\bibfnamefont {S.~M.}\ \bibnamefont {Brewer}}, \bibinfo {author} {\bibfnamefont {J.-S.}\ \bibnamefont {Chen}}, \bibinfo {author} {\bibfnamefont {C.~W.}\ \bibnamefont {Chou}}, \bibinfo {author} {\bibfnamefont {D.~B.}\ \bibnamefont {Hume}},\ and\ \bibinfo {author} {\bibfnamefont {D.~R.}\ \bibnamefont {Leibrandt}},\ }\bibfield  {title} {\bibinfo {title} {Systematic uncertainty due to background-gas collisions in trapped-ion optical clocks},\ }\href {https://doi.org/10.1103/PhysRevA.100.033419} {\bibfield  {journal} {\bibinfo  {journal} {Phys. Rev. A}\ }\textbf {\bibinfo {volume} {100}},\ \bibinfo {pages} {033419} (\bibinfo {year} {2019})}\BibitemShut {NoStop}%
\bibitem [{\citenamefont {Lagakos}\ \emph {et~al.}(1981)\citenamefont {Lagakos}, \citenamefont {Bucaro},\ and\ \citenamefont {Jarzynski}}]{Lagakos1981a}%
  \BibitemOpen
  \bibfield  {author} {\bibinfo {author} {\bibfnamefont {N.}~\bibnamefont {Lagakos}}, \bibinfo {author} {\bibfnamefont {J.~A.}\ \bibnamefont {Bucaro}},\ and\ \bibinfo {author} {\bibfnamefont {J.}~\bibnamefont {Jarzynski}},\ }\bibfield  {title} {\bibinfo {title} {Temperature-induced optical phase shifts in fibers},\ }\href {https://doi.org/10.1364/ao.20.002305} {\bibfield  {journal} {\bibinfo  {journal} {Appl. Opt.}\ }\textbf {\bibinfo {volume} {20}},\ \bibinfo {pages} {2305} (\bibinfo {year} {1981})}\BibitemShut {NoStop}%
\bibitem [{\citenamefont {Lindvall}\ \emph {et~al.}(2025{\natexlab{b}})\citenamefont {Lindvall}, \citenamefont {Pizzocaro}, \citenamefont {Godun}, \citenamefont {Abgrall}, \citenamefont {Akamatsu}, \citenamefont {Amy-Klein}, \citenamefont {Benkler}, \citenamefont {Bhatt}, \citenamefont {Calonico}, \citenamefont {Cantin}, \citenamefont {Cantoni}, \citenamefont {Cerretto}, \citenamefont {Chardonnet}, \citenamefont {Cifuentes~Marin}, \citenamefont {Clivati}, \citenamefont {Condio}, \citenamefont {Curtis}, \citenamefont {Denker}, \citenamefont {Donadello}, \citenamefont {D\"orscher}, \citenamefont {Feng}, \citenamefont {Filzinger}, \citenamefont {Fordell}, \citenamefont {Goti}, \citenamefont {Hanhij\"arvi}, \citenamefont {Hausser}, \citenamefont {Hill}, \citenamefont {Hosaka}, \citenamefont {Huntemann}, \citenamefont {Johnson}, \citenamefont {Keller}, \citenamefont {Klose}, \citenamefont {Kobayashi}, \citenamefont {Koke}, \citenamefont {Kuhl}, \citenamefont {Le~Targat}, \citenamefont {Legero}, \citenamefont
  {Levi}, \citenamefont {Lipphardt}, \citenamefont {Lisdat}, \citenamefont {Liu}, \citenamefont {Lodewyck}, \citenamefont {Lopez}, \citenamefont {Mazouth-Laurol}, \citenamefont {Mehlst\"aubler}, \citenamefont {Mura}, \citenamefont {Nishiyama}, \citenamefont {Nordmann}, \citenamefont {Parsons}, \citenamefont {Petit}, \citenamefont {Pointard}, \citenamefont {Pottie}, \citenamefont {Risaro}, \citenamefont {Robertson}, \citenamefont {Schioppo}, \citenamefont {Shang}, \citenamefont {Stahl}, \citenamefont {Steinel}, \citenamefont {Sterr}, \citenamefont {Tofful}, \citenamefont {T{\o}nnes}, \citenamefont {Tran}, \citenamefont {Tunesi}, \citenamefont {Wallin},\ and\ \citenamefont {Margolis}}]{Lindvall2025a}%
  \BibitemOpen
  \bibfield  {author} {\bibinfo {author} {\bibfnamefont {T.}~\bibnamefont {Lindvall}}, \bibinfo {author} {\bibfnamefont {M.}~\bibnamefont {Pizzocaro}}, \bibinfo {author} {\bibfnamefont {R.~M.}\ \bibnamefont {Godun}}, \bibinfo {author} {\bibfnamefont {M.}~\bibnamefont {Abgrall}}, \bibinfo {author} {\bibfnamefont {D.}~\bibnamefont {Akamatsu}}, \bibinfo {author} {\bibfnamefont {A.}~\bibnamefont {Amy-Klein}}, \bibinfo {author} {\bibfnamefont {E.}~\bibnamefont {Benkler}}, \bibinfo {author} {\bibfnamefont {N.}~\bibnamefont {Bhatt}}, \bibinfo {author} {\bibfnamefont {D.}~\bibnamefont {Calonico}}, \bibinfo {author} {\bibfnamefont {E.}~\bibnamefont {Cantin} \emph{et~al.}},\ }\bibfield  {title} {\bibinfo {title} {Coordinated international comparisons between optical clocks connected via fiber and satellite links},\ }\href {https://doi.org/10.1364/optica.561754} {\bibfield  {journal} {\bibinfo  {journal} {Optica}\ }\textbf {\bibinfo {volume} {12}},\ \bibinfo {pages} {843} (\bibinfo {year} {2025}{\natexlab{b}})} \BibitemShut {NoStop}%
\bibitem [{\citenamefont {Margolis}\ \emph {et~al.}(2024{\natexlab{a}})\citenamefont {Margolis}, \citenamefont {Godun}, \citenamefont {Huntemann}, \citenamefont {Le~Targat}, \citenamefont {Pizzocaro}, \citenamefont {Zawada}, \citenamefont {Abgrall}, \citenamefont {Akamatsu}, \citenamefont {\'{A}lvarez Mart\'{i}nez}, \citenamefont {Amy-Klein}, \citenamefont {Andia}, \citenamefont {Benkler}, \citenamefont {Bhatt}, \citenamefont {Bilicki}, \citenamefont {Bize}, \citenamefont {Bober}, \citenamefont {Calonico}, \citenamefont {Cambier}, \citenamefont {Cantin}, \citenamefont {Chardonnet}, \citenamefont {Cifuentes~Mar\'{i}n}, \citenamefont {Clivati}, \citenamefont {Condio}, \citenamefont {Curtis}, \citenamefont {Czubla}, \citenamefont {Dole\v{z}al}, \citenamefont {D\"orscher}, \citenamefont {Dunst}, \citenamefont {Feng}, \citenamefont {Filzinger}, \citenamefont {Folman}, \citenamefont {Fordell}, \citenamefont {Formichella}, \citenamefont {Foucault}, \citenamefont {Galleani}, \citenamefont {Goti}, \citenamefont
  {Groswasser}, \citenamefont {Gruszczy\'{n}ski}, \citenamefont {Guo}, \citenamefont {Hanhij\"arvi}, \citenamefont {Hausser}, \citenamefont {Hill}, \citenamefont {Hosaka}, \citenamefont {Johnson}, \citenamefont {Keller}, \citenamefont {Klose}, \citenamefont {Kobayashi}, \citenamefont {Koke}, \citenamefont {Kova\v{c}i\'{c}}, \citenamefont {K\c{r}en}, \citenamefont {Kuhl}, \citenamefont {Ledzi\'{n}ski}, \citenamefont {Lema\'{n}ski}, \citenamefont {Levi}, \citenamefont {Lindvall}, \citenamefont {Lisdat}, \citenamefont {Liu}, \citenamefont {Lodewyck}, \citenamefont {Lopez}, \citenamefont {Lorini}, \citenamefont {Lours}, \citenamefont {Ma\v{s}ika}, \citenamefont {Mazouth-Laurol}, \citenamefont {Mehlst\"aubler}, \citenamefont {Moreno}, \citenamefont {Morzy\'{n}ski}, \citenamefont {Naro\.{z}nik}, \citenamefont {Nawrocki}, \citenamefont {Nishiyama}, \citenamefont {Noga\'{s}}, \citenamefont {Nordmann}, \citenamefont {Parsons}, \citenamefont {Pointard}, \citenamefont {Pottie}, \citenamefont {Risaro}, \citenamefont
  {Robertson}, \citenamefont {Romero~Gonz\'{a}lez}, \citenamefont {Schioppo}, \citenamefont {Sesia}, \citenamefont {Shang}, \citenamefont {Signorile}, \citenamefont {Stahl}, \citenamefont {Steinel}, \citenamefont {Sterr}, \citenamefont {Su\'{a}rez~Ram\'{i}rez}, \citenamefont {Tofful}, \citenamefont {T{\o}nnes}, \citenamefont {Tran}, \citenamefont {Tunesi}, \citenamefont {Wallin}, \citenamefont {Waterholter}, \citenamefont {Zarei},\ and\ \citenamefont {Zyskind}}]{Margolis2024b}%
  \BibitemOpen
  \bibfield  {author} {\bibinfo {author} {\bibfnamefont {H.~S.}\ \bibnamefont {Margolis}}, \bibinfo {author} {\bibfnamefont {R.~M.}\ \bibnamefont {Godun}}, \bibinfo {author} {\bibfnamefont {N.}~\bibnamefont {Huntemann}}, \bibinfo {author} {\bibfnamefont {R.}~\bibnamefont {Le~Targat}}, \bibinfo {author} {\bibfnamefont {M.}~\bibnamefont {Pizzocaro}}, \bibinfo {author} {\bibfnamefont {M.}~\bibnamefont {Zawada}}, \bibinfo {author} {\bibfnamefont {M.}~\bibnamefont {Abgrall}}, \bibinfo {author} {\bibfnamefont {D.}~\bibnamefont {Akamatsu}}, \bibinfo {author} {\bibfnamefont {H.}~\bibnamefont {\'{A}lvarez Mart\'{i}nez}}, \bibinfo {author} {\bibfnamefont {A.}~\bibnamefont {Amy-Klein} \emph{et~al.}},\ }\bibfield  {title} {\bibinfo {title} {Robust optical
  clocks for international timescales ({ROCIT})},\ }\href {https://doi.org/10.1088/1742-6596/2889/1/012022} {\bibfield  {journal} {\bibinfo  {journal} {J. Phys. Conf. Ser.}\ }\textbf {\bibinfo {volume} {2889}},\ \bibinfo {pages} {012022} (\bibinfo {year} {2024}{\natexlab{a}})}\BibitemShut {NoStop}%
\bibitem [{Note2()}]{Note2}%
  \BibitemOpen
  \bibinfo {note} {T.~Lindvall and A.~E.~Wallin, Zenodo, 2025, \protect \href {http://doi.org/10.5281/zenodo.17070662}{http://doi.org/10.5281/zenodo.17070662}.}\BibitemShut {Stop}%
\bibitem [{\citenamefont {Denker}\ \emph {et~al.}(2018)\citenamefont {Denker}, \citenamefont {Timmen}, \citenamefont {Voigt}, \citenamefont {Weyers}, \citenamefont {Peik}, \citenamefont {Margolis}, \citenamefont {Delva}, \citenamefont {Wolf},\ and\ \citenamefont {Petit}}]{Denker2018a}%
  \BibitemOpen
  \bibfield  {author} {\bibinfo {author} {\bibfnamefont {H.}~\bibnamefont {Denker}}, \bibinfo {author} {\bibfnamefont {L.}~\bibnamefont {Timmen}}, \bibinfo {author} {\bibfnamefont {C.}~\bibnamefont {Voigt}}, \bibinfo {author} {\bibfnamefont {S.}~\bibnamefont {Weyers}}, \bibinfo {author} {\bibfnamefont {E.}~\bibnamefont {Peik}}, \bibinfo {author} {\bibfnamefont {H.~S.}\ \bibnamefont {Margolis}}, \bibinfo {author} {\bibfnamefont {P.}~\bibnamefont {Delva}}, \bibinfo {author} {\bibfnamefont {P.}~\bibnamefont {Wolf}},\ and\ \bibinfo {author} {\bibfnamefont {G.}~\bibnamefont {Petit}},\ }\bibfield  {title} {\bibinfo {title} {Geodetic methods to determine the relativistic redshift at the level of $10^{-18}$ in the context of international timescales: a review and practical results},\ }\href {https://doi.org/10.1007/s00190-017-1075-1} {\bibfield  {journal} {\bibinfo  {journal} {J. Geod.}\ }\textbf {\bibinfo {volume} {92}},\ \bibinfo {pages} {487} (\bibinfo {year} {2018})}\BibitemShut {NoStop}%
\bibitem [{\citenamefont {Hachisu}\ \emph {et~al.}(2017)\citenamefont {Hachisu}, \citenamefont {Petit}, \citenamefont {Nakagawa}, \citenamefont {Hanado},\ and\ \citenamefont {Ido}}]{Hachisu2017b}%
  \BibitemOpen
  \bibfield  {author} {\bibinfo {author} {\bibfnamefont {H.}~\bibnamefont {Hachisu}}, \bibinfo {author} {\bibfnamefont {G.}~\bibnamefont {Petit}}, \bibinfo {author} {\bibfnamefont {F.}~\bibnamefont {Nakagawa}}, \bibinfo {author} {\bibfnamefont {Y.}~\bibnamefont {Hanado}},\ and\ \bibinfo {author} {\bibfnamefont {T.}~\bibnamefont {Ido}},\ }\bibfield  {title} {\bibinfo {title} {{SI}-traceable measurement of an optical frequency at the low $10^{-16}$ level without a local primary standard},\ }\href {https://doi.org/10.1364/OE.25.008511} {\bibfield  {journal} {\bibinfo  {journal} {Opt. Express}\ }\textbf {\bibinfo {volume} {25}},\ \bibinfo {pages} {8511} (\bibinfo {year} {2017})}\BibitemShut {NoStop}%
\bibitem [{\citenamefont {Pizzocaro}\ \emph {et~al.}(2020)\citenamefont {Pizzocaro}, \citenamefont {Bregolin}, \citenamefont {Barbieri}, \citenamefont {Rauf}, \citenamefont {Levi},\ and\ \citenamefont {Calonico}}]{Pizzocaro2020a}%
  \BibitemOpen
  \bibfield  {author} {\bibinfo {author} {\bibfnamefont {M.}~\bibnamefont {Pizzocaro}}, \bibinfo {author} {\bibfnamefont {F.}~\bibnamefont {Bregolin}}, \bibinfo {author} {\bibfnamefont {P.}~\bibnamefont {Barbieri}}, \bibinfo {author} {\bibfnamefont {B.}~\bibnamefont {Rauf}}, \bibinfo {author} {\bibfnamefont {F.}~\bibnamefont {Levi}},\ and\ \bibinfo {author} {\bibfnamefont {D.}~\bibnamefont {Calonico}},\ }\bibfield  {title} {\bibinfo {title} {Absolute frequency measurement of the $^1${S}$_0${\textendash}$^3${P}$_0$ transition of $^{171}${Yb} with a link to international atomic time},\ }\href {https://doi.org/10.1088/1681-7575/ab50e8} {\bibfield  {journal} {\bibinfo  {journal} {Metrologia}\ }\textbf {\bibinfo {volume} {57}},\ \bibinfo {pages} {035007} (\bibinfo {year} {2020})}\BibitemShut {NoStop}%
\bibitem [{Note3()}]{Note3}%
  \BibitemOpen
  \bibinfo {note} {Circular T is a monthly publication by the Bureau International des Poids et Mesures (BIPM), available at \protect \href {https://www.bipm.org/en/time-ftp/circular-t}{https://www.bipm.org/en/time-ftp/circular-t}.}\BibitemShut {Stop}%
\bibitem [{\citenamefont {Dawkins}\ \emph {et~al.}(2007)\citenamefont {Dawkins}, \citenamefont {McFerran},\ and\ \citenamefont {Luiten}}]{Dawkins2007a}%
  \BibitemOpen
  \bibfield  {author} {\bibinfo {author} {\bibfnamefont {S.~T.}\ \bibnamefont {Dawkins}}, \bibinfo {author} {\bibfnamefont {J.~J.}\ \bibnamefont {McFerran}},\ and\ \bibinfo {author} {\bibfnamefont {A.~N.}\ \bibnamefont {Luiten}},\ }\bibfield  {title} {\bibinfo {title} {{C}onsiderations on the measurement of the stability of oscillators with frequency counters},\ }\href {https://doi.org/10.1109/TUFFC.2007.337} {\bibfield  {journal} {\bibinfo  {journal} {{IEEE} Trans. Ultrason., Ferroelectr., Freq. Control}\ }\textbf {\bibinfo {volume} {54}},\ \bibinfo {pages} {918} (\bibinfo {year} {2007})}\BibitemShut {NoStop}%
\bibitem [{Note4()}]{Note4}%
  \BibitemOpen
  \bibinfo {note} {Files reporting the fractional frequency of EAL per Circular T month. Also list $u_\protect \mathrm {A}$, $u_\protect \mathrm {B}$, and weights of all contributing PSFS. Available at \protect \href {https://webtai.bipm.org/ftp/pub/tai/other-products/etoile/}{https://webtai.bipm.org/ftp/pub/tai/other-products/etoile/}}\BibitemShut {NoStop}%
\bibitem [{\citenamefont {Panfilo}\ and\ \citenamefont {Parker}(2010)}]{Panfilo2010a}%
  \BibitemOpen
  \bibfield  {author} {\bibinfo {author} {\bibfnamefont {G.}~\bibnamefont {Panfilo}}\ and\ \bibinfo {author} {\bibfnamefont {T.~E.}\ \bibnamefont {Parker}},\ }\bibfield  {title} {\bibinfo {title} {A theoretical and experimental analysis of frequency transfer uncertainty, including frequency transfer into {TAI}},\ }\href {https://doi.org/10.1088/0026-1394/47/5/005} {\bibfield  {journal} {\bibinfo  {journal} {Metrologia}\ }\textbf {\bibinfo {volume} {47}},\ \bibinfo {pages} {552} (\bibinfo {year} {2010})}\BibitemShut {NoStop}%
\bibitem [{\citenamefont {Cox}\ \emph {et~al.}(2006)\citenamefont {Cox}, \citenamefont {Ei{\o}}, \citenamefont {Mana},\ and\ \citenamefont {Pennecchi}}]{Cox2006a}%
  \BibitemOpen
  \bibfield  {author} {\bibinfo {author} {\bibfnamefont {M.~G.}\ \bibnamefont {Cox}}, \bibinfo {author} {\bibfnamefont {C.}~\bibnamefont {Ei{\o}}}, \bibinfo {author} {\bibfnamefont {G.}~\bibnamefont {Mana}},\ and\ \bibinfo {author} {\bibfnamefont {F.}~\bibnamefont {Pennecchi}},\ }\bibfield  {title} {\bibinfo {title} {The generalized weighted mean of correlated quantities},\ }\href {https://doi.org/10.1088/0026-1394/43/4/s14} {\bibfield  {journal} {\bibinfo  {journal} {Metrologia}\ }\textbf {\bibinfo {volume} {43}},\ \bibinfo {pages} {S268} (\bibinfo {year} {2006})}\BibitemShut {NoStop}%
\bibitem [{\citenamefont {Margolis}\ \emph {et~al.}(2024{\natexlab{b}})\citenamefont {Margolis}, \citenamefont {Panfilo}, \citenamefont {Petit}, \citenamefont {Oates}, \citenamefont {Ido},\ and\ \citenamefont {Bize}}]{Margolis2024a}%
  \BibitemOpen
  \bibfield  {author} {\bibinfo {author} {\bibfnamefont {H.~S.}\ \bibnamefont {Margolis}}, \bibinfo {author} {\bibfnamefont {G.}~\bibnamefont {Panfilo}}, \bibinfo {author} {\bibfnamefont {G.}~\bibnamefont {Petit}}, \bibinfo {author} {\bibfnamefont {C.}~\bibnamefont {Oates}}, \bibinfo {author} {\bibfnamefont {T.}~\bibnamefont {Ido}},\ and\ \bibinfo {author} {\bibfnamefont {S.}~\bibnamefont {Bize}},\ }\bibfield  {title} {\bibinfo {title} {The {CIPM} list ‘{R}ecommended values of standard frequencies’: 2021 update},\ }\href {https://doi.org/10.1088/1681-7575/ad3afc} {\bibfield  {journal} {\bibinfo  {journal} {Metrologia}\ }\textbf {\bibinfo {volume} {61}},\ \bibinfo {pages} {035005} (\bibinfo {year} {2024}{\natexlab{b}})} \BibitemShut {NoStop}%
\bibitem [{\citenamefont {Margolis}\ \emph {et~al.}(2004)\citenamefont {Margolis}, \citenamefont {Barwood}, \citenamefont {Huang}, \citenamefont {Klein}, \citenamefont {Lea}, \citenamefont {Szymaniec},\ and\ \citenamefont {Gill}}]{Margolis2004a}%
  \BibitemOpen
  \bibfield  {author} {\bibinfo {author} {\bibfnamefont {H.~S.}\ \bibnamefont {Margolis}}, \bibinfo {author} {\bibfnamefont {G.~P.}\ \bibnamefont {Barwood}}, \bibinfo {author} {\bibfnamefont {G.}~\bibnamefont {Huang}}, \bibinfo {author} {\bibfnamefont {H.~A.}\ \bibnamefont {Klein}}, \bibinfo {author} {\bibfnamefont {S.~N.}\ \bibnamefont {Lea}}, \bibinfo {author} {\bibfnamefont {K.}~\bibnamefont {Szymaniec}},\ and\ \bibinfo {author} {\bibfnamefont {P.}~\bibnamefont {Gill}},\ }\bibfield  {title} {\bibinfo {title} {{H}ertz-{L}evel {M}easurement of the {O}ptical {C}lock {F}requency in a {S}ingle $^{88}${S}r$^+$ {I}on},\ }\href {https://doi.org/10.1126/science.1105497} {\bibfield  {journal} {\bibinfo  {journal} {Science}\ }\textbf {\bibinfo {volume} {306}},\ \bibinfo {pages} {1355} (\bibinfo {year} {2004})}\BibitemShut {NoStop}%
\bibitem [{\citenamefont {Madej}\ \emph {et~al.}(2012)\citenamefont {Madej}, \citenamefont {Dub\'e}, \citenamefont {Zhou}, \citenamefont {Bernard},\ and\ \citenamefont {Gertsvolf}}]{Madej2012b}%
  \BibitemOpen
  \bibfield  {author} {\bibinfo {author} {\bibfnamefont {A.~A.}\ \bibnamefont {Madej}}, \bibinfo {author} {\bibfnamefont {P.}~\bibnamefont {Dub\'e}}, \bibinfo {author} {\bibfnamefont {Z.}~\bibnamefont {Zhou}}, \bibinfo {author} {\bibfnamefont {J.~E.}\ \bibnamefont {Bernard}},\ and\ \bibinfo {author} {\bibfnamefont {M.}~\bibnamefont {Gertsvolf}},\ }\bibfield  {title} {\bibinfo {title} {$^{88}\mathrm{Sr}^{+}$ 445-{TH}z {S}ingle-{I}on {R}eference at the ${10}^{-17}$ {L}evel via {C}ontrol and {C}ancellation of {S}ystematic {U}ncertainties and {I}ts {M}easurement against the {SI} {S}econd},\ }\href {https://doi.org/10.1103/PhysRevLett.109.203002} {\bibfield  {journal} {\bibinfo  {journal} {Phys. Rev. Lett.}\ }\textbf {\bibinfo {volume} {109}},\ \bibinfo {pages} {203002} (\bibinfo {year} {2012})}\BibitemShut {NoStop}%
\bibitem [{\citenamefont {Barwood}\ \emph {et~al.}(2014)\citenamefont {Barwood}, \citenamefont {Huang}, \citenamefont {Klein}, \citenamefont {Johnson}, \citenamefont {King}, \citenamefont {Margolis}, \citenamefont {Szymaniec},\ and\ \citenamefont {Gill}}]{Barwood2014a}%
  \BibitemOpen
  \bibfield  {author} {\bibinfo {author} {\bibfnamefont {G.~P.}\ \bibnamefont {Barwood}}, \bibinfo {author} {\bibfnamefont {G.}~\bibnamefont {Huang}}, \bibinfo {author} {\bibfnamefont {H.~A.}\ \bibnamefont {Klein}}, \bibinfo {author} {\bibfnamefont {L.~A.~M.}\ \bibnamefont {Johnson}}, \bibinfo {author} {\bibfnamefont {S.~A.}\ \bibnamefont {King}}, \bibinfo {author} {\bibfnamefont {H.~S.}\ \bibnamefont {Margolis}}, \bibinfo {author} {\bibfnamefont {K.}~\bibnamefont {Szymaniec}},\ and\ \bibinfo {author} {\bibfnamefont {P.}~\bibnamefont {Gill}},\ }\bibfield  {title} {\bibinfo {title} {{A}greement between two $^{88}\mathrm{Sr}^+$ optical clocks to 4 parts in 10$^{17}$},\ }\href {https://doi.org/10.1103/PhysRevA.89.050501} {\bibfield  {journal} {\bibinfo  {journal} {Phys. Rev. A}\ }\textbf {\bibinfo {volume} {89}},\ \bibinfo {pages} {050501} (\bibinfo {year} {2014})}\BibitemShut {NoStop}%
\bibitem [{\citenamefont {Dub\'e}\ \emph {et~al.}(2017)\citenamefont {Dub\'e}, \citenamefont {Bernard},\ and\ \citenamefont {Gertsvolf}}]{Dube2017a}%
  \BibitemOpen
  \bibfield  {author} {\bibinfo {author} {\bibfnamefont {P.}~\bibnamefont {Dub\'e}}, \bibinfo {author} {\bibfnamefont {J.~E.}\ \bibnamefont {Bernard}},\ and\ \bibinfo {author} {\bibfnamefont {M.}~\bibnamefont {Gertsvolf}},\ }\bibfield  {title} {\bibinfo {title} {{A}bsolute frequency measurement of the $^{88}${S}r$^+$ clock transition using a {GPS} link to the {SI} second},\ }\href {https://doi.org/10.1088/1681-7575/aa5e60} {\bibfield  {journal} {\bibinfo  {journal} {Metrologia}\ }\textbf {\bibinfo {volume} {54}},\ \bibinfo {pages} {290} (\bibinfo {year} {2017})}\BibitemShut {NoStop}%
\bibitem [{\citenamefont {Jiang}\ \emph {et~al.}(2009)\citenamefont {Jiang}, \citenamefont {Arora}, \citenamefont {Safronova},\ and\ \citenamefont {Clark}}]{Jiang2009a}%
  \BibitemOpen
  \bibfield  {author} {\bibinfo {author} {\bibfnamefont {D.}~\bibnamefont {Jiang}}, \bibinfo {author} {\bibfnamefont {B.}~\bibnamefont {Arora}}, \bibinfo {author} {\bibfnamefont {M.~S.}\ \bibnamefont {Safronova}},\ and\ \bibinfo {author} {\bibfnamefont {C.~W.}\ \bibnamefont {Clark}},\ }\bibfield  {title} {\bibinfo {title} {{B}lackbody-radiation shift in a $^{88}${S}r$^+$ ion optical frequency standard},\ }\href {https://doi.org/10.1088/0953-4075/42/15/154020} {\bibfield  {journal} {\bibinfo  {journal} {J. Phys. B: At. Mol. Opt. Phys.}\ }\textbf {\bibinfo {volume} {42}},\ \bibinfo {pages} {154020} (\bibinfo {year} {2009})}\BibitemShut {NoStop}%
\bibitem [{\citenamefont {Nemitz}\ \emph {et~al.}(2024)\citenamefont {Nemitz}, \citenamefont {Hachisu}, \citenamefont {Ito}, \citenamefont {Ohtsubo}, \citenamefont {Miyauchi}, \citenamefont {Morikawa}, \citenamefont {Matsubara},\ and\ \citenamefont {Ido}}]{Nemitz2024a}%
  \BibitemOpen
  \bibfield  {author} {\bibinfo {author} {\bibfnamefont {N.}~\bibnamefont {Nemitz}}, \bibinfo {author} {\bibfnamefont {H.}~\bibnamefont {Hachisu}}, \bibinfo {author} {\bibfnamefont {H.}~\bibnamefont {Ito}}, \bibinfo {author} {\bibfnamefont {N.}~\bibnamefont {Ohtsubo}}, \bibinfo {author} {\bibfnamefont {Y.}~\bibnamefont {Miyauchi}}, \bibinfo {author} {\bibfnamefont {M.}~\bibnamefont {Morikawa}}, \bibinfo {author} {\bibfnamefont {K.}~\bibnamefont {Matsubara}},\ and\ \bibinfo {author} {\bibfnamefont {T.}~\bibnamefont {Ido}},\ }\bibfield  {title} {\bibinfo {title} {Hydrogen maser flywheels for optical clocks},\ }\href {https://doi.org/10.1088/1742-6596/2889/1/012019} {\bibfield  {journal} {\bibinfo  {journal} {J. Phys. Conf. Ser.}\ }\textbf {\bibinfo {volume} {2889}},\ \bibinfo {pages} {012019} (\bibinfo {year} {2024})}\BibitemShut {NoStop}%
\bibitem [{\citenamefont {Dick}(1987)}]{Dick1987a}%
  \BibitemOpen
  \bibfield  {author} {\bibinfo {author} {\bibfnamefont {G.~J.}\ \bibnamefont {Dick}},\ }\bibfield  {title} {\bibinfo {title} {{L}ocal oscillator induced instabilities in trapped ion frequency standards},\ }in\ \href {https://apps.dtic.mil/sti/tr/pdf/ADA502386.pdf} {\emph {\bibinfo {booktitle} {Proceedings of the Nineteenth Annual Precise Time and Time Interval (PTTI) Applications and Planning Meeting}}}\ (\bibinfo {address} {Redondo Beach, CA},\ \bibinfo {year} {1987})\ pp.\ \bibinfo {pages} {133}\BibitemShut {NoStop}%
\bibitem [{\citenamefont {Vest{\o}l}\ \emph {et~al.}(2019)\citenamefont {Vest{\o}l}, \citenamefont {{{\AA}gren}}, \citenamefont {Steffen}, \citenamefont {Kierulf},\ and\ \citenamefont {Tarasov}}]{Vestol2019a}%
  \BibitemOpen
  \bibfield  {author} {\bibinfo {author} {\bibfnamefont {O.}~\bibnamefont {Vest{\o}l}}, \bibinfo {author} {\bibfnamefont {J.}~\bibnamefont {{{\AA}gren}}}, \bibinfo {author} {\bibfnamefont {H.}~\bibnamefont {Steffen}}, \bibinfo {author} {\bibfnamefont {H.}~\bibnamefont {Kierulf}},\ and\ \bibinfo {author} {\bibfnamefont {L.}~\bibnamefont {Tarasov}},\ }\bibfield  {title} {\bibinfo {title} {{NKG}2016{LU}: a new land uplift model for {F}ennoscandia and the {B}altic region},\ }\href {https://doi.org/10.1007/s00190-019-01280-8} {\bibfield  {journal} {\bibinfo  {journal} {J. Geod.}\ }\textbf {\bibinfo {volume} {93}},\ \bibinfo {pages} {1759} (\bibinfo {year} {2019})}\BibitemShut {NoStop}%
\bibitem [{\citenamefont {H\"akli}\ \emph {et~al.}(2023)\citenamefont {H\"akli}, \citenamefont {Evers}, \citenamefont {Jivall}, \citenamefont {Nilsson}, \citenamefont {Himle}, \citenamefont {Kollo}, \citenamefont {Liepi{\c{n}}{\v{s}}}, \citenamefont {Par{\v{s}}eli{\={u}}nas}, \citenamefont {Vest{\o}l},\ and\ \citenamefont {Lidberg}}]{Hakli2023a}%
  \BibitemOpen
  \bibfield  {author} {\bibinfo {author} {\bibfnamefont {P.}~\bibnamefont {H\"akli}}, \bibinfo {author} {\bibfnamefont {K.}~\bibnamefont {Evers}}, \bibinfo {author} {\bibfnamefont {L.}~\bibnamefont {Jivall}}, \bibinfo {author} {\bibfnamefont {T.}~\bibnamefont {Nilsson}}, \bibinfo {author} {\bibfnamefont {S.}~\bibnamefont {Himle}}, \bibinfo {author} {\bibfnamefont {K.}~\bibnamefont {Kollo}}, \bibinfo {author} {\bibfnamefont {I.}~\bibnamefont {Liepi{\c{n}}{\v{s}}}}, \bibinfo {author} {\bibfnamefont {E.}~\bibnamefont {Par{\v{s}}eli{\={u}}nas}}, \bibinfo {author} {\bibfnamefont {O.}~\bibnamefont {Vest{\o}l}},\ and\ \bibinfo {author} {\bibfnamefont {M.}~\bibnamefont {Lidberg}},\ }\bibfield  {title} {\bibinfo {title} {{NKG}2020 transformation: An updated transformation between dynamic and static reference frames in the {N}ordic and {B}altic countries},\ }\href {https://doi.org/10.1515/jogs-2022-0155} {\bibfield  {journal} {\bibinfo  {journal} {J. Geod. Sci.}\ }\textbf {\bibinfo {volume} {13}},\ \bibinfo {pages}
  {20220155} (\bibinfo {year} {2023})}\BibitemShut {NoStop}%
\bibitem [{\citenamefont {Blinov}\ \emph {et~al.}(2017)\citenamefont {Blinov}, \citenamefont {Boiko}, \citenamefont {Domnin}, \citenamefont {Kostromin}, \citenamefont {Kupalova},\ and\ \citenamefont {Kupalov}}]{Blinov2017a}%
  \BibitemOpen
  \bibfield  {author} {\bibinfo {author} {\bibfnamefont {I.~Y.}\ \bibnamefont {Blinov}}, \bibinfo {author} {\bibfnamefont {A.~I.}\ \bibnamefont {Boiko}}, \bibinfo {author} {\bibfnamefont {Y.~S.}\ \bibnamefont {Domnin}}, \bibinfo {author} {\bibfnamefont {V.~P.}\ \bibnamefont {Kostromin}}, \bibinfo {author} {\bibfnamefont {O.~V.}\ \bibnamefont {Kupalova}},\ and\ \bibinfo {author} {\bibfnamefont {D.~S.}\ \bibnamefont {Kupalov}},\ }\bibfield  {title} {\bibinfo {title} {Budget of uncertainties in the cesium frequency frame of fountain type},\ }\href {https://doi.org/10.1007/s11018-017-1145-z} {\bibfield  {journal} {\bibinfo  {journal} {Meas. Tech.}\ }\textbf {\bibinfo {volume} {60}},\ \bibinfo {pages} {30} (\bibinfo {year} {2017})}\BibitemShut {NoStop}%
\bibitem [{\citenamefont {Beattie}\ \emph {et~al.}(2020)\citenamefont {Beattie}, \citenamefont {Jian}, \citenamefont {Alcock}, \citenamefont {Gertsvolf}, \citenamefont {Hendricks}, \citenamefont {Szymaniec},\ and\ \citenamefont {Gibble}}]{Beattie2020a}%
  \BibitemOpen
  \bibfield  {author} {\bibinfo {author} {\bibfnamefont {S.}~\bibnamefont {Beattie}}, \bibinfo {author} {\bibfnamefont {B.}~\bibnamefont {Jian}}, \bibinfo {author} {\bibfnamefont {J.}~\bibnamefont {Alcock}}, \bibinfo {author} {\bibfnamefont {M.}~\bibnamefont {Gertsvolf}}, \bibinfo {author} {\bibfnamefont {R.}~\bibnamefont {Hendricks}}, \bibinfo {author} {\bibfnamefont {K.}~\bibnamefont {Szymaniec}},\ and\ \bibinfo {author} {\bibfnamefont {K.}~\bibnamefont {Gibble}},\ }\bibfield  {title} {\bibinfo {title} {First accuracy evaluation of the {NRC}-{FCs}2 primary frequency standard},\ }\href {https://doi.org/10.1088/1681-7575/ab7c54} {\bibfield  {journal} {\bibinfo  {journal} {Metrologia}\ }\textbf {\bibinfo {volume} {57}},\ \bibinfo {pages} {035010} (\bibinfo {year} {2020})}\BibitemShut {NoStop}%
\bibitem [{\citenamefont {Gu\'ena}\ \emph {et~al.}(2012)\citenamefont {Gu\'ena}, \citenamefont {Abgrall}, \citenamefont {Rovera}, \citenamefont {Laurent}, \citenamefont {Chupin}, \citenamefont {Lours}, \citenamefont {Santarelli}, \citenamefont {Rosenbusch}, \citenamefont {Tobar}, \citenamefont {Li}, \citenamefont {Gibble}, \citenamefont {Clairon},\ and\ \citenamefont {Bize}}]{Guena2012a}%
  \BibitemOpen
  \bibfield  {author} {\bibinfo {author} {\bibfnamefont {J.}~\bibnamefont {Gu\'ena}}, \bibinfo {author} {\bibfnamefont {M.}~\bibnamefont {Abgrall}}, \bibinfo {author} {\bibfnamefont {D.}~\bibnamefont {Rovera}}, \bibinfo {author} {\bibfnamefont {P.}~\bibnamefont {Laurent}}, \bibinfo {author} {\bibfnamefont {B.}~\bibnamefont {Chupin}}, \bibinfo {author} {\bibfnamefont {M.}~\bibnamefont {Lours}}, \bibinfo {author} {\bibfnamefont {G.}~\bibnamefont {Santarelli}}, \bibinfo {author} {\bibfnamefont {P.}~\bibnamefont {Rosenbusch}}, \bibinfo {author} {\bibfnamefont {M.}~\bibnamefont {Tobar}}, \bibinfo {author} {\bibfnamefont {R.}~\bibnamefont {Li}}, \bibinfo {author} {\bibfnamefont {K.}~\bibnamefont {Gibble}}, \bibinfo {author} {\bibfnamefont {A.}~\bibnamefont {Clairon}},\ and\ \bibinfo {author} {\bibfnamefont {S.}~\bibnamefont {Bize}},\ }\bibfield  {title} {\bibinfo {title} {{P}rogress in atomic fountains at {LNE}-{SYRTE}},\ }\href {https://doi.org/10.1109/TUFFC.2012.2208} {\bibfield  {journal} {\bibinfo  {journal}
  {{IEEE} Trans. Ultrason., Ferroelectr., Freq. Control}\ }\textbf {\bibinfo {volume} {59}},\ \bibinfo {pages} {391} (\bibinfo {year} {2012})}\BibitemShut {NoStop}%
\bibitem [{\citenamefont {Gu\'ena}\ \emph {et~al.}(2014)\citenamefont {Gu\'ena}, \citenamefont {Abgrall}, \citenamefont {Clairon},\ and\ \citenamefont {Bize}}]{Guena2014a}%
  \BibitemOpen
  \bibfield  {author} {\bibinfo {author} {\bibfnamefont {J.}~\bibnamefont {Gu\'ena}}, \bibinfo {author} {\bibfnamefont {M.}~\bibnamefont {Abgrall}}, \bibinfo {author} {\bibfnamefont {A.}~\bibnamefont {Clairon}},\ and\ \bibinfo {author} {\bibfnamefont {S.}~\bibnamefont {Bize}},\ }\bibfield  {title} {\bibinfo {title} {{C}ontributing to {TAI} with a secondary representation of the {SI} second},\ }\href {https://doi.org/10.1088/0026-1394/51/1/108} {\bibfield  {journal} {\bibinfo  {journal} {Metrologia}\ }\textbf {\bibinfo {volume} {51}},\ \bibinfo {pages} {108} (\bibinfo {year} {2014})}\BibitemShut {NoStop}%
\bibitem [{\citenamefont {Wang}\ \emph {et~al.}(2023)\citenamefont {Wang}, \citenamefont {Ruan}, \citenamefont {Liu}, \citenamefont {Guan}, \citenamefont {Shi}, \citenamefont {Yang}, \citenamefont {Bai}, \citenamefont {Zhang}, \citenamefont {Fan}, \citenamefont {Wu}, \citenamefont {Zhao},\ and\ \citenamefont {Zhang}}]{Wang2023a}%
  \BibitemOpen
  \bibfield  {author} {\bibinfo {author} {\bibfnamefont {X.-L.}\ \bibnamefont {Wang}}, \bibinfo {author} {\bibfnamefont {J.}~\bibnamefont {Ruan}}, \bibinfo {author} {\bibfnamefont {D.-D.}\ \bibnamefont {Liu}}, \bibinfo {author} {\bibfnamefont {Y.}~\bibnamefont {Guan}}, \bibinfo {author} {\bibfnamefont {J.-R.}\ \bibnamefont {Shi}}, \bibinfo {author} {\bibfnamefont {F.}~\bibnamefont {Yang}}, \bibinfo {author} {\bibfnamefont {Y.}~\bibnamefont {Bai}}, \bibinfo {author} {\bibfnamefont {H.}~\bibnamefont {Zhang}}, \bibinfo {author} {\bibfnamefont {S.-C.}\ \bibnamefont {Fan}}, \bibinfo {author} {\bibfnamefont {W.-J.}\ \bibnamefont {Wu}}, \bibinfo {author} {\bibfnamefont {S.-H.}\ \bibnamefont {Zhao}},\ and\ \bibinfo {author} {\bibfnamefont {S.-G.}\ \bibnamefont {Zhang}},\ }\bibfield  {title} {\bibinfo {title} {First evaluation of the primary frequency standard {NTSC-CsF2}},\ }\href {https://doi.org/10.1088/1681-7575/ad023e} {\bibfield  {journal} {\bibinfo  {journal} {Metrologia}\ }\textbf {\bibinfo {volume} {60}},\
  \bibinfo {pages} {065012} (\bibinfo {year} {2023})}\BibitemShut {NoStop}%
\bibitem [{\citenamefont {Levi}\ \emph {et~al.}(2014)\citenamefont {Levi}, \citenamefont {Calonico}, \citenamefont {Calosso}, \citenamefont {Godone}, \citenamefont {Micalizio},\ and\ \citenamefont {Costanzo}}]{Levi2014a}%
  \BibitemOpen
  \bibfield  {author} {\bibinfo {author} {\bibfnamefont {F.}~\bibnamefont {Levi}}, \bibinfo {author} {\bibfnamefont {D.}~\bibnamefont {Calonico}}, \bibinfo {author} {\bibfnamefont {C.~E.}\ \bibnamefont {Calosso}}, \bibinfo {author} {\bibfnamefont {A.}~\bibnamefont {Godone}}, \bibinfo {author} {\bibfnamefont {S.}~\bibnamefont {Micalizio}},\ and\ \bibinfo {author} {\bibfnamefont {G.~A.}\ \bibnamefont {Costanzo}},\ }\bibfield  {title} {\bibinfo {title} {{A}ccuracy evaluation of {ITC}s{F}2: a nitrogen cooled caesium fountain},\ }\href {http://stacks.iop.org/0026-1394/51/i=3/a=270} {\bibfield  {journal} {\bibinfo  {journal} {Metrologia}\ }\textbf {\bibinfo {volume} {51}},\ \bibinfo {pages} {270} (\bibinfo {year} {2014})}\BibitemShut {NoStop}%
\bibitem [{\citenamefont {Fang}\ \emph {et~al.}(2015)\citenamefont {Fang}, \citenamefont {Li}, \citenamefont {Lin}, \citenamefont {Chen}, \citenamefont {Liu}, \citenamefont {Lin}, \citenamefont {Wang}, \citenamefont {Liu}, \citenamefont {Suo},\ and\ \citenamefont {Li}}]{Fang2015a}%
  \BibitemOpen
  \bibfield  {author} {\bibinfo {author} {\bibfnamefont {F.}~\bibnamefont {Fang}}, \bibinfo {author} {\bibfnamefont {M.}~\bibnamefont {Li}}, \bibinfo {author} {\bibfnamefont {P.}~\bibnamefont {Lin}}, \bibinfo {author} {\bibfnamefont {W.}~\bibnamefont {Chen}}, \bibinfo {author} {\bibfnamefont {N.}~\bibnamefont {Liu}}, \bibinfo {author} {\bibfnamefont {Y.}~\bibnamefont {Lin}}, \bibinfo {author} {\bibfnamefont {P.}~\bibnamefont {Wang}}, \bibinfo {author} {\bibfnamefont {K.}~\bibnamefont {Liu}}, \bibinfo {author} {\bibfnamefont {R.}~\bibnamefont {Suo}},\ and\ \bibinfo {author} {\bibfnamefont {T.}~\bibnamefont {Li}},\ }\bibfield  {title} {\bibinfo {title} {{NIM5} {Cs} fountain clock and its evaluation},\ }\href {https://doi.org/10.1088/0026-1394/52/4/454} {\bibfield  {journal} {\bibinfo  {journal} {Metrologia}\ }\textbf {\bibinfo {volume} {52}},\ \bibinfo {pages} {454} (\bibinfo {year} {2015})}\BibitemShut {NoStop}%
\bibitem [{\citenamefont {Hobson}\ \emph {et~al.}(2020)\citenamefont {Hobson}, \citenamefont {Bowden}, \citenamefont {Vianello}, \citenamefont {Silva}, \citenamefont {Baynham}, \citenamefont {Margolis}, \citenamefont {Baird}, \citenamefont {Gill},\ and\ \citenamefont {Hill}}]{Hobson2020b}%
  \BibitemOpen
  \bibfield  {author} {\bibinfo {author} {\bibfnamefont {R.}~\bibnamefont {Hobson}}, \bibinfo {author} {\bibfnamefont {W.}~\bibnamefont {Bowden}}, \bibinfo {author} {\bibfnamefont {A.}~\bibnamefont {Vianello}}, \bibinfo {author} {\bibfnamefont {A.}~\bibnamefont {Silva}}, \bibinfo {author} {\bibfnamefont {C.~F.~A.}\ \bibnamefont {Baynham}}, \bibinfo {author} {\bibfnamefont {H.~S.}\ \bibnamefont {Margolis}}, \bibinfo {author} {\bibfnamefont {P.~E.~G.}\ \bibnamefont {Baird}}, \bibinfo {author} {\bibfnamefont {P.}~\bibnamefont {Gill}},\ and\ \bibinfo {author} {\bibfnamefont {I.~R.}\ \bibnamefont {Hill}},\ }\bibfield  {title} {\bibinfo {title} {A strontium optical lattice clock with $1 \times 10^ {-17}$ uncertainty and measurement of its absolute frequency},\ }\href {https://doi.org/10.1088/1681-7575/abb530} {\bibfield  {journal} {\bibinfo  {journal} {Metrologia}\ }\textbf {\bibinfo {volume} {57}},\ \bibinfo {pages} {065026} (\bibinfo {year} {2020})}\BibitemShut {NoStop}%
\bibitem [{\citenamefont {Jallageas}\ \emph {et~al.}(2018)\citenamefont {Jallageas}, \citenamefont {Devenoges}, \citenamefont {Petersen}, \citenamefont {Morel}, \citenamefont {Bernier}, \citenamefont {Schenker}, \citenamefont {Thomann},\ and\ \citenamefont {Südmeyer}}]{Jallageas2018a}%
  \BibitemOpen
  \bibfield  {author} {\bibinfo {author} {\bibfnamefont {A.}~\bibnamefont {Jallageas}}, \bibinfo {author} {\bibfnamefont {L.}~\bibnamefont {Devenoges}}, \bibinfo {author} {\bibfnamefont {M.}~\bibnamefont {Petersen}}, \bibinfo {author} {\bibfnamefont {J.}~\bibnamefont {Morel}}, \bibinfo {author} {\bibfnamefont {L.~G.}\ \bibnamefont {Bernier}}, \bibinfo {author} {\bibfnamefont {D.}~\bibnamefont {Schenker}}, \bibinfo {author} {\bibfnamefont {P.}~\bibnamefont {Thomann}},\ and\ \bibinfo {author} {\bibfnamefont {T.}~\bibnamefont {Südmeyer}},\ }\bibfield  {title} {\bibinfo {title} {First uncertainty evaluation of the {FoCS-2} primary frequency standard},\ }\href {https://doi.org/10.1088/1681-7575/aab3fa} {\bibfield  {journal} {\bibinfo  {journal} {Metrologia}\ }\textbf {\bibinfo {volume} {55}},\ \bibinfo {pages} {366} (\bibinfo {year} {2018})}\BibitemShut {NoStop}%
\end{thebibliography}
\end{document}